\documentclass[a4paper,11pt]{article}
\usepackage{jheppub}
\usepackage[T1]{fontenc}
\usepackage[utf8]{inputenc}
\usepackage{slashed}
\usepackage{subcaption}

\usepackage{todonotes}
\usepackage{hyperref}
\usepackage{bm}

\usepackage{tikz,environ}

\usepackage{tocloft}

\usetikzlibrary{snakes}
\usetikzlibrary{decorations}
\usetikzlibrary{trees}
\usetikzlibrary{decorations.pathmorphing}
\usetikzlibrary{decorations.markings}
\usetikzlibrary{external}
\usetikzlibrary{intersections}
\usetikzlibrary{shapes,arrows}
\usetikzlibrary{arrows.meta}
\usetikzlibrary{calc}
\usetikzlibrary{shapes.misc}
\usetikzlibrary{decorations.text}
\usetikzlibrary{backgrounds}
\usetikzlibrary{fadings}
\usetikzlibrary{tikzmark,calc,arrows,shapes,decorations.pathreplacing}

\tikzset{
	graviton/.style={decorate,line width=0.1mm, decoration={snake,amplitude=.3mm, segment length=0.8mm}},
	photon/.style={decorate, decoration={snake}, draw=red},
	scalar/.style={postaction={decorate},
	},
	massive/.style={postaction={decorate},
		line width=0.75mm,
	},
	massless/.style={postaction={decorate},
	},
	masslessWithDot/.style={postaction={decorate},
		decoration={
			markings,
			mark=at position 0.5 with {\fill circle (2pt);}}
	},
	massiveWithDot/.style={postaction={decorate},
		line width=0.5mm,
		decoration={
			markings,
			mark=at position 0.5 with {\fill circle (2pt);}}
	},
	massiveWithArrow/.style={postaction={decorate},
		line width=0.75mm,
		decoration={
			markings,
			mark=at position 0.5 with {\arrow{latex}}}
	},
	massiveLin/.style={postaction={decorate},
		double,
		thick,
		fill=white
	},
	massivePhi/.style={postaction={decorate},
		line width=0.75mm,
		dashed
	},
	masslessPhi/.style={postaction={decorate},
		dashed
	},
	unitaryCut/.style={postaction={draw,densely dashed,blue,thin},
		line width = 0.2cm,white
	},
	gluon/.style={decorate, draw=magenta,
		decoration={coil,amplitude=4pt, segment length=5pt}},
	partial ellipse/.style args={#1:#2:#3}{
		insert path={+ (#1:#3) arc (#1:#2:#3)}
	},
	cross/.style={cross out, draw=black, minimum size=2*(#1-\pgflinewidth), inner sep=0pt, outer sep=0pt},
	branchCut/.style={postaction={decorate},
		snake=zigzag,
		decoration = {snake=zigzag,segment length = 2mm, amplitude = 2mm}	
	}
	cross/.default={1pt}
}

\colorlet{mred}{black!30!red}
\colorlet{mgreen}{black!30!green}
\colorlet{mblue}{black!30!blue}
\colorlet{morange}{blue!70!red}

\newcommand{\treet}{
\begin{tikzpicture}[scale=1]
		\coordinate (e1) at (-0.5,0.);
		\coordinate (e2) at (.5,0.);
		\coordinate (e3) at (.5,-1.);
		\coordinate (e4) at (-0.5,-1.);	
		
		\coordinate (v1) at (0,0);
		\coordinate (v2) at (0,-1.);
		
		\draw[massive] (e1) -- (v1) -- (e2) ;
		\draw[massive] (e3) -- (v2) -- (e4) ;
		
		\draw[massless] (v1) -- (v2) ;

        \node[below left]  at (e4) {$p_1$};
        \node[above left]  at (e1) {$p_2$};
        \node[above right] at (e2) {$p_3$};
        \node[below right] at (e3) {$p_4$};
\end{tikzpicture}}

\newcommand{\btriOF}{
\begin{tikzpicture}[scale=1]
		\coordinate (e1) at (-0.5,0.);
		\coordinate (e2) at (1.5,0.);
		\coordinate (e3) at (1.5,-1.);
		\coordinate (e4) at (-0.5,-1.);	
		
		\coordinate (v1) at (0,0);
		\coordinate (v2) at (1,0);
		\coordinate (v3) at (.5,-1);
		
		\draw[massive] (e1) -- (v1) ;
		\draw[massive] (v1) -- (v2) ;
		\draw[massive] (v2) -- (e2) ;
		\draw[massive] (e4) -- (v3) ;
		\draw[massive] (v3) -- (e3) ;
		
		\draw[massless] (v1) -- (v3) ;
		\draw[massless] (v2) -- (v3) ;

        \node[below left]  at (e4) {$p_1$};
        \node[above left]  at (e1) {$p_2$};
        \node[above right] at (e2) {$p_3$};
        \node[below right] at (e3) {$p_4$};
\end{tikzpicture}}

\newcommand{\lintriOF}{
\begin{tikzpicture}[scale=1]
		\coordinate (e1) at (-0.5,0.);
		\coordinate (e2) at (1.5,0.);
		\coordinate (e3) at (1.5,-1.);
		\coordinate (e4) at (-0.5,-1.);	
		
		\coordinate (v1) at (0,0);
		\coordinate (v2) at (1,0);
		\coordinate (v3) at (.5,-1);
		
		\draw[massiveLin] (e1) -- (v1) ;
		\draw[massiveLin] (v1) -- (v2) ;
		\draw[massiveLin] (v2) -- (e2) ;
		\draw[massiveLin] (e4) -- (v3) ;
		\draw[massiveLin] (v3) -- (e3) ;
		
		\draw[massless] (v1) -- (v3) ;
		\draw[massless] (v2) -- (v3) ;

        \node[below left]  at (e4) {$p_1$};
        \node[above left]  at (e1) {$p_2$};
        \node[above right] at (e2) {$p_3$};
        \node[below right] at (e3) {$p_4$};
\end{tikzpicture}}

\newcommand{\btriTT}{
\begin{tikzpicture}[scale=1]
		\coordinate (e1) at (-0.5,0.);
		\coordinate (e2) at (1.5,0.);
		\coordinate (e3) at (1.5,-1.);
		\coordinate (e4) at (-0.5,-1.);	
		
		\coordinate (v1) at (.5,0);
		\coordinate (v3) at (0,-1);
		\coordinate (v4) at (1,-1);
		
		\draw[massive] (e1) -- (v1) ;
		\draw[massive] (v1) -- (e2) ;
		\draw[massive] (e4) -- (v3) ;
		\draw[massive] (v3) -- (v4) ;
		\draw[massive] (v4) -- (e3) ;
		
		\draw[massless] (v1) -- (v3) ;
		\draw[massless] (v1) -- (v4) ;

        \node[below left]  at (e4) {$p_1$};
        \node[above left]  at (e1) {$p_2$};
        \node[above right] at (e2) {$p_3$};
        \node[below right] at (e3) {$p_4$};
\end{tikzpicture}}

\newcommand{\lintriTT}{
\begin{tikzpicture}[scale=1]
		\coordinate (e1) at (-0.5,0.);
		\coordinate (e2) at (1.5,0.);
		\coordinate (e3) at (1.5,-1.);
		\coordinate (e4) at (-0.5,-1.);	
		
		\coordinate (v1) at (.5,0);
		\coordinate (v3) at (0,-1);
		\coordinate (v4) at (1,-1);
		
		\draw[massiveLin] (e1) -- (v1) ;
		\draw[massiveLin] (v1) -- (e2) ;
		\draw[massiveLin] (e4) -- (v3) ;
		\draw[massiveLin] (v3) -- (v4) ;
		\draw[massiveLin] (v4) -- (e3) ;
		
		\draw[massless] (v1) -- (v3) ;
		\draw[massless] (v1) -- (v4) ;

        \node[below left]  at (e4) {$p_1$};
        \node[above left]  at (e1) {$p_2$};
        \node[above right] at (e2) {$p_3$};
        \node[below right] at (e3) {$p_4$};
\end{tikzpicture}}

\newcommand{\lintriTTNoLab}{
\begin{tikzpicture}[scale=1]
		\coordinate (e1) at (-0.5,0.);
		\coordinate (e2) at (1.5,0.);
		\coordinate (e3) at (1.5,-1.);
		\coordinate (e4) at (-0.5,-1.);	
		
		\coordinate (v1) at (.5,0);
		\coordinate (v3) at (0,-1);
		\coordinate (v4) at (1,-1);
		
		\draw[massiveLin] (e1) -- (v1) ;
		\draw[massiveLin] (v1) -- (e2) ;
		\draw[massiveLin] (e4) -- (v3) ;
		\draw[massiveLin] (v3) -- (v4) ;
		\draw[massiveLin] (v4) -- (e3) ;
		
		\draw[massless] (v1) -- (v3) ;
		\draw[massless] (v1) -- (v4) ;

        \node[below left]  at (e4) {$\, $};
        \node[above left]  at (e1) {$\, $};
        \node[above right] at (e2) {$\, $};
        \node[below right] at (e3) {$\, $};
\end{tikzpicture}}

\newcommand{\bubbleWithDot}{
\begin{tikzpicture}[scale=1]
		\coordinate (e1) at (0.5,0.);
		\coordinate (e2) at (1.5,0.);
		\coordinate (e3) at (0.5,-1.);
		\coordinate (e4) at (1.5,-1.);	
		
		\coordinate (v1) at (1,0);

		\coordinate (v2) at (1,-1);

		\draw[massiveLin] (e1) -- (v1) ;
		\draw[massiveLin] (v1) -- (e2) ;
		\draw[massiveLin] (e4) -- (v2) ;
		\draw[massiveLin] (v2) -- (e3) ;
		
		\draw[massless] (v1) to[out=0,in=0] (v2) ;
		\draw[masslessWithDot] (v1) to[out=180,in=180] (v2) ;
\end{tikzpicture}}

\newcommand{\bbox}{
\begin{tikzpicture}[scale=1]
		\coordinate (e1) at (-0.5,0.);
		\coordinate (e2) at (1.5,0.);
		\coordinate (e3) at (1.5,-1.);
		\coordinate (e4) at (-0.5,-1.);	
		
		\coordinate (v1) at (0,0);
		\coordinate (v2) at (1,0);
		\coordinate (v3) at (0,-1);
		\coordinate (v4) at (1,-1);
		
		\draw[massive] (e1) -- (v1) ;
		\draw[massive] (v1) -- (v2) ;
		\draw[massive] (v2) -- (e2) ;
		\draw[massive] (e4) -- (v3) ;
		\draw[massive] (v3) -- (v4) ;
		\draw[massive] (v4) -- (e3) ;
		
		\draw[massless] (v1) -- (v3) ;
		\draw[massless] (v2) -- (v4) ;

        \node[below left]  at (e4) {$p_1$};
        \node[above left]  at (e1) {$p_2$};
        \node[above right] at (e2) {$p_3$};
        \node[below right] at (e3) {$p_4$};
\end{tikzpicture}}

\newcommand{\bxbox}{
\begin{tikzpicture}[scale=1]
		\coordinate (e1) at (-0.5,0.);
		\coordinate (e2) at (1.5,0.);
		\coordinate (e3) at (1.5,-1.);
		\coordinate (e4) at (-0.5,-1.);	
		
		\coordinate (v1) at (0,0);
		\coordinate (v2) at (1,0);
		\coordinate (v3) at (0,-1);
		\coordinate (v4) at (1,-1);
		
		\draw[massless] (v2) -- (v3) ;
	    \draw[line width = 0.2cm,white ] (v1) -- (v4) ;
		\draw[massless] (v1) -- (v4) ;

		\draw[massive] (e1) -- (v1) ;
		\draw[massive] (v1) -- (v2) ;
		\draw[massive] (v2) -- (e2) ;
		\draw[massive] (e4) -- (v3) ;
		\draw[massive] (v3) -- (v4) ;
		\draw[massive] (v4) -- (e3) ;
		
		\node[below left]  at (e4) {$p_1$};
        \node[above left]  at (e1) {$p_2$};
        \node[above right] at (e2) {$p_3$};
        \node[below right] at (e3) {$p_4$};
\end{tikzpicture}}

\newcommand{\linTree}{ 
\begin{tikzpicture}[scale=1]
		\coordinate (e1) at (-0.5,0.);
		\coordinate (e2) at (0.5,0.);
		\coordinate (e3) at (0.5,-1.);
		\coordinate (e4) at (-0.5,-1.);	
		
		\coordinate (v1) at (0,0);
		\coordinate (v2) at (0,-1);
		
		\draw[massiveLin] (e1) -- (v1) ;
		\draw[massiveLin] (v1) -- (e2) ;
		\draw[massiveLin] (v2) -- (e3) ;
		\draw[massiveLin] (e4) -- (v2) ;
		
		\draw[massless] (v1) -- (v2) ;
\end{tikzpicture}}

\newcommand{\linBox}{
\begin{tikzpicture}[scale=1]
		\coordinate (e1) at (-0.5,0.);
		\coordinate (e2) at (1.5,0.);
		\coordinate (e3) at (1.5,-1.);
		\coordinate (e4) at (-0.5,-1.);	
		
		\coordinate (v1) at (0,0);
		\coordinate (v2) at (1,0);
		\coordinate (v3) at (0,-1);
		\coordinate (v4) at (1,-1);
		
		\draw[massiveLin] (e1) -- (v1) ;
		\draw[massiveLin] (v1) -- (v2) ;
		\draw[massiveLin] (v2) -- (e2) ;
		\draw[massiveLin] (e4) -- (v3) ;
		\draw[massiveLin] (v3) -- (v4) ;
		\draw[massiveLin] (v4) -- (e3) ;
		
		\draw[massless] (v1) -- (v3) ;
		\draw[massless] (v2) -- (v4) ;
\end{tikzpicture}}

\newcommand{\linBoxPropLab}{
\begin{tikzpicture}[scale=1]
		\coordinate (e1) at (-0.5,0.);
		\coordinate (e2) at (1.5,0.);
		\coordinate (e3) at (1.5,-1.);
		\coordinate (e4) at (-0.5,-1.);	
		
		\coordinate (v1) at (0,0);
		\coordinate (v2) at (1,0);
		\coordinate (v3) at (0,-1);
		\coordinate (v4) at (1,-1);
		
		\draw[massiveLin] (e1) -- (v1) ;
		\draw[massiveLin] (v1) -- (v2) node [midway,fill=white] {2};
		\draw[massiveLin] (v2) -- (e2) ;
		\draw[massiveLin] (e4) -- (v3) ;
		\draw[massiveLin] (v3) -- (v4) node [midway,fill=white] {1};
		\draw[massiveLin] (v4) -- (e3) ;
		
		\draw[massless] (v1) -- (v3) node [midway,fill=white] {3};
		\draw[massless] (v2) -- (v4) node [midway,fill=white] {4};
\end{tikzpicture}}

\newcommand{\linCutBox}{
\begin{tikzpicture}[scale=1]
		\coordinate (e1) at (-0.5,0.);
		\coordinate (e2) at (1.5,0.);
		\coordinate (e3) at (1.5,-1.);
		\coordinate (e4) at (-0.5,-1.);	
		
		\coordinate (v1) at (0,0);
		\coordinate (v2) at (1,0);
		\coordinate (v3) at (0,-1);
		\coordinate (v4) at (1,-1);
		
		\draw[massiveLin] (e1) -- (v1) ;
		\draw[massiveLin] (v1) -- (v2) ;
		\draw[massiveLin] (v2) -- (e2) ;
		\draw[massiveLin] (e4) -- (v3) ;
		\draw[massiveLin] (v3) -- (v4) ;
		\draw[massiveLin] (v4) -- (e3) ;
		
		\draw[massless] (v1) -- (v3) ;
		\draw[massless] (v2) -- (v4) ;
		\draw[line width = 0.2cm, white] (0.5,0.25) -- (0.5,-1.25);
	    \draw[densely dashed, blue] (0.5,0.25) -- (0.5,-1.25);
\end{tikzpicture}}

\newcommand{\linxBox}{
\begin{tikzpicture}[scale=1]
		\coordinate (e1) at (-0.5,0.);
		\coordinate (e2) at (1.5,0.);
		\coordinate (e3) at (1.5,-1.);
		\coordinate (e4) at (-0.5,-1.);	
		
		\coordinate (v1) at (0,0);
		\coordinate (v2) at (1,0);
		\coordinate (v3) at (0,-1);
		\coordinate (v4) at (1,-1);
		
		\draw[massless] (v2) -- (v3) ;
	    \draw[line width = 0.2cm,white ] (v1) -- (v4) ;
		\draw[massless] (v1) -- (v4) ;

		\draw[massiveLin] (e1) -- (v1) ;
		\draw[massiveLin] (v1) -- (v2) ;
		\draw[massiveLin] (v2) -- (e2) ;
		\draw[massiveLin] (e4) -- (v3) ;
		\draw[massiveLin] (v3) -- (v4) ;
		\draw[massiveLin] (v4) -- (e3) ;
\end{tikzpicture}}

\newcommand{\trippleCutIX}{
\begin{tikzpicture}[scale=1]
	    \coordinate (e1) at (-0.5,0.);
	    \coordinate (e2) at (2.5,0.);
	    \coordinate (e3) at (2.5,-1.);
	    \coordinate (e4) at (-0.5,-1.);	
	    \coordinate (t1) at (0,0);
	    \coordinate (t2) at (1,0);
	    \coordinate (t3) at (2,0);
	    \coordinate (b1) at (0,-1);
	    \coordinate (b2) at (1,-1);
	    \coordinate (b3) at (2,-1);
	
	    \draw[massiveLin] (e1) -- (t1) -- (t2) -- (t3) -- (e2);
	    \draw[massiveLin] (e4) -- (b1) -- (b2) -- (b3) -- (e3);
	    \draw[name path=rung1,massless] (t1) -- (b1) ;
	    \draw[name path=rung2,massless] (t2) -- (b3) ;
	    \draw[name path=rung3,massless] (t3) -- (b2) ;
	
	    \fill[name intersections={of=rung2 and rung3, name=i,total=\t},white] foreach \s in {1,...,\t}{(i-\s) circle (5pt)};
	    \draw[massless] (t3) -- (b2) ;
	    \draw[line width = 0.2cm,white ] (0.25,0.25) -- (1.75,-1.25);
	    \draw[densely dashed,blue] (0.25,0.25) -- (1.75,-1.25);
\end{tikzpicture}}

\newcommand{\linNDot}{
\begin{tikzpicture}
        \coordinate (e1) at (-0.5,0);
        \coordinate (e2) at (1.5,0);
        \coordinate (e3) at (1.5,-1);
        \coordinate (e4) at (-0.5,-1);	

        \coordinate (v1) at (0,0);
        \coordinate (v2) at (0,-1);
        \coordinate (v3) at (1,0);
        \coordinate (v4) at (1,-1);

        \draw[massless] (v1) -- (v2);
        \draw[massless] (v3) -- (v4);
        \draw[massless] (v1) --  (v4);

        \draw[massiveLin] (e1) --  (v1) -- (v3) --(e2);

        \draw[massiveLin] (e4) --  (v2) -- (v4) --(e3);
        \fill (0.5,-0.5) circle[radius=2pt];
\end{tikzpicture}}

\newcommand{\trippleCutNDot}{
\begin{tikzpicture}
        \coordinate (e1) at (-0.5,0);
        \coordinate (e2) at (1.5,0);
        \coordinate (e3) at (1.5,-1);
        \coordinate (e4) at (-0.5,-1);	

        \coordinate (v1) at (0,0);
        \coordinate (v2) at (0,-1);
        \coordinate (v3) at (1,0);
        \coordinate (v4) at (1,-1);

        \draw[massless] (v1) -- (v2);
        \draw[massless] (v3) -- (v4);
        \draw[massless] (v1) --  (v4);
        I
        \draw[massiveLin] (e1) --  (v1) -- (v3) --(e2);

        \draw[massiveLin] (e4) --  (v2) -- (v4) --(e3);
        \draw[line width = 0.2cm,white ] (0.5,0.25) -- (0.5,-1.25);
        \draw[densely dashed,blue] (0.5,0.25) -- (0.5,-1.25);
        \fill (0.75,-0.75) circle[radius=2pt];
\end{tikzpicture}}

\newcommand{\linTriTri}{
\begin{tikzpicture}
		\coordinate (e1) at (0,0.);
		\coordinate (e2) at (2,0.);
		\coordinate (e3) at (0.5,-1.);
		\coordinate (e4) at (1.5,-1.);	
		
		\coordinate (v1) at (0.5,0);
		\coordinate (v2) at (1,0);
		\coordinate (v3) at (1.5,0);

		\coordinate (v4) at (1,-1);

		\draw[massiveLin] (e1) -- (v1) ;
		\draw[massiveLin] (v1) -- (v2) ;
		\draw[massiveLin] (v2) -- (v3) ;
		\draw[massiveLin] (v3) -- (e2) ;
		\draw[massiveLin] (e4) -- (v4) ;
		\draw[massiveLin] (e3) -- (v4) ;
		
		\draw[massless] (v1) -- (v4) ;
		\draw[massless] (v2) -- (v4) ;
		\draw[massless] (v3) -- (v4) ;
\end{tikzpicture}}

\newcommand{\sunriseDiag}{
\begin{tikzpicture}[scale=1]
		\coordinate (e1) at (0.5,0.);
		\coordinate (e2) at (1.5,0.);
		\coordinate (e3) at (0.5,-1.);
		\coordinate (e4) at (1.5,-1.);	
		
		\coordinate (v1) at (1,0);

		\coordinate (v2) at (1,-1);

		\draw[massiveLin] (e1) -- (v1) ;
		\draw[massiveLin] (v1) -- (e2) ;
		\draw[massiveLin] (e4) -- (v2) ;
		\draw[massiveLin] (v2) -- (e3) ;
		
		\draw[massless] (v1) to[out=0,in=0] (v2) ;
		\draw[massless] (v1) -- (v2) ;
		\draw[massless] (v1) to[out=180,in=180] (v2) ;
\end{tikzpicture}}

\newcommand{\BNDIntA}{
\begin{tikzpicture}
        \coordinate (e1) at (0,0);
        \coordinate (e2) at (2,0);
        \coordinate (e3) at (0.5,-1.);
        \coordinate (e4) at (1.5,-1.);	

        \coordinate (v1) at (0.5,0);
        \coordinate (v2) at (1.5,0);
        \coordinate (v3) at (1,-1);

        \draw[massless] (v1) -- (v4) ;
        \draw[massless] (v2) -- (v4) ;
        \draw[massless] (v3) -- (v4) ;
        \draw[massless] (v1) to[out=90,in=90] (v2) ;

        \draw[massiveLin] (e1) -- (v1) ;
        \draw[massiveLin] (v1) -- (v2) ;
        \draw[massiveLin] (v2) -- (e2) ;
        \draw[massiveLin] (e3) -- (v3) ;
        \draw[massiveLin] (v3) -- (e4) ;
\end{tikzpicture}}

\newcommand{\BNDIntB}{
\begin{tikzpicture}
        \coordinate (e1) at (0,0);
        \coordinate (e2) at (2,0);
        \coordinate (e3) at (0.5,-1.);
        \coordinate (e4) at (1.5,-1.);	

        \coordinate (v1) at (0.5,0);
        \coordinate (v2) at (1.5,0);
        \coordinate (v3) at (1,-1);

        \draw[massless] (v1) -- (v4) ;
        \draw[massless] (v2) -- (v4) ;
        \draw[massless] (v3) -- (v4) ;
        \draw[massless] (v2) to[out=-45,in=25] (v3) ;

        \draw[massiveLin] (e1) -- (v1) ;
        \draw[massiveLin] (v1) -- (v2) ;
        \draw[massiveLin] (v2) -- (e2) ;
        \draw[massiveLin] (e3) -- (v3) ;
        \draw[massiveLin] (v3) -- (e4) ;
\end{tikzpicture}}

\newcommand{\BNDIntC}{
\begin{tikzpicture}
        \coordinate (e1) at (0,0);
        \coordinate (e2) at (2,0);
        \coordinate (e3) at (0,-1);
        \coordinate (e4) at (2,-1);	

        \coordinate (v1) at (0.5,0);
        \coordinate (v2) at (1.5,0);
        \coordinate (v3) at (1,-0.5);
        \coordinate (v4) at (0.5,-1);
        \coordinate (v5) at (1.5,-1);

        \draw[massless] (v1) -- (v3) ;
        \draw[massless] (v2) -- (v3) ;
        \draw[massless] (v4) -- (v3) ;
        \draw[massless] (v5) -- (v3) ;

        \draw[massiveLin] (e1) -- (v1) ;
        \draw[massiveLin] (v1) -- (v2) ;
        \draw[massiveLin] (v2) -- (e2) ;
        \draw[massiveLin] (e3) -- (v4) ;
        \draw[massiveLin] (v4) -- (v5) ;
        \draw[massiveLin] (v5) -- (e4) ;
\end{tikzpicture}}

\newcommand{\BNDIntD}{
\begin{tikzpicture}
        \coordinate (e1) at (0,0);
        \coordinate (e2) at (2,0);
        \coordinate (e3) at (0.5,-1);
        \coordinate (e4) at (1.5,-1);	

        \coordinate (v1) at (0.5,0);
        \coordinate (v2) at (1.5,0);
        \coordinate (v3) at (1,-0.5);
        \coordinate (v4) at (1,-1);

        \draw[massless] (v1) -- (v3) ;
        \draw[massless] (v2) -- (v3) ;
        \draw[massless] (v4) to[out=0,in=0] (v3) ;
        \draw[massless] (v4) to[out=180,in=180] (v3) ;

        \draw[massiveLin] (e1) -- (v1) ;
        \draw[massiveLin] (v1) -- (v2) ;
        \draw[massiveLin] (v2) -- (e2) ;
        \draw[massiveLin] (e3) -- (v4) ;
        \draw[massiveLin] (v4) -- (e4) ;
\end{tikzpicture}}

\newcommand{\BNDIntE}{
\begin{tikzpicture}
        \coordinate (e1) at (0.5,0);
        \coordinate (e2) at (1.5,0);
        \coordinate (e3) at (0.5,-1);
        \coordinate (e4) at (1.5,-1);	

        \coordinate (v1) at (1,0);
        \coordinate (v3) at (1,-0.5);
        \coordinate (v4) at (1,-1);

        \draw[massless] (v1) to[out=0,in=0] (v3) ;
        \draw[massless] (v1) to[out=180,in=180] (v3) ;
        \draw[massless] (v3) to[out=0,in=0] (v4) ;
        \draw[massless] (v3) to[out=180,in=180] (v4) ;

        \draw[massiveLin] (e1) -- (v1) ;
        \draw[massiveLin] (v1) -- (e2) ;
        \draw[massiveLin] (e3) -- (v4) ;
        \draw[massiveLin] (v4) -- (e4) ;
\end{tikzpicture}}

\newcommand{\twoPtCutGeneric}{ \begin{tikzpicture}[scale=0.8]
	\coordinate (e1) at (-1.5,0.);
	\coordinate (e2) at (1.5,0.);
	\coordinate (e3) at (1.5,-1.);
	\coordinate (e4) at (-1.5,-1.);	
	
	\draw[massiveLin] (e1) -- (e2) ;
	\draw[massiveLin] (e3) -- (e4) ;
	\draw[fill=lightgray, opacity=1] (-0.7,-0.5) ellipse (0.45 and 0.65) node {$I_{\mathrm{L}}$};
	\draw[fill=lightgray, opacity=1] (0.7,-0.5) ellipse (0.45 and 0.65) node {$I_{\mathrm{R}}$};
	\draw[line width = 0.2cm,white ] (0,0.25) -- (0,-1.25);
	\draw[densely dashed,blue] (0,0.25) -- (0,-1.25);
	\end{tikzpicture}}

\newcommand{\twoPtCutIII}{ \begin{tikzpicture}[scale=1]
		\coordinate (e1) at (-0.5,0.);
		\coordinate (e2) at (2.5,0.);
		\coordinate (e3) at (2.5,-1.);
		\coordinate (e4) at (-0.5,-1.);	
		
		\coordinate (v1) at (0,0);
		\coordinate (v2) at (1,0);
		\coordinate (v3) at (2,0);
		\coordinate (v4) at (2,-1);
		\coordinate (v5) at (1,-1);
		\coordinate (v6) at (0,-1);
		
		\draw[massiveLin] (e1) -- (v1) ;
		\draw[massiveLin] (v1) -- (v2) ;
		\draw[massiveLin] (v2) -- (v3) ;
		\draw[massiveLin] (v3) -- (e2) ;
		\draw[massiveLin] (e4) -- (v6) ;
		\draw[massiveLin] (v6) -- (v5) ;
		\draw[massiveLin] (v5) -- (v4) ;
		\draw[massiveLin] (v4) -- (e3) ;
		
		\draw[massless] (v1) -- (v6) ;
		\draw[massless] (v2) -- (v5) ;
		\draw[massless] (v3) -- (v4) ;
		\draw[line width = 0.2cm,white ] (0.5,0.25) -- (0.5,-1.25);
	\draw[densely dashed,blue] (0.5,0.25) -- (0.5,-1.25);
\end{tikzpicture}}

\newcommand{\linearIII}{ \begin{tikzpicture}[scale=1]
		\coordinate (e1) at (-0.5,0.);
		\coordinate (e2) at (2.5,0.);
		\coordinate (e3) at (2.5,-1.);
		\coordinate (e4) at (-0.5,-1.);	
		
		\coordinate (v1) at (0,0);
		\coordinate (v2) at (1,0);
		\coordinate (v3) at (2,0);
		\coordinate (v4) at (2,-1);
		\coordinate (v5) at (1,-1);
		\coordinate (v6) at (0,-1);
		
		\draw[massiveLin] (e1) -- (v1) ;
		\draw[massiveLin] (v1) -- (v2) ;
		\draw[massiveLin] (v2) -- (v3) ;
		\draw[massiveLin] (v3) -- (e2) ;
		\draw[massiveLin] (e4) -- (v6) ;
		\draw[massiveLin] (v6) -- (v5) ;
		\draw[massiveLin] (v5) -- (v4) ;
		\draw[massiveLin] (v4) -- (e3) ;
		
		\draw[massless] (v1) -- (v6) ;
		\draw[massless] (v2) -- (v5) ;
		\draw[massless] (v3) -- (v4) ;
\end{tikzpicture}}

\newcommand{\intIII}{ 
\begin{tikzpicture}[scale=1]
		\coordinate (e1) at (-0.5,0.);
		\coordinate (e2) at (2.5,0.);
		\coordinate (e3) at (2.5,-1.);
		\coordinate (e4) at (-0.5,-1.);	
		
		\coordinate (v1) at (0,0);
		\coordinate (v2) at (1,0);
		\coordinate (v3) at (2,0);
		\coordinate (v4) at (2,-1);
		\coordinate (v5) at (1,-1);
		\coordinate (v6) at (0,-1);
		
		\draw[massive] (e1) -- (v1) ;
		\draw[massive] (v1) -- (v2) ;
		\draw[massive] (v2) -- (v3) ;
		\draw[massive] (v3) -- (e2) ;
		\draw[massive] (e4) -- (v6) ;
		\draw[massive] (v6) -- (v5) ;
		\draw[massive] (v5) -- (v4) ;
		\draw[massive] (v4) -- (e3) ;
		
		\draw[massless] (v1) -- (v6) ;
		\draw[massless] (v2) -- (v5) ;
		\draw[massless] (v3) -- (v4) ;
		
		\node[below left]  at (e4) {$p_1$};
        \node[above left]  at (e1) {$p_2$};
        \node[above right] at (e2) {$p_3$};
        \node[below right] at (e3) {$p_4$};
		
\end{tikzpicture}}

\newcommand{\intIX}{
\begin{tikzpicture}[scale=1]
	    \coordinate (e1) at (-0.5,0.);
	    \coordinate (e2) at (2.5,0.);
	    \coordinate (e3) at (2.5,-1.);
	    \coordinate (e4) at (-0.5,-1.);	

	    \coordinate (t1) at (0,0);
	    \coordinate (t2) at (1,0);
	    \coordinate (t3) at (2,0);
	    \coordinate (b1) at (0,-1);
	    \coordinate (b2) at (1,-1);
	    \coordinate (b3) at (2,-1);
	
	    \draw[massive] (e1) -- (t1) -- (t2) -- (t3) -- (e2);
	    \draw[massive] (e4) -- (b1) -- (b2) -- (b3) -- (e3);
	    \draw[name path=rung1,massless] (t1) -- (b1) ;
	    \draw[name path=rung2,massless] (t2) -- (b3) ;
	    \draw[name path=rung3,massless] (t3) -- (b2) ;
	
	    \fill[name intersections={of=rung2 and rung3, name=i,total=\t},white] foreach \s in {1,...,\t}{(i-\s) circle (5pt)};
	    \draw[massless] (t3) -- (b2) ;
	    \node[below left]  at (e4) {$p_1$};
        \node[above left]  at (e1) {$p_2$};
        \node[above right] at (e2) {$p_3$};
        \node[below right] at (e3) {$p_4$};
\end{tikzpicture}}

\newcommand{\intXI}{
\begin{tikzpicture}[scale=1]
	    \coordinate (e1) at (-0.5,0.);
	    \coordinate (e2) at (2.5,0.);
	    \coordinate (e3) at (2.5,-1.);
	    \coordinate (e4) at (-0.5,-1.);	

	    \coordinate (t1) at (0,0);
	    \coordinate (t2) at (1,0);
	    \coordinate (t3) at (2,0);
	    \coordinate (b1) at (0,-1);
	    \coordinate (b2) at (1,-1);
	    \coordinate (b3) at (2,-1);
	
	    \draw[massive] (e1) -- (t1) -- (t2) -- (t3) -- (e2);
	    \draw[massive] (e4) -- (b1) -- (b2) -- (b3) -- (e3);
	    \draw[name path=rung1,massless] (t3) -- (b3) ;
	    \draw[name path=rung2,massless] (t1) -- (b2) ;
	    \draw[name path=rung3,massless] (t2) -- (b1) ;
	
	    \fill[name intersections={of=rung2 and rung3, name=i,total=\t},white] foreach \s in {1,...,\t}{(i-\s) circle (5pt)};
	    \draw[massless] (t1) -- (b2) ;
	    \node[below left]  at (e4) {$p_1$};
        \node[above left]  at (e1) {$p_2$};
        \node[above right] at (e2) {$p_3$};
        \node[below right] at (e3) {$p_4$};
\end{tikzpicture}}

\newcommand{\intH}{ 
\begin{tikzpicture}[scale=1]
		\coordinate (e1) at (-0.5,0);
		\coordinate (e2) at (-0.5,2);
		\coordinate (e3) at (1.5,2);
		\coordinate (e4) at (1.5,0);	
		
		\coordinate (b1) at (0,0);
		\coordinate (b2) at (1,0);
		\coordinate (m1) at (0,1);
		\coordinate (m2) at (1,1);
		\coordinate (t1) at (0,2);
		\coordinate (t2) at (1,2);
		
		\draw[massive] (e1) -- (b1) ;
		\draw[massive] (e4) -- (b2) ;
		\draw[massive] (e2) -- (t1) ;
		\draw[massive] (e3) -- (t2) ;
		\draw[massive] (t1) -- (t2) ;
		\draw[massive] (b1) -- (b2) ;
		
		\draw[massless] (b1) -- (t1) ;
		\draw[massless] (b2) -- (t2) ;
		\draw[massless] (m1) -- (m2) ;
		
		\node[below left]  at (e1) {$p_1$};
        \node[above left]  at (e2) {$p_2$};
        \node[above right] at (e3) {$p_3$};
        \node[below right] at (e4) {$p_4$};
		
\end{tikzpicture}}

\newcommand{\intYInp}{ 
\begin{tikzpicture}[scale=1]
		\coordinate (e1) at (-0.5,0);
		\coordinate (e2) at (-0.5,2);
		\coordinate (e3) at (1.5,2);
		\coordinate (e4) at (1.5,0);	
		
		\coordinate (b1) at (0,0);
		\coordinate (b2) at (1,0);
		\coordinate (m1) at (0,1);
		\coordinate (tm) at (.5,2);
		\coordinate (t1) at (0,2);
		\coordinate (t2) at (1,2);
		
		\draw[massive] (e1) -- (b1) ;
		\draw[massive] (e4) -- (b2) ;
		\draw[massive] (e2) -- (t1) ;
		\draw[massive] (e3) -- (t2) ;
		\draw[massive] (t1) -- (t2) ;
		\draw[massive] (b1) -- (b2) ;
		
		\draw[name path=rung1,massless] (b1) -- (t1) ;
	    \draw[name path=rung2,massless] (b2) -- (tm) ;
	    \draw[name path=rung3,massless] (m1) -- (t2) ;
        
        \fill[name intersections={of=rung2 and rung3, name=i,total=\t},white] foreach \s in {1,...,\t}{(i-\s) circle (5pt)};
        
        \draw[massless] (b2) -- (tm) ;
		
		\node[below left]  at (e1) {$p_1$};
        \node[above left]  at (e2) {$p_2$};
        \node[above right] at (e3) {$p_3$};
        \node[below right] at (e4) {$p_4$};
		
\end{tikzpicture}}

\newcommand{\intYIFlnp}{ 
\begin{tikzpicture}[scale=1]
		\coordinate (e1) at (-0.5,0);
		\coordinate (e2) at (-0.5,2);
		\coordinate (e3) at (1.5,2);
		\coordinate (e4) at (1.5,0);	
		
		\coordinate (b1) at (0,0);
		\coordinate (b2) at (1,0);
		\coordinate (m1) at (0,1);
		\coordinate (bm) at (.5,0);
		\coordinate (t1) at (0,2);
		\coordinate (t2) at (1,2);
		
		\draw[massive] (e1) -- (b1) ;
		\draw[massive] (e4) -- (b2) ;
		\draw[massive] (e2) -- (t1) ;
		\draw[massive] (e3) -- (t2) ;
		\draw[massive] (t1) -- (t2) ;
		\draw[massive] (b1) -- (b2) ;
		
		\draw[name path=rung1,massless] (b1) -- (t1) ;
	    \draw[name path=rung2,massless] (b2) -- (m1) ;
	    \draw[name path=rung3,massless] (bm) -- (t2) ;
        
        \fill[name intersections={of=rung2 and rung3, name=i,total=\t},white] foreach \s in {1,...,\t}{(i-\s) circle (5pt)};
        
        \draw[massless] (bm) -- (t2) ;
		
		\node[below left]  at (e1) {$p_1$};
        \node[above left]  at (e2) {$p_2$};
        \node[above right] at (e3) {$p_3$};
        \node[below right] at (e4) {$p_4$};
		
\end{tikzpicture}}

\newcommand{\xMushroomTop}{\begin{tikzpicture}[scale=1]
        
        \coordinate (e1) at (-0.5,0);
		\coordinate (e2) at (-0.5,1.5);
		\coordinate (e3) at (1.5,1.5);
		\coordinate (e4) at (1.5,0);	

        \coordinate (t1) at (-.25,1.5);
        \coordinate (t2) at (0,1.5);
        \coordinate (t3) at (1,1.5);
        \coordinate (t4) at (1.25,1.5);
        \coordinate (b1) at (0,0);
        \coordinate (b2) at (1,0);

        \draw[massive] (e1) -- (e4);
        \draw[massive] (e2) -- (e3);

        \draw[name path=rung1,massless] (b1) -- (t3) ;
	    \draw[name path=rung2,massless] (b2) -- (t2) ;
        
        \fill[name intersections={of=rung1 and rung2, name=i,total=\t},white] foreach \s in {1,...,\t}{(i-\s) circle (5pt)};
        \draw[massless] (b1) -- (t3) ;
        \draw[massless] (t1) to[out=90,in=90]  (t4);
        
        \node[below left]  at (e1) {$p_1$};
        \node[above left]  at (e2) {$p_2$};
        \node[above right] at (e3) {$p_3$};
        \node[below right] at (e4) {$p_4$};
\end{tikzpicture}}

\newcommand{\xMushroomBottom}{\begin{tikzpicture}[scale=1]
        
        \coordinate (e1) at (-0.5,0);
		\coordinate (e2) at (-0.5,1.5);
		\coordinate (e3) at (1.5,1.5);
		\coordinate (e4) at (1.5,0);	

        \coordinate (b1) at (-.25,0);
        \coordinate (b2) at (0,0);
        \coordinate (b3) at (1,0);
        \coordinate (b4) at (1.25,0);
        \coordinate (t1) at (0,1.5);
        \coordinate (t2) at (1,1.5);

        \draw[massive] (e1) -- (e4);
        \draw[massive] (e2) -- (e3);

        \draw[name path=rung1,massless] (t1) -- (b3) ;
	    \draw[name path=rung2,massless] (t2) -- (b2) ;
        
        \fill[name intersections={of=rung1 and rung2, name=i,total=\t},white] foreach \s in {1,...,\t}{(i-\s) circle (5pt)};
        \draw[massless] (t1) -- (b3) ;
        \draw[massless] (b1) to[out=-90,in=-90]  (b4);
        
        \node[below left]  at (e1) {$p_1$};
        \node[above left]  at (e2) {$p_2$};
        \node[above right] at (e3) {$p_3$};
        \node[below right] at (e4) {$p_4$};
\end{tikzpicture}}

\newcommand{\softkinA}{
\begin{tikzpicture}
        \coordinate (e1) at (-1,-1);	
        \coordinate (e2) at (-1, 1);
        \coordinate (e3) at ( 1, 1);
        \coordinate (e4) at ( 1,-1);
      
        \coordinate (v1) at (0,0);
      
        \draw[massiveWithArrow] (-.75,-.75) -- (-1,-1);
        \draw[massive] (e1) -- (v1);
        \draw[massiveWithArrow] (-.75,.75) -- (-1,1);
        \draw[massive] (e2) -- (v1);
        \draw[massiveWithArrow] (.75,-.75) -- (1,-1);
        \draw[massive] (e4) -- (v1);
        \draw[massiveWithArrow] (.75,.75) -- (1,1);
        \draw[massive] (e3) -- (v1);
      
        \draw[fill=lightgray, opacity=1] (0,0) circle (0.65);
      
        \node[below left]  at (e1) {$p_1$};
        \node[above left]  at (e2) {$p_2$};
        \node[above right] at (e3) {$p_3=-p_2-q$};
        \node[below right] at (e4) {$p_4=-p_1+q$};
\end{tikzpicture}}

\newcommand{\softkinB}{
\begin{tikzpicture}
        \coordinate (e1) at (-1,-1);	
        \coordinate (e2) at (-1, 1);
        \coordinate (e3) at ( 1, 1);
        \coordinate (e4) at ( 1,-1);
      
        \coordinate (v1) at (0,0);
      
        \draw[massiveWithArrow] (e1) -- (v1);
        \draw[massiveWithArrow] (e2) -- (v1);
        \draw[massiveWithArrow] (.75,-.75) -- (1,-1);
        \draw[massive] (e4) -- (v1);
        \draw[massiveWithArrow] (.75,.75) -- (1,1);
        \draw[massive] (e3) -- (v1);
      
        \draw[fill=lightgray, opacity=1] (0,0) circle (0.65);
      
        \node[below left]  at (e1) {$\overline{p}_1 - \frac{q}{2}$};
        \node[above left]  at (e2) {$\overline{p}_2 + \frac{q}{2}$};
        \node[above right] at (e3) {$\overline{p}_2 - \frac{q}{2}$};
        \node[below right] at (e4) {$\overline{p}_1 + \frac{q}{2}$};
\end{tikzpicture}}

\newcommand{\kmocvirtual}{
\begin{tikzpicture}
        \coordinate (e1) at (-1,-1);	
        \coordinate (e2) at (-1, 1);
        \coordinate (e3) at ( 1, 1);
        \coordinate (e4) at ( 1,-1);
      
        \coordinate (v1) at (0,0);
      
        \draw[massive] (e1) -- (v1);
        \draw[massive] (e2) -- (v1);
        \draw[massive] (e4) -- (v1);
        \draw[massive] (e3) -- (v1);
      
        \draw[fill=lightgray, opacity=1] (0,0) circle (0.65);
        \node at (v1) {$\mathcal{M}$};
      
        \node[below left]  at (e1) {$p_1$};
        \node[above left]  at (e2) {$p_2$};
        \node[above right] at (e3) {$p_3$};
        \node[below right] at (e4) {$p_4$};
\end{tikzpicture}}

\newcommand{\kmocreal}[2]{
\begin{tikzpicture}
        \coordinate (e1) at (-2.25,-1);	
	    	\coordinate (e2) at (-2.25, 1);
	    	\coordinate (e3) at ( 2.25, 1);
	    	\coordinate (e4) at ( 2.25,-1);
    
	    	\coordinate (v1) at (-1.25, 0);
	    	\coordinate (v2) at ( 1.25, 0);
	    	
	    	\coordinate (m1) at (0, 1.00);
	    	\coordinate (m2) at (0,-1.00);
    
	    	\draw[massive] (e1) -- (v1);
	    	\draw[massive] (e2) -- (v1);
	    	\draw[massive] (e3) -- (v2);
	    	\draw[massive] (e4) -- (v2);
    
	    	\draw[massive, name path=rung1] (v1) to[out=90, in=90](v2);
	    	\draw[massive, name path=rung2] (v1) to[out=-90, in=-90](v2);
	    	\draw[graviton, name path=rung3, line width=0.25mm] (v1) to (v2);
	    	\draw[line width=0.25mm, dashed, name path=rung4, white] (0, 1) to (0, -1);
	    	
            \draw[fill=lightgray, opacity=1] (v1) circle (0.65);
            \draw[fill=lightgray, opacity=1] (v2) circle (0.65);
	    	\node  at (v1) {$\mathcal{M}$};
	    	\node  at (v2) {$\mathcal{M}^*$};
    
	    	\fill[name intersections={of=rung1 and rung4, name=i,total=\t},white] foreach \s in {1,...,\t}{(i-\s) circle (5pt)};
	    	\fill[name intersections={of=rung2 and rung4, name=i,total=\t},white] foreach \s in {1,...,\t}{(i-\s) circle (5pt)};
	    	\fill[name intersections={of=rung3 and rung4, name=i,total=\t},white] foreach \s in {1,...,\t}{(i-\s) circle (5pt)};
    
	    	\draw[line width=0.3mm, dashed, name path=rung4, blue] (m1) to (m2);
    
	        \node[below left]  at (e1) {$p_1$};
	        \node[above left]  at (e2) {$p_2$};
	        \node[above right] at (e3) {$p_3$};
	        \node[below right] at (e4) {$p_4$};
	    	\node[fill=white, opacity=1]        at (0,0) {$\ell_X$};
	    	\node[above, fill=white, opacity=1] at (m1) {#1}; %$\ell_2-p_2$
	    	\node[below, fill=white, opacity=1] at (m2) {#2}; %$\ell_1-p_1$
\end{tikzpicture}}

\newcommand{\kmocvirtualtree}{
\begin{tikzpicture}
                \coordinate (e1) at (-1,-1);	
	     	\coordinate (e2) at (-1, 1);
	     	\coordinate (e3) at ( 1, 1);
	     	\coordinate (e4) at ( 1,-1);
	     	
	     	\coordinate (v1) at (0,0);
     
	     	\draw[massive] (e1) -- (v1);
	     	\draw[massive] (e2) -- (v1);
	     	\draw[massive] (e4) -- (v1);
	     	\draw[massive] (e3) -- (v1);
	     	
            \draw[fill=lightgray, opacity=1] (0,0) circle (0.65);
	     	\node[below left]  at (e1) {$p_1$};
	     	\node[above left]  at (e2) {$p_2$};
	     	\node[above right] at (e3) {$p_3$};
	     	\node[below right] at (e4) {$p_4$};
\end{tikzpicture}}

\newcommand{\kmocvirtualtreenolab}{
\begin{tikzpicture}
                \coordinate (e1) at (-1,-1);	
	     	\coordinate (e2) at (-1, 1);
	     	\coordinate (e3) at ( 1, 1);
	     	\coordinate (e4) at ( 1,-1);
	     	
	     	\coordinate (v1) at (0,0);
     
	     	\draw[massive] (e1) -- (v1);
	     	\draw[massive] (e2) -- (v1);
	     	\draw[massive] (e4) -- (v1);
	     	\draw[massive] (e3) -- (v1);
	     	
            \draw[fill=lightgray, opacity=1] (0,0) circle (0.65);
\end{tikzpicture}}

\newcommand{\kmocvirtualnlo}{
\begin{tikzpicture}
            \coordinate (e1) at (-1,-1);	
	     	\coordinate (e2) at (-1, 1);
	     	\coordinate (e3) at ( 1, 1);
	     	\coordinate (e4) at ( 1,-1);
	     	
	     	\coordinate (v1) at (0,0);
     
	     	\draw[massive] (e1) -- (v1);
	     	\draw[massive] (e2) -- (v1);
	     	\draw[massive] (e4) -- (v1);
	     	\draw[massive] (e3) -- (v1);
	     	
            \draw[fill=lightgray, opacity=1] (0,0) circle (0.65);
            \draw[fill=white, opacity=1] (0,0) circle (0.35);
	     	\node[below left]  at (e1) {$p_1$};
	     	\node[above left]  at (e2) {$p_2$};
	     	\node[above right] at (e3) {$p_3$};
	     	\node[below right] at (e4) {$p_4$};
\end{tikzpicture}}

\newcommand{\kmocvirtualnlonolab}{
\begin{tikzpicture}
                \coordinate (e1) at (-1,-1);	
	     	\coordinate (e2) at (-1, 1);
	     	\coordinate (e3) at ( 1, 1);
	     	\coordinate (e4) at ( 1,-1);
	     	
	     	\coordinate (v1) at (0,0);
     
	     	\draw[massive] (e1) -- (v1);
	     	\draw[massive] (e2) -- (v1);
	     	\draw[massive] (e4) -- (v1);
	     	\draw[massive] (e3) -- (v1);
	     	
            \draw[fill=lightgray, opacity=1] (0,0) circle (0.65);
            \draw[fill=white, opacity=1] (0,0) circle (0.35);
\end{tikzpicture}}

\newcommand{\kmocvirtualnnlonolab}{
\begin{tikzpicture}
                \coordinate (e1) at (-1,-1);	
	     	\coordinate (e2) at (-1, 1);
	     	\coordinate (e3) at ( 1, 1);
	     	\coordinate (e4) at ( 1,-1);
	     	
	     	\coordinate (v1) at (0,0);
     
	     	\draw[massive] (e1) -- (v1);
	     	\draw[massive] (e2) -- (v1);
	     	\draw[massive] (e4) -- (v1);
	     	\draw[massive] (e3) -- (v1);
	     	
            \draw[fill=lightgray, opacity=1] (0,0) circle (0.65);
            \draw[fill=white, opacity=1] (-0.25,0) ellipse (0.2 and 0.4);
            \draw[fill=white, opacity=1] ( 0.25,0) ellipse (0.2 and 0.4);
\end{tikzpicture}}

\newcommand{\kmocrealnlo}{
\begin{tikzpicture}
                \coordinate (e1) at (-2,-1);	
	    	\coordinate (e2) at (-2, 1);
	    	\coordinate (e3) at ( 2, 1);
	    	\coordinate (e4) at ( 2,-1);
    
	    	\coordinate (v1) at (-1, 0);
	    	\coordinate (v2) at ( 1, 0);
	    	
	    	\coordinate (m1) at (0, 1.00);
	    	\coordinate (m2) at (0,-1.00);
    
	    	\draw[massive] (e1) -- (v1);
	    	\draw[massive] (e2) -- (v1);
	    	\draw[massive] (e3) -- (v2);
	    	\draw[massive] (e4) -- (v2);
    
	    	\draw[massive, name path=rung1] (v1) to[out=90, in=90](v2);
	    	\draw[massive, name path=rung2] (v1) to[out=-90, in=-90](v2);
	    	\draw[line width=0.25mm, dashed, name path=rung4, white] (m1) to (m2);
	    	
            \draw[fill=lightgray, opacity=1] (v1) circle (0.65);
            \draw[fill=lightgray, opacity=1] (v2) circle (0.65);
	    	\fill[name intersections={of=rung1 and rung4, name=i,total=\t},white] foreach \s in {1,...,\t}{(i-\s) circle (5pt)};
	    	\fill[name intersections={of=rung2 and rung4, name=i,total=\t},white] foreach \s in {1,...,\t}{(i-\s) circle (5pt)};
    
	    	\draw[line width=0.3mm, dashed, name path=rung4, blue] (m1) to (m2);
         	
	        \node[below left]  at (e1) {$p_1$};
	        \node[above left]  at (e2) {$p_2$};
	        \node[above right] at (e3) {$p_3$};
	        \node[below right] at (e4) {$p_4$};
	    	\node[above, fill=white, opacity=1] at (m1) {$\ell_2-p_2$};
	    	\node[below, fill=white, opacity=1] at (m2) {$\ell_1-p_1$};
\end{tikzpicture}}

\newcommand{\kmocrealnnlotreeloop}{
\begin{tikzpicture}
                \coordinate (e1) at (-2,-1);	
	    	\coordinate (e2) at (-2, 1);
	    	\coordinate (e3) at ( 2, 1);
	    	\coordinate (e4) at ( 2,-1);
    
	    	\coordinate (v1) at (-1, 0);
	    	\coordinate (v2) at ( 1, 0);
	    	
	    	\coordinate (m1) at (0, 1.00);
	    	\coordinate (m2) at (0,-1.00);
    
	    	\draw[massive] (e1) -- (v1);
	    	\draw[massive] (e2) -- (v1);
	    	\draw[massive] (e3) -- (v2);
	    	\draw[massive] (e4) -- (v2);
    
	    	\draw[massive, name path=rung1] (v1) to[out=90, in=90](v2);
	    	\draw[massive, name path=rung2] (v1) to[out=-90, in=-90](v2);
	    	\draw[line width=0.25mm, dashed, name path=rung4, white] (m1) to (m2);
	    	
            \draw[fill=lightgray, opacity=1] (v1) circle (0.65);
            \draw[fill=lightgray, opacity=1] (v2) circle (0.65);
            \draw[fill=white, opacity=1] (v2) circle (0.35);
	    	\fill[name intersections={of=rung1 and rung4, name=i,total=\t},white] foreach \s in {1,...,\t}{(i-\s) circle (5pt)};
	    	\fill[name intersections={of=rung2 and rung4, name=i,total=\t},white] foreach \s in {1,...,\t}{(i-\s) circle (5pt)};
    
	    	\draw[line width=0.3mm, dashed, name path=rung4, blue] (m1) to (m2);
         	
	        \node[below left]  at (e1) {$p_1$};
	        \node[above left]  at (e2) {$p_2$};
	        \node[above right] at (e3) {$p_3$};
	        \node[below right] at (e4) {$p_4$};
	    	\node[above, fill=white, opacity=1] at (m1) {$\ell_2-p_2$};
	    	\node[below, fill=white, opacity=1] at (m2) {$\ell_1-p_1$};
\end{tikzpicture}}

\newcommand{\kmocrealnnlolooptree}{
\begin{tikzpicture}
            \coordinate (e1) at (-2,-1);	
	    	\coordinate (e2) at (-2, 1);
	    	\coordinate (e3) at ( 2, 1);
	    	\coordinate (e4) at ( 2,-1);
    
	    	\coordinate (v1) at (-1, 0);
	    	\coordinate (v2) at ( 1, 0);
	    	
	    	\coordinate (m1) at (0, 1.00);
	    	\coordinate (m2) at (0,-1.00);
    
	    	\draw[massive] (e1) -- (v1);
	    	\draw[massive] (e2) -- (v1);
	    	\draw[massive] (e3) -- (v2);
	    	\draw[massive] (e4) -- (v2);
    
	    	\draw[massive, name path=rung1] (v1) to[out=90, in=90](v2);
	    	\draw[massive, name path=rung2] (v1) to[out=-90, in=-90](v2);
	    	\draw[line width=0.25mm, dashed, name path=rung4, white] (m1) to (m2);
	    	
            \draw[fill=lightgray, opacity=1] (v1) circle (0.65);
            \draw[fill=lightgray, opacity=1] (v2) circle (0.65);
            \draw[fill=white, opacity=1] (v1) circle (0.35);
	    	\fill[name intersections={of=rung1 and rung4, name=i,total=\t},white] foreach \s in {1,...,\t}{(i-\s) circle (5pt)};
	    	\fill[name intersections={of=rung2 and rung4, name=i,total=\t},white] foreach \s in {1,...,\t}{(i-\s) circle (5pt)};
    
	    	\draw[line width=0.3mm, dashed, name path=rung4, blue] (m1) to (m2);
         	
	        \node[below left]  at (e1) {$p_1$};
	        \node[above left]  at (e2) {$p_2$};
	        \node[above right] at (e3) {$p_3$};
	        \node[below right] at (e4) {$p_4$};
	    	\node[above, fill=white, opacity=1] at (m1) {$\ell_2-p_2$};
	    	\node[below, fill=white, opacity=1] at (m2) {$\ell_1-p_1$};
\end{tikzpicture}}

\newcommand{\kmocrealnnlotreetree}{
\begin{tikzpicture}
            \coordinate (e1) at (-2.25,-1);	
	    	\coordinate (e2) at (-2.25, 1);
	    	\coordinate (e3) at ( 2.25, 1);
	    	\coordinate (e4) at ( 2.25,-1);
    
	    	\coordinate (v1) at (-1.25, 0);
	    	\coordinate (v2) at ( 1.25, 0);
	    	
	    	\coordinate (m1) at (0, 1.00);
	    	\coordinate (m2) at (0,-1.00);
    
	    	\draw[massive] (e1) -- (v1);
	    	\draw[massive] (e2) -- (v1);
	    	\draw[massive] (e3) -- (v2);
	    	\draw[massive] (e4) -- (v2);
    
	    	\draw[massive, name path=rung1] (v1) to[out=90, in=90](v2);
	    	\draw[massive, name path=rung2] (v1) to[out=-90, in=-90](v2);
	    	\draw[graviton, name path=rung3, line width=0.25mm] (v1) to (v2);
	    	\draw[line width=0.25mm, dashed, name path=rung4, white] (0, 1) to (0, -1);
	    	
            \draw[fill=lightgray, opacity=1] (v1) circle (0.65);
            \draw[fill=lightgray, opacity=1] (v2) circle (0.65);
	    	\fill[name intersections={of=rung1 and rung4, name=i,total=\t},white] foreach \s in {1,...,\t}{(i-\s) circle (5pt)};
	    	\fill[name intersections={of=rung2 and rung4, name=i,total=\t},white] foreach \s in {1,...,\t}{(i-\s) circle (5pt)};
	    	\fill[name intersections={of=rung3 and rung4, name=i,total=\t},white] foreach \s in {1,...,\t}{(i-\s) circle (5pt)};
    
	    	\draw[line width=0.3mm, dashed, name path=rung4, blue] (m1) to (m2);
    
	        \node[below left]  at (e1) {$p_1$};
	        \node[above left]  at (e2) {$p_2$};
	        \node[above right] at (e3) {$p_3$};
	        \node[below right] at (e4) {$p_4$};
	    	\node[fill=white, opacity=1]        at (0,0) {$\ell_X$};
	    	\node[above, fill=white, opacity=1] at (m1) {$\ell_2-p_2$};
	    	\node[below, fill=white, opacity=1] at (m2) {$\ell_1-p_1$};
\end{tikzpicture}}

\newcommand{\wcut}{
\begin{tikzpicture}
		\coordinate (e1) at (-0.75, 0.25);
		\coordinate (e2) at ( 2.75, 0.25);
		\coordinate (e3) at ( 2.00,-1.25);
		\coordinate (e4) at ( 0.00,-1.25);	
		
		\coordinate (v1) at (0,0);
		\coordinate (v2) at (1,0);
		\coordinate (v3) at (2,0);
		\coordinate (v4) at (1,-1);
		
		\draw[massive] (e1) -- (v1);
		\draw[massive] (v1) -- (v2);
		\draw[massive] (v2) -- (v3);
		\draw[massive] (v3) -- (e2);
		\draw[massive] (e3) -- (v4);
		\draw[massive] (e4) -- (v4);
		\draw[graviton] (v1) -- (v4);
		\draw[graviton] (v2) -- (v4);
		\draw[graviton] (v3) -- (v4);

		\draw[fill=lightgray, opacity=1] (v1) circle (0.2);
                \draw[fill=lightgray, opacity=1] (v2) circle (0.2);
                \draw[fill=lightgray, opacity=1] (v3) circle (0.2);
                \draw[fill=lightgray, opacity=1] (v4) ellipse (0.4 and 0.2);
		\end{tikzpicture}
		}
\newcommand{\ncut}{
\begin{tikzpicture}
		\coordinate (e1) at (-0.75, 0.25);
		\coordinate (e2) at ( 1.75, 0.25);
		\coordinate (e3) at ( 1.75,-1.25);
		\coordinate (e4) at (-0.75,-1.25);	
		
		\coordinate (v1) at (0,0);
		\coordinate (v2) at (0,-1);
		\coordinate (v3) at (1,0);
		\coordinate (v4) at (1,-1);

		\draw[massive] (e1) -- (v1);
		\draw[massive]  (v1) -- (v3);
		\draw[massive] (e2) -- (v3);
		
		\draw[massive] (e4) -- (v2);
		\draw[massive]  (v2) -- (v4);
		\draw[massive] (e3) -- (v4);
		
		\draw[graviton] (v1) -- (v2);
		\draw[graviton] (v3) -- (v4);
		\draw[graviton] (v1) --  (v4);

                \draw[fill=lightgray, opacity=1] (v1) circle (0.2);
                \draw[fill=lightgray, opacity=1] (v2) circle (0.2);
                \draw[fill=lightgray, opacity=1] (v3) circle (0.2);
                \draw[fill=lightgray, opacity=1] (v4) circle (0.2);
		\end{tikzpicture}
}

\newcommand{\hcut}{
\begin{tikzpicture}
		\coordinate (e1) at (-0.75, 0.25);
		\coordinate (e2) at ( 1.75, 0.25);
		\coordinate (e3) at ( 1.75,-2.25);
		\coordinate (e4) at (-0.75,-2.25);	
	
		\coordinate (v1) at (0,0);
		\coordinate (v2) at (0,-1);
		\coordinate (v3) at (0,-2);
		\coordinate (v4) at (1,0);
		\coordinate (v5) at (1,-1);
		\coordinate (v6) at (1,-2);
		
		\draw[massive] (e1) -- (v1);
		\draw[massive] (v1) -- (v4);
		\draw[massive] (e2) -- (v4);
		
		\draw[massive] (e4) -- (v3);
		\draw[massive] (v3) -- (v6);
		\draw[massive] (e3) -- (v6);
		
		\draw[graviton] (v1) -- (v2);
		\draw[graviton] (v2) -- (v3);
		\draw[graviton] (v4) -- (v5);
		\draw[graviton] (v5) -- (v6);
		\draw[graviton] (v2) -- (v5);

		\draw[fill=lightgray, opacity=1] (v1) circle (0.2);
		\draw[fill=lightgray, opacity=1] (v4) circle (0.2);
        \draw[fill=lightgray, opacity=1] (0.5, -1) ellipse (0.6 and 0.2);
        \draw[fill=lightgray, opacity=1] (v3) circle (0.2);
		\draw[fill=lightgray, opacity=1] (v6) circle (0.2);
\end{tikzpicture}}

\newcommand{\threecomptoncut}{
\begin{tikzpicture}
		\coordinate (e1) at (-0.75, 0.25);
		\coordinate (e2) at ( 1.75, 0.25);
		\coordinate (e3) at ( 1.50,-1.25);
		\coordinate (e4) at (-0.50,-1.25);	
		
		\coordinate (v1) at (0,0);
		\coordinate (v2) at (0.5,-1);
		\coordinate (v3) at (1,0);

		\draw[massive] (e1) -- (v1);
		\draw[massive]  (v1) -- (v3);
		\draw[massive] (e2) -- (v3);
		
		\draw[massive] (e4) -- (v2);
		\draw[massive] (e3) -- (v2);
		
		\draw[graviton] (v1) -- (v2);
		\draw[graviton] (v3) -- (v2);
		\draw[graviton] (v1) to[out=100,in=80]  (v3);

        \draw[fill=lightgray, opacity=1] (v1) circle (0.2);
        \draw[fill=lightgray, opacity=1] (v2) ellipse (0.4 and 0.2);
        \draw[fill=lightgray, opacity=1] (v3) circle (0.2);	
\end{tikzpicture}}

\newcommand{\threegravitoncut}{
\begin{tikzpicture}
\coordinate (e2) at (-1.00, 0.75);
		\coordinate (e3) at ( 1.00, 0.75);
		\coordinate (e4) at ( 1.00,-0.75);
		\coordinate (e1) at (-1.00,-0.75);	
		
		\coordinate (v1) at (0,-0.5);
		\coordinate (v2) at (0, 0.5);
		
		\draw[massive] (e1) -- (v1);
		\draw[massive] (v1) -- (e4);
		\draw[massive] (e3) -- (v2);
		\draw[massive] (e2) -- (v2);
		\draw[graviton] (v1) -- (v2);
		\draw[graviton] (v1) to[out=15,in=-15] (v2);
		\draw[graviton] (v2) to[out=165,in=-165] (v1);

        \draw[fill=lightgray, opacity=1] (v1) ellipse (0.4 and 0.2);
        \draw[fill=lightgray, opacity=1] (v2) ellipse (0.4 and 0.2);
\end{tikzpicture}}

\newcommand{\graphsixteen}{
\begin{tikzpicture}
        \coordinate (e1) at (-1, 0);
		\coordinate (e2) at ( 1, 0);
		\coordinate (e3) at ( 1,-1);
		\coordinate (e4) at (-1,-1);

		\coordinate (v1) at (-0.5,0);
		\coordinate (v2) at (-0.5,-1);
		\coordinate (v3) at (0.5,0);
		\coordinate (v4) at (0.5,-1);
		\coordinate (v5) at (-1,0);
		\coordinate (v6) at (1,0);
		
		\draw[massive] (e1) -- (v1);
		\draw[massive]  (v1) -- (v3);
		\draw[massive] (e2) -- (v3);

		\draw[massive] (e4) -- (v2);
		\draw[massive]  (v2) -- (v4);
		\draw[massive] (e3) -- (v4);
		
		\draw[massless] (v1) -- (v2);
		\draw[massless] (v3) -- (v4);

                \draw[fill=white, opacity=1] (0.5,-0.5) circle (0.2);
\end{tikzpicture}}

\newcommand{\graphscaleless}{
\begin{tikzpicture}
        \coordinate (e1) at (-1, 0);
		\coordinate (e2) at ( 2, 0);
		\coordinate (e3) at ( 2,-1);
		\coordinate (e4) at (-1,-1);

		\coordinate (v1) at (0,0);
		\coordinate (v2) at (0.5,-1);
		\coordinate (v3) at (1,0);
		\coordinate (v4) at (0.5, 0);
		\coordinate (v5) at (-0.5,0);
		\coordinate (v6) at (1.5,0);
		
		\draw[massive] (e1) -- (v1);
		\draw[massive]  (v1) -- (v3);
		\draw[massive] (e2) -- (v3);
		
		\draw[massive] (e4) -- (v2);
		\draw[massive] (e3) -- (v2);
		
		\draw[massless] (v2) -- (v4);
		\draw[massless] (1.5, 0) arc (0:180:1cm and 0.75cm);
		\draw[massless] (1.0, 0) arc (0:180:0.5cm and 0.38cm);
\end{tikzpicture}}

\newcommand{\graphquantumbuble}{
\begin{tikzpicture}
		\coordinate (e1) at (-0.75, 0.75);
		\coordinate (e2) at ( 0.75, 0.75);
		\coordinate (e3) at ( 0.75,-0.75);
		\coordinate (e4) at (-0.75,-0.75);

		\coordinate (v1) at (0, 0.75);
		\coordinate (v2) at (0,-0.75);
		\coordinate (v3) at (0, 0.00);
		
		\draw[massive] (e1) -- (v1);
		\draw[massive] (e2) -- (v1);
		\draw[massive] (e4) -- (v2);
		\draw[massive] (v2) -- (e3);
		
		\draw[massless] (v1) -- (v2);
                \draw[fill=white, opacity=1] (v3) circle (0.4);
		\draw[massless] (0.4,0) -- (-0.4,0);
 
\end{tikzpicture}}

\newcommand{\graphquantumtriangle}{
\begin{tikzpicture}
		\coordinate (e1) at (-1, 0.75);
		\coordinate (e2) at ( 1, 0.75);
		\coordinate (e3) at ( 1,-0.75);
		\coordinate (e4) at (-1,-0.75);

		\coordinate (v1) at (0.5, 0.75);
		\coordinate (v4) at (-0.5, 0.75);
		\coordinate (v2) at (0,-0.75);
		\coordinate (v3) at (0, 0.00);
		\coordinate (v5) at (0, -0.35);
		
		\draw[massive] (e1) -- (v1);
		\draw[massive] (e2) -- (v1);
		\draw[massive] (e4) -- (v2);
		\draw[massive] (v2) -- (e3);
		
		\draw[massless] (v1) -- (v5);
		\draw[massless] (v4) -- (v5);
		\draw[massless] (v3) -- (v2);
                \draw[fill=white, opacity=1] (v3) circle (0.35);
 
\end{tikzpicture}}

\newcommand{\linMushroom}{\begin{tikzpicture}[scale=1]

\draw[fill=white,draw=none] (-0.25,-1.5) rectangle (2,0.5);

\coordinate (e1) at (-0.25,0);
\coordinate (e2) at (2,0);
\coordinate (e3) at (2,-1);
\coordinate (e4) at (-0.25,-1);	

\coordinate (v1) at (0,0);
\coordinate (v2) at (0.5,0);
\coordinate (v3) at (1,0);
\coordinate (v4) at (1.5,0);
\coordinate (v5) at (0.5,-1);
\coordinate (v6) at (1.5,-1);
\draw[massiveLin] (e1) -- (v1);
\draw[massiveLin] (v1) -- (v2);
\draw[massiveLin] (v2) -- (v3);
\draw[massiveLin] (v3) -- (v4) ;
\draw[massiveLin] (v4) -- (e2);
\draw[massiveLin] (e4) -- (v5);
\draw[massiveLin] (v5) -- (v6);
\draw[massiveLin] (v6) -- (e3);

\draw[massless] (v2) -- (v5);
\draw[massless] (v4) -- (v6);
\draw[massless] (v1) to[out=90,in=90]  (v3) ;
\end{tikzpicture}}

\newcommand{\linMushroomWithLabels}{\begin{tikzpicture}[scale=2]

\coordinate (e1) at (-0.25,0);
\coordinate (e2) at (2,0);
\coordinate (e3) at (2,-1);
\coordinate (e4) at (-0.25,-1);	

\coordinate (v1) at (0,0);
\coordinate (v2) at (0.5,0);
\coordinate (v3) at (1,0);
\coordinate (v4) at (1.5,0);
\coordinate (v5) at (0.5,-1);
\coordinate (v6) at (1.5,-1);
\draw[massiveLin] (e1) -- (v1);
\draw[massiveLin] (v1) -- (v2) node [midway,fill=white] {1};
\draw[massiveLin] (v2) -- (v3) node [midway,fill=white] {2};
\draw[massiveLin] (v3) -- (v4) node [midway,fill=white] {3};
\draw[massiveLin] (v4) -- (e2);
\draw[massiveLin] (e4) -- (v5);
\draw[massiveLin] (v5) -- (v6);
\draw[massiveLin] (v6) -- (e3);

\draw[massless] (v2) -- (v5) node [midway, left,fill=white] {$\ell_2\uparrow$};
\draw[massless] (v4) -- (v6);
\draw[massless] (v1) to[out=90,in=90] node [midway, above,fill=white] {$\stackrel{\ell_1}{\leftarrow}$} (v3) ;
\end{tikzpicture}}

\newcommand{\linMushroomPFI}{\begin{tikzpicture}[scale=1]

\draw[fill=white,draw=none] (-0.25,-1.5) rectangle (2,0.5);

\coordinate (e1) at (-0.25,0);
\coordinate (e2) at (2,0);
\coordinate (e3) at (2,-1);
\coordinate (e4) at (-0.25,-1);	

\coordinate (v1) at (0,0);
\coordinate (v2) at (0.5,0);
\coordinate (v3) at (1,0);
\coordinate (v4) at (1.5,0);
\coordinate (v5) at (0.5,-1);
\coordinate (v6) at (1.5,-1);

\draw[massless] (v2) -- (v5);
\draw[massless] (v4) -- (v6);
\draw (v3) arc (0:180:0.25) ;

\draw[massiveLin] (e1) -- (v1);
\draw[massiveLin] (v1) -- (v2);
\draw[massiveLin] (v2) -- (v3);
\draw[massiveLin] (v3) -- (v4);
\draw[massiveLin] (v4) -- (e2);
\draw[massiveLin] (e4) -- (v5);
\draw[massiveLin] (v5) -- (v6);
\draw[massiveLin] (v6) -- (e3);

\fill (1.25,0) circle[radius=2pt];

\end{tikzpicture}}

\newcommand{\linMushroomPFII}{\begin{tikzpicture}[scale=1]

\draw[fill=white,draw=none] (-0.25,-1.5) rectangle (2,0.5);

\coordinate (e1) at (-0.25,0);
\coordinate (e2) at (2,0);
\coordinate (e3) at (2,-1);
\coordinate (e4) at (-0.25,-1);	

\coordinate (v1) at (0,0);
\coordinate (v2) at (0.5,0);
\coordinate (v3) at (1,0);
\coordinate (v4) at (1.5,0);
\coordinate (v5) at (0.5,-1);
\coordinate (v6) at (1.5,-1);

\draw[massless] (v2) -- (v5);
\draw[massless] (v4) -- (v6);
\draw (v2) arc (0:180:0.25) ;

\draw[massiveLin] (e1) -- (v1);
\draw[massiveLin] (v1) -- (v2);
\draw[massiveLin] (v2) -- (v3);
\draw[massiveLin] (v3) -- (v4);
\draw[massiveLin] (v4) -- (e2);
\draw[massiveLin] (e4) -- (v5);
\draw[massiveLin] (v5) -- (v6);
\draw[massiveLin] (v6) -- (e3);

\fill (1,0) circle[radius=2pt];

\end{tikzpicture}}

\newcommand{\MushroomToEmbedd}{\begin{tikzpicture}[scale=1]
	
	\draw[fill=white,draw=none] (0,-1.5) rectangle (2,0.5);
	
	\coordinate (e1) at (0,0);
	\coordinate (e2) at (2,0);
	\coordinate (e3) at (2,-1);
	\coordinate (e4) at (0,-1);	
	
	\coordinate (v1) at (0,0);
	\coordinate (v2) at (0.5,0);
	\coordinate (v3) at (1,0);
	\coordinate (v4) at (1.5,0);
	\coordinate (v5) at (0.5,-1);
	\coordinate (v6) at (1.5,-1);

	\draw[massless] (v2) -- (v5);
    \draw[massless] (v4) -- (v6);
    \draw (v3) arc (0:180:0.25) ;

	\draw[massiveLin] (e1) -- (v1);
	\draw[massiveLin] (v1) -- (v2);
	\draw[massiveLin] (v2) -- (v3);
	\draw[massiveLin] (v3) -- (v4);
	\draw[massiveLin] (v4) -- (e2);
	\draw[massiveLin] (e4) -- (v5);
	\draw[massiveLin] (v5) -- (v6);
	\draw[massiveLin] (v6) -- (e3);

	\end{tikzpicture}}

\newcommand{\IYlin}{\begin{tikzpicture}[scale=1]
	
	\draw[fill=white,draw=none] (-0.25,-1.5) rectangle (2,0.5);
	
	\coordinate (e1) at (-0.25,0);
	\coordinate (e2) at (2,0);
	\coordinate (e3) at (2,-1);
	\coordinate (e4) at (-0.25,-1);	

	\coordinate (t1) at (0.2,0);
	\coordinate (t2) at (0.8,0);
	\coordinate (t3) at (1.5,0);
	
	\coordinate (c1) at (0.5,-0.5);
	
	\coordinate (b1) at (0.5,-1);
	\coordinate (b2) at (1.5,-1);

	\draw[massless] (t1) -- (c1);
	\draw[massless] (t2) -- (c1) node[midway,right] {\scriptsize\rotatebox[origin=c]{145}{$\uparrow$} $\ell_1$};
	\draw[massless] (c1) -- (b1) node[midway,left] {\scriptsize $\ell_2 \uparrow$};
	\draw[massless] (t3) --  (b2);
	
	\draw[massiveLin] (e1) -- (t1);
	\draw[massiveLin] (t1) -- (t2);
	\draw[massiveLin] (t2) -- (t3);
	\draw[massiveLin] (t3) -- (e2);

	\draw[massiveLin] (e4) -- (b1);
	\draw[massiveLin] (b1) -- (b2);
	\draw[massiveLin] (b2) -- (e3);

\end{tikzpicture}}

\newcommand{\kinPlotX}{
\begin{tikzpicture}
	\draw[-latex] (0,-2) -- (0,2);
	\draw[-latex] (-2,0) -- (2,0);
			\draw[thick,mgreen] (0,-0.2) --(1.5,-0.2) ;
			\draw[thick,mred] (-1.5,0.2) --(0,0.2) ;
	\draw[thick,mblue] (0,-1.7) arc (-90:0.2:1.5) ;
	\draw[thick,mblue] (-1.5,0.2) arc (180:90.2:1.5) ;
	\draw[snake=zigzag,segment length=3pt]                        (-1.5,0) -- (0,0);
	\draw[snake=zigzag,segment length=3pt]                        (0,0) -- (1.5,0);
	\fill (-1.5,0) circle[radius=2pt];
	\fill (0,0) circle[radius=2pt];
	\fill (1.5,0) circle[radius=2pt];
	\node (a) at (1.5,1.5) {$x$};
	\draw (a.north west) -- (a.south west) -- (a.south east);
	\draw (0.25,-1.70) node[below]{\scriptsize$-\imath$};
	\draw (0.25,1.70) node[above]{\scriptsize$+\imath$};
	\draw (0,0) node[yshift=0.3cm,xshift=0.2cm] {\scriptsize $0$};
	\draw (1.5,0) node[xshift=.2cm,yshift=0.3cm] {\scriptsize $1$};
	\draw (-1.5,0) node[yshift=-0.3cm,xshift=-0.3cm] {\scriptsize $-1$};
	\draw (0,-1.7) node[cross=2pt] {};
	\draw (0,1.7) node[cross=2pt] {};
	\end{tikzpicture}
}

\newcommand{\kinPlotY}{
\begin{tikzpicture}
\tikzfading[name=fade out, inner color=transparent!0,
         outer color=transparent!100]
    \draw[-latex] (0,-2) -- (0,2);
    \draw[-latex] (-3,0) -- (3,0);
    \draw[thick,mblue] (0,0)--(1,0.2) ;
    \draw[thick,mblue] (-1,-0.2)--(0,0) ;
    \draw[thick,mred,path fading=west] (-3,-0.2) --(-1,-0.2) ;
    \draw[thick,mgreen,path fading=east] (1,0.2) --node[above]{{$u$-channel}}(3,0.2) ;
    \draw[snake=zigzag,segment length=3pt,path fading=west]                        (-1,0) -- (-3,0);
    \draw[snake=zigzag,segment length=3pt,path fading=east]                        (3,0) -- (1,0);
    \fill (-1,0) circle[radius=2pt];
    \fill (1,0) circle[radius=2pt];
    \draw (0,0) node[cross=2pt] {};
    \node (a) at (1.5,1.5) {$y$};
    \draw (a.north west) -- (a.south west) -- (a.south east);
    \draw[draw=none]  (-3,0)  -- node[below,yshift=-0.25cm]{{\scriptsize\color{mred}$u$-channel}}(-1,0)  ;
    \draw[draw=none]  (1,0)  -- node[above,yshift=0.25cm]{{\scriptsize\color{mgreen}$s$-channel}}(3,0)  ;
    \draw[draw=none]  (-1,0)  --node[above,yshift=0.25cm,fill=white]{{\scriptsize\color{mblue} Eucl. region}}(1,0)  ;
    \draw (0.15,0) node[yshift=-0.2cm] {\scriptsize $0$};
    \draw (1,0) node[xshift=.1cm,yshift=-0.2cm] {\scriptsize $1$};
    \draw (-1,0) node[xshift=-.2cm,yshift=0.2cm] {\scriptsize $-1$};
    \end{tikzpicture}
}
\newcommand{\IIIdiagAppendix}{
\begin{tikzpicture}[scale=2]
		\coordinate (e1) at (-0.5,0.);
		\coordinate (e2) at (2.5,0.);
		\coordinate (e3) at (2.5,-1.);
		\coordinate (e4) at (-0.5,-1.);	
		
		\coordinate (v1) at (0,0);
		\coordinate (v2) at (1,0);
		\coordinate (v3) at (2,0);
		\coordinate (v4) at (2,-1);
		\coordinate (v5) at (1,-1);
		\coordinate (v6) at (0,-1);
		
		\draw[massiveLin] (e1) -- (v1) ;
		\draw[massiveLin] (v1) -- (v2) node [midway,fill=white] {1};
		\draw[massiveLin] (v2) -- (v3) node [midway,fill=white] {3};
		\draw[massiveLin] (v3) -- (e2) ;
		\draw[massiveLin] (e4) -- (v6) ;
		\draw[massiveLin] (v6) -- (v5) node [midway,fill=white] {2};
		\draw[massiveLin] (v5) -- (v4) node [midway,fill=white] {4};
		\draw[massiveLin] (v4) -- (e3) ;
		
		\draw[massless] (v1) -- (v6) node [midway,fill=white] {5};
		\draw[massless] (v2) -- (v5) node [midway,fill=white] {7};
		\draw[massless] (v3) -- (v4) node [midway,fill=white] {6};
		\end{tikzpicture}
}

\newcommand{\HdiagAppendix}{
\begin{tikzpicture}[scale=2]
		
		\coordinate (e1) at (-0.5,0);
		\coordinate (e2) at (1.5,0);
		\coordinate (e3) at (1.5,-2);
		\coordinate (e4) at (-0.5,-2);	
		
		\coordinate (v1) at (0,0);
		\coordinate (v2) at (0,-1);
		\coordinate (v3) at (0,-2);
		\coordinate (v4) at (1,0);
		\coordinate (v5) at (1,-1);
		\coordinate (v6) at (1,-2);
		
		\draw[massiveLin] (e1) -- (v1);
		\draw[massiveLin]  (v1) -- (v4) node [midway,fill=white] {3};
		\draw[massiveLin] (e2) -- (v4);
		
		\draw[massiveLin] (e4) -- (v3);
		\draw[massiveLin]  (v3) -- (v6) node [midway,fill=white] {2};
		\draw[massiveLin] (e3) -- (v6);
		
		\draw[massless] (v1) -- (v2) node [midway,fill=white] {9};
		\draw[massless] (v2) -- (v3) node [midway,fill=white] {5};
		\draw[massless] (v4) -- (v5) node [midway,fill=white] {6};
		\draw[massless] (v5) -- (v6) node [midway,fill=white] {8};
		\draw[massless] (v2) -- (v5) node [midway,fill=white] {7};
		
		\end{tikzpicture}
}
\newcommand{\IXdiagAppendix}{
	\begin{tikzpicture}[scale=2]
	\coordinate (e1) at (-0.5,0.);
	\coordinate (e2) at (2.5,0.);
	\coordinate (e3) at (2.5,-1.);
	\coordinate (e4) at (-0.5,-1.);	
	\coordinate (t1) at (0,0);
	\coordinate (t2) at (1,0);
	\coordinate (t3) at (2,0);
	\coordinate (b1) at (0,-1);
	\coordinate (b2) at (1,-1);
	\coordinate (b3) at (2,-1);
	
	\draw[massiveLin] (e1) -- (t1) -- (t2) node [midway,fill=white] {1}-- (t3) node [midway,fill=white] {3}-- (e2);
	\draw[massiveLin] (e4) -- (b1) -- (b2) node [midway,fill=white] {2}-- (b3) node [midway,fill=white] {4}-- (e3);
	
	\draw[name path=rung1,massless] (t1) -- (b1) node [midway,fill=white] {5};
	\draw[name path=rung2,massless] (t2) -- (b3) node [near start,fill=white] {7};
	\draw[name path=rung3,massless] (t3) -- (b2) ;
	
	\fill[name intersections={of=rung2 and rung3, name=i,total=\t},white] foreach \s in {1,...,\t}{(i-\s) circle (5pt)};
	\draw[massless] (t3) -- (b2) node [near end,fill=white] {6};
	\end{tikzpicture}
}

\newcommand{\iteratedtwocut}{
\begin{tikzpicture}
            \coordinate (e1) at (-3,-1);	
	    	\coordinate (e2) at (-3, 1);
	    	\coordinate (e3) at ( 3, 1);
	    	\coordinate (e4) at ( 3,-1);
    
	    	\coordinate (v1) at (-2, 0);
	    	\coordinate (v2) at ( 2, 0);
	    	\coordinate (v0) at ( 0, 0);
	    	
	    	\coordinate (m1) at (-1, 1.00);
	    	\coordinate (m2) at (-1,-1.00);
	    	\coordinate (m3) at ( 1, 1.00);
	    	\coordinate (m4) at ( 1,-1.00);
    
	    	\draw[massive] (e1) -- (v1);
	    	\draw[massive] (e2) -- (v1);
	    	\draw[massive] (e3) -- (v2);
	    	\draw[massive] (e4) -- (v2);
    
	    	\draw[massive, name path=rung1] (v1) to[out=90, in=90](v0);
	    	\draw[massive, name path=rung2] (v1) to[out=-90, in=-90](v0);
	    	\draw[line width=0.25mm, dashed, name path=rung4, white] (m1) to (m2);
	    	\draw[massive, name path=rung6] (v0) to[out=90, in=90](v2);
	    	\draw[massive, name path=rung7] (v0) to[out=-90, in=-90](v2);
	    	\draw[line width=0.25mm, dashed, name path=rung5, white] (m3) to (m4);
	    	
            \draw[fill=lightgray, opacity=1] (v1) circle (0.65);
            \draw[fill=lightgray, opacity=1] (v2) circle (0.65);
            \draw[fill=lightgray, opacity=1] (v0) circle (0.65);
	    	\fill[name intersections={of=rung1 and rung4, name=i,total=\t},white] foreach \s in {1,...,\t}{(i-\s) circle (5pt)};
	    	\fill[name intersections={of=rung2 and rung4, name=i,total=\t},white] foreach \s in {1,...,\t}{(i-\s) circle (5pt)};
	    	\fill[name intersections={of=rung5 and rung6, name=i,total=\t},white] foreach \s in {1,...,\t}{(i-\s) circle (5pt)};
	    	\fill[name intersections={of=rung5 and rung7, name=i,total=\t},white] foreach \s in {1,...,\t}{(i-\s) circle (5pt)};
	    	\draw[line width=0.3mm, dashed, name path=rung4, blue] (m1) to (m2);
	    	\draw[line width=0.3mm, dashed, name path=rung5, blue] (m3) to (m4);
         	
	        \node[below left]  at (e1) {$p_1$};
	        \node[above left]  at (e2) {$p_2$};
	        \node[above right] at (e3) {$p_3$};
	        \node[below right] at (e4) {$p_4$};
	    	\node[above, fill=white, opacity=1] at (m1) {$\ell_4-p_2$};
	    	\node[below, fill=white, opacity=1] at (m2) {$\ell_3-p_1$};
	    	\node[above, fill=white, opacity=1] at (m3) {$\ell_2-p_2$};
	    	\node[below, fill=white, opacity=1] at (m4) {$\ell_1-p_1$};
\end{tikzpicture}}

\newcommand{\cutRuleAmpLoopTree}{
\begin{tikzpicture}
            \coordinate (e1) at (-2,-1);	
	    	\coordinate (e2) at (-2, 1);
	    	\coordinate (e3) at ( 2, 1);
	    	\coordinate (e4) at ( 2,-1);
    
	    	\coordinate (v1) at (-1, 0);
	    	\coordinate (v2) at ( 1, 0);
	    	
	    	\coordinate (m1) at (0, 1.00);
	    	\coordinate (m2) at (0,-1.00);
    
	    	\draw[massive] (e1) -- (v1);
	    	\draw[massive] (e2) -- (v1);
	    	\draw[massive] (e3) -- (v2);
	    	\draw[massive] (e4) -- (v2);
    
	    	\draw[massive, name path=rung1] (v1) to[out=90, in=90](v2);
	    	\draw[massive, name path=rung2] (v1) to[out=-90, in=-90](v2);
	    	\draw[line width=0.25mm, dashed, name path=rung4, white] (m1) to (m2);
	    	
            \draw[fill=lightgray, opacity=1] (v1) circle (0.65);
            \draw[fill=lightgray, opacity=1] (v2) circle (0.65);
            \draw[fill=white, opacity=1] (v1) circle (0.35);
	    	\fill[name intersections={of=rung1 and rung4, name=i,total=\t},white] foreach \s in {1,...,\t}{(i-\s) circle (5pt)};
	    	\fill[name intersections={of=rung2 and rung4, name=i,total=\t},white] foreach \s in {1,...,\t}{(i-\s) circle (5pt)};
    
	    	\draw[line width=0.3mm, dashed, name path=rung4, blue] (m1) to (m2);
        
\end{tikzpicture}}

\newcommand{\cutRuleAmpTreeLoop}{
\begin{tikzpicture}
                \coordinate (e1) at (-2,-1);	
	    	\coordinate (e2) at (-2, 1);
	    	\coordinate (e3) at ( 2, 1);
	    	\coordinate (e4) at ( 2,-1);
    
	    	\coordinate (v1) at (-1, 0);
	    	\coordinate (v2) at ( 1, 0);
	    	
	    	\coordinate (m1) at (0, 1.00);
	    	\coordinate (m2) at (0,-1.00);
    
	    	\draw[massive] (e1) -- (v1);
	    	\draw[massive] (e2) -- (v1);
	    	\draw[massive] (e3) -- (v2);
	    	\draw[massive] (e4) -- (v2);
    
	    	\draw[massive, name path=rung1] (v1) to[out=90, in=90](v2);
	    	\draw[massive, name path=rung2] (v1) to[out=-90, in=-90](v2);
	    	\draw[line width=0.25mm, dashed, name path=rung4, white] (m1) to (m2);
	    	
            \draw[fill=lightgray, opacity=1] (v1) circle (0.65);
            \draw[fill=lightgray, opacity=1] (v2) circle (0.65);
            \draw[fill=white, opacity=1] (v2) circle (0.35);
	    	\fill[name intersections={of=rung1 and rung4, name=i,total=\t},white] foreach \s in {1,...,\t}{(i-\s) circle (5pt)};
	    	\fill[name intersections={of=rung2 and rung4, name=i,total=\t},white] foreach \s in {1,...,\t}{(i-\s) circle (5pt)};
    
	    	\draw[line width=0.3mm, dashed, name path=rung4, blue] (m1) to (m2);
\end{tikzpicture}}

\newcommand{\cutRuleAmpTreeTree}{
\begin{tikzpicture}
            \coordinate (e1) at (-2.25,-1);	
	    	\coordinate (e2) at (-2.25, 1);
	    	\coordinate (e3) at ( 2.25, 1);
	    	\coordinate (e4) at ( 2.25,-1);
    
	    	\coordinate (v1) at (-1.25, 0);
	    	\coordinate (v2) at ( 1.25, 0);
	    	
	    	\coordinate (m1) at (0, 1.00);
	    	\coordinate (m2) at (0,-1.00);
    
	    	\draw[massive] (e1) -- (v1);
	    	\draw[massive] (e2) -- (v1);
	    	\draw[massive] (e3) -- (v2);
	    	\draw[massive] (e4) -- (v2);
    
	    	\draw[massive, name path=rung1] (v1) to[out=90, in=90](v2);
	    	\draw[massive, name path=rung2] (v1) to[out=-90, in=-90](v2);
	    	\draw[graviton, name path=rung3, line width=0.25mm] (v1) to (v2);
	    	\draw[line width=0.25mm, dashed, name path=rung4, white] (0, 1) to (0, -1);
	    	
            \draw[fill=lightgray, opacity=1] (v1) circle (0.65);
            \draw[fill=lightgray, opacity=1] (v2) circle (0.65);
	    	\fill[name intersections={of=rung1 and rung4, name=i,total=\t},white] foreach \s in {1,...,\t}{(i-\s) circle (5pt)};
	    	\fill[name intersections={of=rung2 and rung4, name=i,total=\t},white] foreach \s in {1,...,\t}{(i-\s) circle (5pt)};
	    	\fill[name intersections={of=rung3 and rung4, name=i,total=\t},white] foreach \s in {1,...,\t}{(i-\s) circle (5pt)};
    
	    	\draw[line width=0.3mm, dashed, name path=rung4, blue] (m1) to (m2);
    
\end{tikzpicture}}

\newcommand{\DiagIIIPhiVirt}{
	\begin{tikzpicture}[xscale=0.75,yscale=1.5]
	\coordinate (e1) at (-0.5,0.);
	\coordinate (e2) at (2.5,0.);
	\coordinate (e3) at (2.5,-1.);
	\coordinate (e4) at (-0.5,-1.);	
	
	\coordinate (v1) at (0,0);
	\coordinate (v2) at (1,0);
	\coordinate (v3) at (2,0);
	\coordinate (v4) at (2,-1);
	\coordinate (v5) at (1,-1);
	\coordinate (v6) at (0,-1);
	
	\draw[massivePhi] (e1) -- (e2);
	\draw[massivePhi] (e3) -- (e4);
	
	\draw[masslessPhi] (v1) -- (v6);
	\draw[masslessPhi] (v2) -- (v5);
	\draw[masslessPhi] (v3) -- (v4);
	\end{tikzpicture}
}

\newcommand{\DiagIIIPhiVirtCutLeft}{
	\begin{tikzpicture}[xscale=0.75,yscale=1.5]
	\coordinate (e1) at (-0.5,0.);
	\coordinate (e2) at (2.5,0.);
	\coordinate (e3) at (2.5,-1.);
	\coordinate (e4) at (-0.5,-1.);	
	
	\coordinate (v1) at (0,0);
	\coordinate (v2) at (1,0);
	\coordinate (v3) at (2,0);
	\coordinate (v4) at (2,-1);
	\coordinate (v5) at (1,-1);
	\coordinate (v6) at (0,-1);
	
	\draw[massivePhi] (e1) -- (e2);
	\draw[massivePhi] (e3) -- (e4);
	
	\draw[masslessPhi] (v1) -- (v6);
	\draw[masslessPhi] (v2) -- (v5);
	\draw[masslessPhi] (v3) -- (v4);
	
	\draw[unitaryCut] (-0.25,0.25) -- (-0.25,-1.25);
	\end{tikzpicture}}
\newcommand{\DiagIIIPhiCutLeft}{
	\begin{tikzpicture}[xscale=0.75,yscale=1.5]
	\coordinate (e1) at (-0.5,0.);
	\coordinate (e2) at (2.5,0.);
	\coordinate (e3) at (2.5,-1.);
	\coordinate (e4) at (-0.5,-1.);	
	
	\coordinate (v1) at (0,0);
	\coordinate (v2) at (1,0);
	\coordinate (v3) at (2,0);
	\coordinate (v4) at (2,-1);
	\coordinate (v5) at (1,-1);
	\coordinate (v6) at (0,-1);
	
	\draw[massivePhi] (e1) -- (e2);
	\draw[massivePhi] (e3) -- (e4);
	
	\draw[masslessPhi] (v1) -- (v6);
	\draw[masslessPhi] (v2) -- (v5);
	\draw[masslessPhi] (v3) -- (v4);
	
	\draw[unitaryCut] (0.5,0.25) -- (0.5,-1.25);
	\end{tikzpicture}
}
\newcommand{\DiagIIIPhiCutRight}{
	\begin{tikzpicture}[xscale=0.75,yscale=1.5]
	\coordinate (e1) at (-0.5,0.);
	\coordinate (e2) at (2.5,0.);
	\coordinate (e3) at (2.5,-1.);
	\coordinate (e4) at (-0.5,-1.);	
	
	\coordinate (v1) at (0,0);
	\coordinate (v2) at (1,0);
	\coordinate (v3) at (2,0);
	\coordinate (v4) at (2,-1);
	\coordinate (v5) at (1,-1);
	\coordinate (v6) at (0,-1);
	
	\draw[massivePhi] (e1) -- (e2);
	\draw[massivePhi] (e3) -- (e4);
	
	\draw[masslessPhi] (v1) -- (v6);
	\draw[masslessPhi] (v2) -- (v5);
	\draw[masslessPhi] (v3) -- (v4);
	
	\draw[unitaryCut] (1.5,0.25) -- (1.5,-1.25);
	\end{tikzpicture}}
\newcommand{\DiagIIIPhiVirtCutRight}{
	\begin{tikzpicture}[xscale=0.75,yscale=1.5]
	\coordinate (e1) at (-0.5,0.);
	\coordinate (e2) at (2.5,0.);
	\coordinate (e3) at (2.5,-1.);
	\coordinate (e4) at (-0.5,-1.);	
	
	\coordinate (v1) at (0,0);
	\coordinate (v2) at (1,0);
	\coordinate (v3) at (2,0);
	\coordinate (v4) at (2,-1);
	\coordinate (v5) at (1,-1);
	\coordinate (v6) at (0,-1);
	
	\draw[massivePhi] (e1) -- (e2);
	\draw[massivePhi] (e3) -- (e4);
		
	\draw[masslessPhi] (v1) -- (v6);
	\draw[masslessPhi] (v2) -- (v5);
	\draw[masslessPhi] (v3) -- (v4);
	
	\draw[unitaryCut] (2.25,0.25) -- (2.25,-1.25);
	\end{tikzpicture}}
\newcommand{\DiagIIIPhiCutNE}{
	\begin{tikzpicture}[xscale=0.75,yscale=1.5]
	\coordinate (e1) at (-0.5,0.);
	\coordinate (e2) at (2.5,0.);
	\coordinate (e3) at (2.5,-1.);
	\coordinate (e4) at (-0.5,-1.);	
	
	\coordinate (v1) at (0,0);
	\coordinate (v2) at (1,0);
	\coordinate (v3) at (2,0);
	\coordinate (v4) at (2,-1);
	\coordinate (v5) at (1,-1);
	\coordinate (v6) at (0,-1);
	
	\draw[massivePhi] (e1) -- (e2);
	\draw[massivePhi] (e3) -- (e4);
	
	\draw[masslessPhi] (v1) -- (v6);
	\draw[masslessPhi] (v2) -- (v5);
	\draw[masslessPhi] (v3) -- (v4);
	
	\draw[unitaryCut] (1.75,0.25) -- (0.25,-1.25);
	\end{tikzpicture}}
\newcommand{\DiagIIIPhiCutSE}{
	\begin{tikzpicture}[xscale=0.75,yscale=1.5]
	\coordinate (e1) at (-0.5,0.);
	\coordinate (e2) at (2.5,0.);
	\coordinate (e3) at (2.5,-1.);
	\coordinate (e4) at (-0.5,-1.);	
	
	\coordinate (v1) at (0,0);
	\coordinate (v2) at (1,0);
	\coordinate (v3) at (2,0);
	\coordinate (v4) at (2,-1);
	\coordinate (v5) at (1,-1);
	\coordinate (v6) at (0,-1);
	
	\draw[massivePhi] (e1) -- (e2);
	\draw[massivePhi] (e3) -- (e4);
	
	\draw[masslessPhi] (v1) -- (v6);
	\draw[masslessPhi] (v2) -- (v5);
	\draw[masslessPhi] (v3) -- (v4);
	
	\draw[unitaryCut] (0.25,0.25) -- (1.75,-1.25);
	\end{tikzpicture}
}
\newcommand{\DiagIIIVirt}{
	\begin{tikzpicture}[xscale=0.75,yscale=1.5]
	\coordinate (e1) at (-0.5,0.);
	\coordinate (e2) at (2.5,0.);
	\coordinate (e3) at (2.5,-1.);
	\coordinate (e4) at (-0.5,-1.);	
	
	\coordinate (v1) at (0,0);
	\coordinate (v2) at (1,0);
	\coordinate (v3) at (2,0);
	\coordinate (v4) at (2,-1);
	\coordinate (v5) at (1,-1);
	\coordinate (v6) at (0,-1);
	
	\draw[massive] (e1) -- (v1);
	\draw[massive] (v1) -- (v2);
	\draw[massive] (v2) -- (v3);
	\draw[massive] (v3) -- (e2);
	\draw[massive] (e4) -- (v6);
	\draw[massive] (v6) -- (v5);
	\draw[massive] (v5) -- (v4);
	\draw[massive] (v4) -- (e3);
	
	\draw[massless] (v1) -- (v6);
	\draw[massless] (v2) -- (v5);
	\draw[massless] (v3) -- (v4);
	\end{tikzpicture}}

\newcommand{\DiagIIICutNE}{
	\begin{tikzpicture}[xscale=0.75,yscale=1.5]
	\coordinate (e1) at (-0.5,0.);
	\coordinate (e2) at (2.5,0.);
	\coordinate (e3) at (2.5,-1.);
	\coordinate (e4) at (-0.5,-1.);	
	
	\coordinate (v1) at (0,0);
	\coordinate (v2) at (1,0);
	\coordinate (v3) at (2,0);
	\coordinate (v4) at (2,-1);
	\coordinate (v5) at (1,-1);
	\coordinate (v6) at (0,-1);
	
	\draw[massive] (e1) -- (v1);
	\draw[massive] (v1) -- (v2);
	\draw[massive] (v2) -- (v3);
	\draw[massive] (v3) -- (e2);
	\draw[massive] (e4) -- (v6);
	\draw[massive] (v6) -- (v5);
	\draw[massive] (v5) -- (v4);
	\draw[massive] (v4) -- (e3);
	
	\draw[massless] (v1) -- (v6);
	\draw[massless] (v2) -- (v5);
	\draw[massless] (v3) -- (v4);
	\draw[unitaryCut] (1.75,0.25) -- (0.25,-1.25);
	\end{tikzpicture}}
\newcommand{\DiagIIICutLeft}{
	\begin{tikzpicture}[xscale=0.75,yscale=1.5]
	\coordinate (e1) at (-0.5,0.);
	\coordinate (e2) at (2.5,0.);
	\coordinate (e3) at (2.5,-1.);
	\coordinate (e4) at (-0.5,-1.);	
	
	\coordinate (v1) at (0,0);
	\coordinate (v2) at (1,0);
	\coordinate (v3) at (2,0);
	\coordinate (v4) at (2,-1);
	\coordinate (v5) at (1,-1);
	\coordinate (v6) at (0,-1);
	
	\draw[massive] (e1) -- (v1);
	\draw[massive] (v1) -- (v2);
	\draw[massive] (v2) -- (v3);
	\draw[massive] (v3) -- (e2);
	\draw[massive] (e4) -- (v6);
	\draw[massive] (v6) -- (v5);
	\draw[massive] (v5) -- (v4);
	\draw[massive] (v4) -- (e3);
	
	\draw[massless] (v1) -- (v6);
	\draw[massless] (v2) -- (v5);
	\draw[massless] (v3) -- (v4);
	\draw[unitaryCut] (0.5,0.25) -- (0.5,-1.25);
	\end{tikzpicture}}

\newcommand{\DiagIXVirt}{
	\begin{tikzpicture}[xscale=0.75,yscale=1.5]
	\coordinate (e1) at (-0.5,0.);
	\coordinate (e2) at (2.5,0.);
	\coordinate (e3) at (2.5,-1.);
	\coordinate (e4) at (-0.5,-1.);	
	\coordinate (t1) at (0,0);
	\coordinate (t2) at (1,0);
	\coordinate (t3) at (2,0);
	\coordinate (b1) at (0,-1);
	\coordinate (b2) at (1,-1);
	\coordinate (b3) at (2,-1);
	
	\draw[massive] (e1) -- (t1) -- (t2) -- (t3) -- (e2);
	\draw[massive] (e4) -- (b1) -- (b2) -- (b3) -- (e3);
	
	\draw[name path=rung1,massless] (t1) -- (b1) ;
	\draw[name path=rung2,massless] (t2) -- (b3) ;
	\draw[name path=rung3,massless] (t3) -- (b2) ;
	
	\fill[name intersections={of=rung2 and rung3, name=i,total=\t},white] foreach \s in {1,...,\t}{(i-\s) circle (5pt)};
	\draw[massless] (t3) -- (b2);
	\end{tikzpicture}
}
\newcommand{\DiagIXCutLeft}{
	\begin{tikzpicture}[xscale=0.75,yscale=1.5]
	\coordinate (e1) at (-0.5,0.);
	\coordinate (e2) at (2.5,0.);
	\coordinate (e3) at (2.5,-1.);
	\coordinate (e4) at (-0.5,-1.);	
	\coordinate (t1) at (0,0);
	\coordinate (t2) at (1,0);
	\coordinate (t3) at (2,0);
	\coordinate (b1) at (0,-1);
	\coordinate (b2) at (1,-1);
	\coordinate (b3) at (2,-1);
	
	\draw[massive] (e1) -- (t1) -- (t2) -- (t3) -- (e2);
	\draw[massive] (e4) -- (b1) -- (b2) -- (b3) -- (e3);
	
	\draw[name path=rung1,massless] (t1) -- (b1) ;
	\draw[name path=rung2,massless] (t2) -- (b3) ;
	\draw[name path=rung3,massless] (t3) -- (b2) ;
	
	\fill[name intersections={of=rung2 and rung3, name=i,total=\t},white] foreach \s in {1,...,\t}{(i-\s) circle (5pt)};
	\draw[massless] (t3) -- (b2);
	\draw[unitaryCut] (0.5,0.25) -- (0.5,-1.25);
	\end{tikzpicture}
}
\newcommand{\DiagIXCutNE}{
	\begin{tikzpicture}[xscale=0.75,yscale=1.5]
	\coordinate (e1) at (-0.5,0.);
	\coordinate (e2) at (2.5,0.);
	\coordinate (e3) at (2.5,-1.);
	\coordinate (e4) at (-0.5,-1.);	
	\coordinate (t1) at (0,0);
	\coordinate (t2) at (1,0);
	\coordinate (t3) at (2,0);
	\coordinate (b1) at (0,-1);
	\coordinate (b2) at (1,-1);
	\coordinate (b3) at (2,-1);
	
	\draw[massive] (e1) -- (t1) -- (t2) -- (t3) -- (e2);
	\draw[massive] (e4) -- (b1) -- (b2) -- (b3) -- (e3);
	
	\draw[name path=rung1,massless] (t1) -- (b1) ;
	\draw[name path=rung2,massless] (t2) -- (b3) ;
	\draw[name path=rung3,massless] (t3) -- (b2) ;
	
	\fill[name intersections={of=rung2 and rung3, name=i,total=\t},white] foreach \s in {1,...,\t}{(i-\s) circle (5pt)};
	\draw[massless] (t3) -- (b2);
	\draw[unitaryCut] (1.75,0.25) -- (0.25,-1.25);
	\end{tikzpicture}
}

\newcommand{\DiagHVirt}{
	\begin{tikzpicture}[yscale=0.75,xscale=1.5]
	
	\coordinate (e1) at (-0.5,0);
	\coordinate (e2) at (1.5,0);
	\coordinate (e3) at (1.5,-2);
	\coordinate (e4) at (-0.5,-2);	
	
	\coordinate (v1) at (0,0);
	\coordinate (v2) at (0,-1);
	\coordinate (v3) at (0,-2);
	\coordinate (v4) at (1,0);
	\coordinate (v5) at (1,-1);
	\coordinate (v6) at (1,-2);
	
	\draw[massive] (e1) -- (v1);
	\draw[massive] (v1) -- (v4);
	\draw[massive] (e2) -- (v4);
	\draw[massive] (e4) -- (v3);
	\draw[massive] (v3) -- (v6);
	\draw[massive] (e3) -- (v6);
	
	\draw[massless] (v1) -- (v2);
	\draw[massless] (v2) -- (v3);
	\draw[massless] (v4) -- (v5);
	\draw[massless] (v5) -- (v6);
	\draw[massless] (v2) -- (v5);
	
	\end{tikzpicture}
}

\newcommand{\DiagHCutCentre}{
	\begin{tikzpicture}[yscale=0.75,xscale=1.5]
	
	\coordinate (e1) at (-0.5,0);
	\coordinate (e2) at (1.5,0);
	\coordinate (e3) at (1.5,-2);
	\coordinate (e4) at (-0.5,-2);	
	
	\coordinate (v1) at (0,0);
	\coordinate (v2) at (0,-1);
	\coordinate (v3) at (0,-2);
	\coordinate (v4) at (1,0);
	\coordinate (v5) at (1,-1);
	\coordinate (v6) at (1,-2);
	
	\draw[massive] (e1) -- (v1);
	\draw[massive] (v1) -- (v4);
	\draw[massive] (e2) -- (v4);
	\draw[massive] (e4) -- (v3);
	\draw[massive] (v3) -- (v6);
	\draw[massive] (e3) -- (v6);
	
	\draw[massless] (v1) -- (v2);
	\draw[massless] (v2) -- (v3);
	\draw[massless] (v4) -- (v5);
	\draw[massless] (v5) -- (v6);
	\draw[massless] (v2) -- (v5);
	
	\draw[unitaryCut] (0.5,0.25) -- (0.5,-2.25);
	\end{tikzpicture}
}

\newcommand{\DiagHPhiVirtCutLeft}{
	\begin{tikzpicture}[yscale=0.75,xscale=1.5]
	
	\coordinate (e1) at (-0.5,0);
	\coordinate (e2) at (1.5,0);
	\coordinate (e3) at (1.5,-2);
	\coordinate (e4) at (-0.5,-2);	
	
	\coordinate (v1) at (0,0);
	\coordinate (v2) at (0,-1);
	\coordinate (v3) at (0,-2);
	\coordinate (v4) at (1,0);
	\coordinate (v5) at (1,-1);
	\coordinate (v6) at (1,-2);
	
	\draw[massivePhi] (e1) -- (e2);
	\draw[massivePhi] (e3) -- (e4);
	
	\draw[masslessPhi] (v1) -- (v2);
	\draw[masslessPhi] (v2) -- (v3);
	\draw[masslessPhi] (v4) -- (v5);
	\draw[masslessPhi] (v5) -- (v6);
	\draw[masslessPhi] (v2) -- (v5);
	
	\draw[unitaryCut] (-0.25,0.25) -- (-0.25,-2.25);
	\end{tikzpicture}}
\newcommand{\DiagHPhiCutCentre}{
	\begin{tikzpicture}[yscale=0.75,xscale=1.5]
	
	\coordinate (e1) at (-0.5,0);
	\coordinate (e2) at (1.5,0);
	\coordinate (e3) at (1.5,-2);
	\coordinate (e4) at (-0.5,-2);	
	
	\coordinate (v1) at (0,0);
	\coordinate (v2) at (0,-1);
	\coordinate (v3) at (0,-2);
	\coordinate (v4) at (1,0);
	\coordinate (v5) at (1,-1);
	\coordinate (v6) at (1,-2);
	
	\draw[massivePhi] (e1) -- (e2);
	\draw[massivePhi] (e3) -- (e4);
	
	\draw[masslessPhi] (v1) -- (v2);
	\draw[masslessPhi] (v2) -- (v3);
	\draw[masslessPhi] (v4) -- (v5);
	\draw[masslessPhi] (v5) -- (v6);
	\draw[masslessPhi] (v2) -- (v5);
	
	\draw[unitaryCut] (0.5,0.25) -- (0.5,-2.25);
	\end{tikzpicture}}
	
\newcommand{\DiagHPhiVirtCutRight}{
	\begin{tikzpicture}[yscale=0.75,xscale=1.5]
	
	\coordinate (e1) at (-0.5,0);
	\coordinate (e2) at (1.5,0);
	\coordinate (e3) at (1.5,-2);
	\coordinate (e4) at (-0.5,-2);	
	
	\coordinate (v1) at (0,0);
	\coordinate (v2) at (0,-1);
	\coordinate (v3) at (0,-2);
	\coordinate (v4) at (1,0);
	\coordinate (v5) at (1,-1);
	\coordinate (v6) at (1,-2);
	
	\draw[massivePhi] (e1) -- (e2);
	\draw[massivePhi] (e3) -- (e4);
	
	\draw[masslessPhi] (v1) -- (v2);
	\draw[masslessPhi] (v2) -- (v3);
	\draw[masslessPhi] (v4) -- (v5);
	\draw[masslessPhi] (v5) -- (v6);
	\draw[masslessPhi] (v2) -- (v5);
	
	\draw[unitaryCut] (1.25,0.25) -- (1.25,-2.25);
	\end{tikzpicture}
}

\newcommand{\buboxs}{
\begin{tikzpicture}
	    \coordinate (e1) at (-1, 0);
		\coordinate (e2) at ( 1, 0);
		\coordinate (e3) at ( 1,-1);
		\coordinate (e4) at (-1,-1);	
		
		\coordinate (v1) at (-0.5,0);
		\coordinate (v2) at (-0.5,-1);
		\coordinate (v3) at (0.5,0);
		\coordinate (v4) at (0.5,-1);
		\coordinate (v5) at (-1,0);
		\coordinate (v6) at (1,0);
		
		\draw[massive] (e1) -- (v1);
		\draw[massive]  (v1) -- (v3);
		\draw[massive] (e2) -- (v3);

		\draw[massive] (e4) -- (v2);
		\draw[massive]  (v2) -- (v4);
		\draw[massive] (e3) -- (v4);
		
		\draw[massless] (v1) -- (v2);
		\draw[massless] (v3) -- (v4);

        \draw[fill=white, opacity=1] (0.5,-0.5) circle (0.2);
		\node[left]  at (e1) {$p_2$};
	    \node[right]  at (e2) {$p_3$};
	    \node[right] at (e3) {$p_4$};
	    \node[left] at (e4) {$p_1$};
		\end{tikzpicture}
		}
		
\newcommand{\buboxu}{
\begin{tikzpicture}
	    \coordinate (e1) at (-1, 0);
		\coordinate (e2) at ( 1, 0);
		\coordinate (e3) at ( 1,-1);
		\coordinate (e4) at (-1,-1);	
		
		\coordinate (v1) at (-0.5,0);
		\coordinate (v2) at (-0.5,-1);
		\coordinate (v3) at (0.5,0);
		\coordinate (v4) at (0.5,-1);
		\coordinate (v5) at (-1,0);
		\coordinate (v6) at (1,0);
		
		\draw[massive] (e1) -- (v1);
		\draw[massive] (v1) -- (v3);
		\draw[massive] (e2) -- (v3);

		\draw[massive] (e4) -- (v2);
		\draw[massive] (v2) -- (v4);
		\draw[massive] (e3) -- (v4);
		
		\draw[massless] (v1) -- (v2);
		\draw[massless] (v3) -- (v4);

        \draw[fill=white, opacity=1] (0.5,-0.5) circle (0.2);
		\node[left]  at (e1) {$p_3$};
	    \node[right]  at (e2) {$p_2$};
	    \node[right] at (e3) {$p_4$};
	    \node[left] at (e4) {$p_1$};
		\end{tikzpicture}
		}
		
\newcommand{\buboxcut}{
\begin{tikzpicture}
	    \coordinate (e1) at (-1, 0);
		\coordinate (e2) at ( 1, 0);
		\coordinate (e3) at ( 1,-1);
		\coordinate (e4) at (-1,-1);	
		
		\coordinate (v1) at (-0.5,0);
		\coordinate (v2) at (-0.5,-1);
		\coordinate (v3) at (0.5,0);
		\coordinate (v4) at (0.5,-1);
		\coordinate (v5) at (-1,0);
		\coordinate (v6) at (1,0);
		
		\draw[massive] (e1) -- (v1);
		\draw[massive]  (v1) -- (v3);
		\draw[massive] (e2) -- (v3);

		\draw[massive] (e4) -- (v2);
		\draw[massive]  (v2) -- (v4);
		\draw[massive] (e3) -- (v4);
		
		\draw[massless] (v1) -- (v2);
		\draw[massless] (v3) -- (v4);
		\draw[unitaryCut] (0,0.25) -- (0,-1.25);

        \draw[fill=white, opacity=1] (0.5,-0.5) circle (0.2);
		\node[left]  at (e1) {$p_2$};
	    \node[right]  at (e2) {$p_3$};
	    \node[right] at (e3) {$p_4$};
	    \node[left] at (e4) {$p_1$};
		\end{tikzpicture}
		}
		
\newcommand{\arrowIn}{
    \tikz \draw[-stealth] (0,1pt) -- (0,-1pt);
    }
		
\newcommand{\ladderlabelsleft}[3]{
		\begin{tikzpicture}
		\coordinate (e1) at (-0.5, 0);
		\coordinate (e2) at ( 2.5, 0);
		\coordinate (e3) at ( 2.5,-1);
		\coordinate (e4) at (-0.5,-1);	
		
		\coordinate (v1) at (0,0);
		\coordinate (v2) at (1,0);
		\coordinate (v3) at (2,0);
		\coordinate (v4) at (2,-1);
		\coordinate (v5) at (1,-1);
		\coordinate (v6) at (0,-1);
		
		\draw[massive] (e1) -- (v1);
		\draw[massive] (v1) -- (v2);
		\draw[massive] (v2) -- (v3);
		\draw[massive] (v3) -- (e2);
		\draw[massive] (e4) -- (v6);
		\draw[massive] (v6) -- (v5);
		\draw[massive] (v5) -- (v4);
		\draw[massive] (v4) -- (e3);
		
		\draw[massless] (v1) -- (v6) node[pos=0.25]{\arrowIn};
		\draw[massless] (v2) -- (v5) node[pos=0.25]{\arrowIn};
		\draw[massless] (v3) -- (v4) node[pos=0.25]{\arrowIn};
		
		\draw[unitaryCut] (0.5,0.25) -- (0.5,-1.25);
		
		\node[below left]  at (v1) {#1};
	    \node[below right] at (v2) {#2};
	    \node[below right] at (v3) {#3};
		\end{tikzpicture}
}

\newcommand{\ladderlabelsright}[3]{
		\begin{tikzpicture}
		\coordinate (e1) at (-0.5, 0);
		\coordinate (e2) at ( 2.5, 0);
		\coordinate (e3) at ( 2.5,-1);
		\coordinate (e4) at (-0.5,-1);	
		
		\coordinate (v1) at (0,0);
		\coordinate (v2) at (1,0);
		\coordinate (v3) at (2,0);
		\coordinate (v4) at (2,-1);
		\coordinate (v5) at (1,-1);
		\coordinate (v6) at (0,-1);
		
		\draw[massive] (e1) -- (v1);
		\draw[massive] (v1) -- (v2);
		\draw[massive] (v2) -- (v3);
		\draw[massive] (v3) -- (e2);
		\draw[massive] (e4) -- (v6);
		\draw[massive] (v6) -- (v5);
		\draw[massive] (v5) -- (v4);
		\draw[massive] (v4) -- (e3);
		
		\draw[massless] (v1) -- (v6) node[pos=0.25]{\arrowIn};
		\draw[massless] (v2) -- (v5) node[pos=0.25]{\arrowIn};
		\draw[massless] (v3) -- (v4) node[pos=0.25]{\arrowIn};
		
		\draw[unitaryCut] (1.5,0.25) -- (1.5,-1.25);
		
		\node[below left]  at (v1) {#1};
	    \node[below left] at (v2) {#2};
	    \node[below right] at (v3) {#3};
		\end{tikzpicture}
}

\newcommand{\ladderlabelsNWSE}[3]{
		\begin{tikzpicture}
		\coordinate (e1) at (-0.5, 0);
		\coordinate (e2) at ( 2.5, 0);
		\coordinate (e3) at ( 2.5,-1);
		\coordinate (e4) at (-0.5,-1);	
		
		\coordinate (v1) at (0,0);
		\coordinate (v2) at (1,0);
		\coordinate (v3) at (2,0);
		\coordinate (v4) at (2,-1);
		\coordinate (v5) at (1,-1);
		\coordinate (v6) at (0,-1);
		
		\draw[massive] (e1) -- (v1);
		\draw[massive] (v1) -- (v2);
		\draw[massive] (v2) -- (v3);
		\draw[massive] (v3) -- (e2);
		\draw[massive] (e4) -- (v6);
		\draw[massive] (v6) -- (v5);
		\draw[massive] (v5) -- (v4);
		\draw[massive] (v4) -- (e3);
		
		\draw[massless] (v1) -- (v6) node[pos=0.25]{\arrowIn};
		\draw[massless] (v2) -- (v5) node[pos=0.25]{\arrowIn};
		\draw[massless] (v3) -- (v4) node[pos=0.25]{\arrowIn};
		
		\draw[unitaryCut] (0.25,0.25) -- (1.75,-1.25);
		
		\node[below left]  at (v1) {#1};
	    \node[below right] at (v2) {#2};
	    \node[below right] at (v3) {#3};
		\end{tikzpicture}
}

\newcommand{\ladderlabelsSWNE}[3]{
		\begin{tikzpicture}
		\coordinate (e1) at (-0.5, 0);
		\coordinate (e2) at ( 2.5, 0);
		\coordinate (e3) at ( 2.5,-1);
		\coordinate (e4) at (-0.5,-1);	
		
		\coordinate (v1) at (0,0);
		\coordinate (v2) at (1,0);
		\coordinate (v3) at (2,0);
		\coordinate (v4) at (2,-1);
		\coordinate (v5) at (1,-1);
		\coordinate (v6) at (0,-1);
		
		\draw[massive] (e1) -- (v1);
		\draw[massive] (v1) -- (v2);
		\draw[massive] (v2) -- (v3);
		\draw[massive] (v3) -- (e2);
		\draw[massive] (e4) -- (v6);
		\draw[massive] (v6) -- (v5);
		\draw[massive] (v5) -- (v4);
		\draw[massive] (v4) -- (e3);
		
		\draw[massless] (v1) -- (v6) node[pos=0.25]{\arrowIn};
		\draw[massless] (v2) -- (v5) node[pos=0.25]{\arrowIn};
		\draw[massless] (v3) -- (v4) node[pos=0.25]{\arrowIn};
		
		\draw[unitaryCut] (0.25,-1.25) -- (1.75,0.25);
		
		\node[below left]  at (v1) {#1};
	    \node[below left] at (v2) {#2};
	    \node[below right] at (v3) {#3};
		\end{tikzpicture}
}

\graphicspath{{figures/}}
\usepackage[export]{adjustbox}

\newcommand{\nocontentsline}[3]{}
\newcommand{\tocless}[2]{\bgroup\let\addcontentsline=\nocontentsline#1{#2}\egroup}

\newcommand{\tblue}[1]{\textcolor{blue}{#1}}
\newcommand{\tred}[1]{\textcolor{red}{#1}}

\newcommand{\calO}{\mathcal{O}}
\newcommand{\calN}{\mathcal{N}}
\newcommand{\calM}{\mathcal{M}}
\newcommand{\calI}{\mathcal{I}}

\newcommand{\qperp}{ \bm{q}_{\perp}}
\newcommand{\qperpsq}{\bm{q}^{2}_{\perp}}
\newcommand{\bperp}{\bm{b}_{\perp}}
\newcommand{\bperpsq}{\bm{b}^2_{\perp}}

\newcommand{\nn}{\nonumber}
\renewcommand{\imath}{\mathrm{i}}

\newcommand{\cons}{{\rm cons.}}
\newcommand{\eik}{{\rm eik}}

\def\RT{\mathrm{III}}

\def\IX{\mathrm{IX}}

\def\H{\mathrm{H}}

\def\EulerGamma{\gamma_{E}}
\def\relfbar{y}
\def\sqrtmQSq{\sqrt{-q^2}}

\let\Im\relax

\DeclareMathOperator{\Im}{Im}

\begin{document}
\title{Radiative Classical Gravitational Observables at $\mathcal{O}(G^3)$ from Scattering Amplitudes}

\author[1]{Enrico Herrmann,}
\affiliation[1]{Mani L. Bhaumik Institute for Theoretical Physics,\\
UCLA Department of Physics and Astronomy, Los Angeles, CA 90095, USA}
\emailAdd{eh10@g.ucla.edu}

\author[2]{Julio~Parra-Martinez,}
\affiliation[2]{Walter Burke Institute for Theoretical Physics, \\
California Institute of Technology, Pasadena, CA 91125, USA}
\emailAdd{jparram@caltech.edu}

\author[3]{Michael S. Ruf,}
\affiliation[3]{Physikalisches Institut, Albert-Ludwigs Universität Freiburg, D-79104 Freiburg, Germany}
\emailAdd{michael.ruf@physik.uni-freiburg.de}

\author[4]{Mao Zeng}
\affiliation[4]{Rudolf Peierls Centre for Theoretical Physics, \\
University of Oxford, Parks Road, Oxford OX1 3PU, United Kingdom
}
\emailAdd{mao.zeng@physics.ox.ac.uk}

%================================================
\abstract{
We compute classical gravitational observables for the scattering of two spinless black holes in general relativity and $\calN {=} 8$ supergravity in the formalism of Kosower, Maybee, and O'Connell (KMOC). We focus on the gravitational impulse with radiation reaction and the radiated momentum in black hole scattering at $\calO(G^3)$ to all orders in the velocity. These classical observables require the construction and evaluation of certain loop-level quantities which are greatly simplified by harnessing recent advances from scattering amplitudes and collider physics. In particular, we make use of generalized unitarity to construct the relevant loop integrands, employ reverse unitarity, the method of regions, integration-by-parts (IBP), and (canonical) differential equations to simplify and evaluate all loop and phase-space integrals to obtain the classical gravitational observables of interest to two-loop order. The KMOC formalism naturally incorporates radiation effects which enables us to explore these classical quantities beyond the conservative two-body dynamics. From the impulse and the radiated momentum, we extract the scattering angle and the radiated energy. Finally, we discuss universality of the impulse in the high-energy limit and the relation to the eikonal phase. 
}
%\preprint{Preprint-Number}

%================================================

\maketitle

%
%================================================
%
\section{Introduction}
%
%================================================
%
%
The increasing experimental success of current gravitational wave astronomy~\cite{Abbott:2016blz,TheLIGOScientific:2017qsa} combined with the design specifications of future detectors \cite{Punturo_2010,Audley:2017drz,Reitze:2019iox} require theoretical predictions for the classical general relativistic two-body problem to keep up with the experimental accuracy~\cite{Purrer:2019jcp}.

An important tool for the generation of waveform templates, used for detection and parameter estimation, are fast and reliable semi-analytic models of the binary merger. Prominent examples of are the ones provided by the effective one-body (EOB) formalism \cite{Buonanno_1999}, which take as input numerical simulations combined with analytic results of the late ringdown and early inspiralling phase of the merger. In the latter phase, the two bodies are still widely separated and in a weak-field, slow motion regime---amenable to perturbation theory. Traditionally, this phase is analyzed in the Post-Newtonian (PN) approximation, which is an expansion both in Newton's constant, $G$, as well as the relative velocity $v$ of the constituents. Both are linked by the virial theorem for a bound system.

Besides the gravitational inspiral, one can also consider scattering (or hyperbolic) events in which compact objects fly past one another while interacting gravitationally. In such events, the deflection of the objects' trajectories is accompanied by the emission of gravitational radiation (dubbed gravitational Bremsstrahlung in analogy to the electromagnetic case). Even though hyperbolic motion events currently appear to be out of experimental reach \cite{Kocsis:2006hq,Mukherjee:2020hnm}, it has been suggested \cite{Damour:2016gwp} that scattering observables, nonetheless, can be used as input to determine parameters in EOB models which are subsequently applied to the bound state problem. Furthermore, in certain favorable circumstances, it is even possible to directly link bound and unbound observables via analytic continuation \cite{Kalin:2019rwq,Kalin:2019inp,Bini:2020hmy}. For scattering kinematics, as opposed to virialized bound state systems, the velocity and $G$ are not necessarily linked expansion parameters and it is therefore possible to explore perturbation theory in $G$ only, to all orders in $v$, the so-called Post-Minkowskian (PM) expansion. The leading order waveform in this regime has been constructed in papers by Peters, Kovacs, Thorne, and Crowley in the 1970s \cite{Peters:1970mx,Kovacs:1974aa,Crowley:1977us,Kovacs:1977uw,Kovacs:1978eu}, and was recently revisited in Refs.~\cite{DeVittori:2012da,Grobner:2020fnb,Capozziello:2008mn,Mogull:2020sak,Jakobsen:2021smu,Mougiakakos:2021ckm}. 

From the previous comments, it should come as no surprise that the classical gravitational two-body problem has attracted a renewed broad interest ranging from the classical general relativity community to effective field theorists and scattering amplitude practitioners. In our work, we focus on scattering amplitude tools to study classical black hole interactions in hyperbolic orbits. Recent years have seen a number of applications of on-shell techniques, pioneered in Quantum Field Theory, to the classical general relativistic two-body problem. Notable examples include generalized unitarity \cite{Bern:1994zx,Bern:1994cg,Britto:2004nc}, the double copy \cite{Bern:2008qj,Bern:2010ue,Bern:2012uf,Bern:2017ucb, Bern:2018jmv,Bern:2019prr}, and effective field theory ideas \cite{Goldberger:2004jt,Porto:2016pyg,Cheung:2018wkq} that have produced new results for the dynamics of spinless \cite{Damour:2016gwp,Foffa:2016rgu,Blumlein:2019zku,Blumlein:2020znm,Bjerrum-Bohr:2018xdl,Cheung:2018wkq,Bern:2019nnu,Foffa:2019hrb,Bern:2019crd,Cheung:2020gyp,Kalin:2020mvi,Kalin:2020fhe,Cristofoli:2020uzm,Bini:2020uiq,Bini:2020rzn,Loebbert:2020aos,Bern:2021dqo} and spinning \cite{Vaidya:2014kza,Vines:2017hyw,Guevara:2017csg,Vines:2018gqi,Guevara:2018wpp,Chung:2018kqs,Guevara:2019fsj,Chung:2019duq,Damgaard:2019lfh,Aoude:2020onz,Bern:2020buy,Guevara:2020xjx,Levi:2020kvb,Levi:2020uwu} black holes, including finite-size effects \cite{Cheung:2020sdj,Haddad:2020que,Kalin:2020lmz,Brandhuber:2019qpg,Huber:2019ugz,AccettulliHuber:2020oou,Bern:2020uwk,Cheung:2020gbf,Aoude:2020ygw}.  Relativistic covariance, innate to these scattering amplitude techniques, allows for the extraction of results to all orders in velocity, i.e.\ the Post-Minkowskian expansion. Most of the initial applications of scattering amplitudes tools in the classical gravitational context involved auxiliary quantities and one derives either the eikonal phase \cite{Amati:1990xe,Ciafaloni:2015xsr,Ciafaloni:2018uwe,KoemansCollado:2019ggb,Parra-Martinez:2020dzs} or a conservative Hamiltonian \cite{Cheung:2018wkq,Cristofoli:2019neg} via an EFT matching procedure \cite{Cheung:2018wkq}. 

In a beautiful paper, Kosower, Maybee, and O'Connell (KMOC) \cite{Kosower:2018adc} pointed out how to directly relate classical observables to scattering amplitudes using a version of the in-in formalism. In their original work (and in its extension to spinning objects \cite{Maybee:2019jus}) the formalism was verified at leading order (LO) and next-to-leading order (NLO) by comparing their formulae for the electromagnetic impulse to expressions obtained by solving classical equations of motion. Beyond leading order, however, these checks were performed at the level of unintegrated expressions, and the full evaluation of the corresponding loop and phase-space integrals was left as a problem for the future. Naively, the solution to this problem poses considerable technical challenges. In our previous letter \cite{Herrmann:2021lqe}, we have emphasized the similarities between the KMOC setup and cross-section calculations in traditional particle-physics settings. We also emphasized how collider physics ideas, like reverse unitarity \cite{Anastasiou:2002yz,Anastasiou:2002qz,Anastasiou:2003yy,Anastasiou:2015yha}, integration-by-parts reduction \cite{Tkachov:1981wb,Chetyrkin:1981qh}, and (canonical) differential equations \cite{Kotikov:1990kg,Bern:1992em,Gehrmann:1999as,Henn:2013pwa,Henn:2014qga} are ideally suited to render the KMOC formalism a practical computational tool to derive state-of-the-art results for the relativistic two-body dynamics. As an example we computed the radiated momentum in a binary black hole encounter \cite{Herrmann:2021lqe}.

It is the aim of this work to elaborate on the technical details of our recent letter,  and to obtain additional gravitational observables. More concretely, we calculate the $\calO(G^3)$ gravitational impulse, i.e.\ the momentum change between the initial and final state of one of the scattering black holes. This verifies the classical calculations of Portilla \cite{Portilla:1979xx,Portilla:1980uz} and Westpfahl \cite{Westpfahl:1985tsl,Westpfahl_1987} done several decades ago, and extends them to one higher order using very different methods. From the impulse, it is possible to extract the radiative scattering angle at $\calO(G^3)$. Inspired by a computation in maximal supergravity \cite{DiVecchia:2020ymx}, the radiative GR angle has been obtained earlier by Damour \cite{Damour:2020tta} from a linear response computation and was later confirmed (subject to certain assumptions) by DiVecchia et al. \cite{DiVecchia:2021ndb}, using eikonal methods. This scattering angle, which includes radiation reaction corrections, also cleared up some of the confusions arising in the high-energy limit of the conservative result of Refs.~\cite{Bern:2019nnu,Bern:2019crd}. 

Since the technical bottleneck involves the evaluation of loop integrals, we give a detailed account of the computation of all relevant master integrals that appear in the KMOC setup at $\calO(G^3)$, i.e.\ at two loops. As will be reviewed in the main text, the KMOC formalism includes both virtual loop amplitudes as well as phase-space integrals over products of lower order amplitudes. Since we are interested in classical physics, these amplitudes are expanded in the $\hbar\to0$ limit, which is equivalent to the \emph{soft region} in the language of expansion by regions \cite{Beneke:1997zp}. We therefore lay out how to expand, reduce, and evaluate all relevant two-loop soft integrals via differential equations that were previously adapted to the Post-Minkowskian expansion in classical gravity \cite{Parra-Martinez:2020dzs}. One of the key advantages of the KMOC formalism, together with the integration tools described here, is that we can treat \emph{inclusive} observables such as the total radiated momentum or the impulse, that only depend on a small number of kinematic scales in an efficient and streamlined fashion without having to go through the multi-scale gravitational waveform where subsequent integrations are technically challenging.

The remainder of this work is organized as follows. In section \ref{sec:KMOC}, we briefly recall some of the features of the KMOC formalism and summarize how to represent the gravitational impulse (subsec.~\ref{subsec:impulse}) as well as the radiated momentum (subsec.~\ref{subsec:radiated_momentum}) in terms of scattering amplitudes and their unitarity cuts. In section \ref{sec:setup_review}, we give a broad review of the methods that have already been successfully applied to the conservative two-body dynamics. We discuss the scattering kinematics, the relevant classical regions together with a brief reminder of generalized unitarity that allows us to efficiently derive the relevant loop integrands. Starting from these integrands, we recall the classical expansion (i.e.\ the soft expansion in the method of regions \cite{Beneke:1997zp}) in subsection \ref{subsec:ibp_de}, before reducing all integrals to a set of independent masters with the help of integration-by-parts relations. In order to obtain analytic expressions for the resulting master integrals, we also review the differential equations of Ref.~\cite{Parra-Martinez:2020dzs}. Section \ref{sec:soft_region_methods} comments on the novelties of the soft region and introduces reverse unitarity in subsection \ref{subsec:rev_unitarity} to treat phase-space integrals (involving on-shell delta functions) on the same footing as virtual ones. Section \ref{sec:soft_integrals} is devoted to the evaluation of the virtual and cut master integrals in the soft region. Evaluation of these integrals require the knowledge of differential equations and their boundary conditions at certain convenient kinematic points. We furthermore discuss the analytic continuation of these solutions to the relevant physical scattering regions. Next, we comment on several general properties of the KMOC observables and simplified properties in terms of scattering amplitudes in section \ref{sec:KMOC_discussion_expansion}. In section \ref{sec:result} we present the final results for the gravitational impulse in $\calN=8$ supergravity and general relativity and we investigate universality in the high-energy limit and the relation to the eikonal calculation in Refs.~\cite{DiVecchia:2021ndb,DiVecchia:2021bdo}. We close with our conclusions and an outlook to future directions. Appendices \ref{app:FT_collection}, \ref{app:angle_impulse}, \ref{app:cutting_rules}, and \ref{app:masters_notation} respectively include details on relevant Fourier transformation identities, the relation between the impulse and the scattering angle, unitarity relations and cutting rules to determine certain phase-space integrals from the imaginary part of virtual diagrams, as well as our conventions for the soft master integrals. The results of all soft master integrals, as well as our conventions are attached to this \texttt{arXiv} submission as computer readable files.

\medskip
\noindent
\textbf{Note added: } In the course of this work, we learned from an independent computation of several of the two-loop soft master integrals by Di Vecchia, Heissenberg, Russo, and Veneziano \cite{DiVecchia:2021bdo} in the context of the eikonal approach to classical gravitational scattering. We are grateful for discussions and comparisons as well as for coordinating publication.

%================================================
%
\section{Gravitational observables via the KMOC formalism}
\label{sec:KMOC}
%
%================================================
%
To begin, let us briefly review the key features of the KMOC formalism. For a detailed discussion, the reader is referred to the original article \cite{Kosower:2018adc}. The basic idea of KMOC is to set up a thought experiment for the scattering of two wavepackets evolving from the asymptotic past to the asymptotic future and to measure the change of some observable, $\Delta O$. In the asymptotic past, the wavepackets are represented by $|{\text{in}} \rangle$, an \emph{in} quantum state constructed from two-particle momentum eigenstates $|p_1,p_2\rangle_{\text{in}}$ with wavefunctions $\phi_i(p_i)$, which are well separated by an impact parameter $b^\mu$
\begin{align}
\label{eq:wavefuncs}
 | {\text{in}} \rangle = \int \mathrm{d}\Phi_2(p_1,p_2) \, 
					\phi_1(p_1)\phi_2(p_2) e^{\imath \, b\cdot p_1/\hbar} \, |p_1,p_2\rangle_{\text{in}}\,.
\end{align}
From now on we will drop the \emph{in} subscript in the momentum eigenstates and leave it implicit.  For convenience, we have introduced the Lorentz-invariant multi-particle on-shell phase-space measure\footnote{Following \cite{Kosower:2018adc}, we introduce the notation 
$\hat{\mathrm{d}} x  \equiv  \mathrm{d} x/(2\pi)$ and 
$\hat \delta(x) \equiv (2\pi)\delta(x)$.}
\begin{equation}
\label{eq:1P_phase_space_all_pi}
 \mathrm{d}\Phi_{n}(p_1,\cdots\!,p_n) = 
 \prod_{i}  \mathrm{d}\Phi_1(p_i) \quad \text{with} 
 \quad
 \mathrm{d}\Phi_1(p_i) = \hat{\mathrm{d}}^D p_i\,  
                        \Theta(\pm p_i^0)\,\hat\delta(p^2_i-m^2_i)\,.
\end{equation}
We work in an all-outgoing convention for the momenta $p_i$, in which physical incoming (outgoing) states have negative (positive) energy components. The sign in the Heavyside function, $\Theta$, is chosen accordingly.

The incoming state $| {\text{in}} \rangle$ will evolve to an \emph{out} state in the asymptotic future, $|\text{out}\rangle$, which might contain additional particles produced during the interaction. The change in an observable, $O$, can be simply obtained by evaluating the difference of the expectation value of the corresponding Hermitean operator, $\mathbb{O}$, between \emph{in} and \emph{out} states 
\begin{equation}
\label{eq:kmoc_start}
 \Delta O = \langle {\rm out} | \mathbb{O} | {\rm out} \rangle  - \langle {\rm in} | \mathbb{O} | {\rm in} \rangle\,.
\end{equation}
In quantum mechanics, the \emph{out} states are related to the \emph{in} states by the time evolution operator, i.e.\ the S-matrix: $|\text{out}\rangle = S |\text{in} \rangle$ and we can write 
\begin{equation}
 \Delta O = \imath \int  \mathrm{d}\Phi_4(p_1,\cdots\!,p_4)  \, 
					\phi_1(p_1)\phi_2(p_2) \phi_2(p_3)^*\phi_1(p_4)^* \hat\delta^{(D)}(\textstyle\sum_i p_i) e^{\imath \, b\cdot (p_1+p_4)/\hbar} \, \, {\cal I}_O
		 \,.
\end{equation}
The stripped kernel $\calI_O$ is related to the matrix elements via
\begin{equation}
 \widetilde {\cal I}_O =\widehat\delta^{(D)}(\textstyle\sum_i p_i)\,\calI_O
    =-\imath\, \langle p_4,p_3|S^\dagger[\mathbb{O},S]|p_1,p_2\rangle\,,
\end{equation}
where, to arrive at this expression, we have used the unitarity of the S-matrix, $S^\dagger S=1$. Following \cite{Kosower:2018adc}, $\widetilde {\cal I}_O$ can be related to scattering amplitudes by writing $S=1+\imath T$ such that
\begin{equation}
 \widetilde {\cal I}_O =
        \widetilde {\cal I}_{O,\,{\rm v}} 
        + \widetilde {\cal I}_{O,\,{\rm r}} 
    = \,\langle p_4,p_3|[\mathbb{O},T]|p_1,p_2\rangle 
        -i \, \langle p_4,p_3|T^\dagger[\mathbb{O},T]|p_1,p_2\rangle\,,
\end{equation}
which we conveniently separate into two contributions $\widetilde {\cal I}_{O, {\rm v}}$ and  $\widetilde {\cal I}_{O,\, {\rm r}}$, preemptively called \emph{virtual} and \emph{real}, respectively. The reason for this nomenclature becomes apparent when one evaluates the expectation values. For the \emph{virtual} part the result is simply
\begin{align}
\label{eq:kmocvirtual}
    {\cal I}_{O,\,{\rm v}} 
    = 
    \Delta \mathbb{O} \big[{\cal M}(p_1,p_2,p_3,p_4)\big]
    =
    \Delta \mathbb{O} \left[
    \raisebox{-32pt}{
      \!\!\!\scalebox{0.8}{\kmocvirtual}
  } \right] \,,
\end{align}
where $\Delta \mathbb{O}$ is a measurement function acting on the scattering amplitude, $\calM$, defined as
\begin{equation}
 \langle p_4,p_3|T|p_1,p_2\rangle = \widehat\delta^{(D)}\big(\textstyle\sum_i p_i\big) \, \calM(p_1,p_2,p_3,p_4)\,.
\end{equation}
On the other hand, to evaluate the \emph{real} contribution, $\calI_{O,\,{\rm r}}$, we insert a complete set of states 
\begin{equation}
\hspace{-.5cm}
    \langle p_4,\!p_3|T^\dagger[\mathbb{O},T]|p_1,\!p_2\rangle \!= \!\sum\limits_X \int \mathrm{d}\Phi_{2+|X|}(r_1,\!r_2,\!X)\, \langle p_4,\!p_3|T^\dagger|r_1,\!r_2,\!X\rangle \langle r_1,\!r_2,\!X|[\mathbb{O},T]|p_1,\!p_2\rangle \,,
\hspace{-.4cm}    
\end{equation}
which includes a sum over an arbitrary (possibly empty) set of intermediate ``messenger'' particles, $X$. Here $\mathrm{d}\Phi_{2+|X|}(r_1,r_2,X)$ is the multiparticle on-shell phase-space measure of the massive scalars with momenta $r_1$ and $r_2$ together with the massless messengers in the set $X$. The \emph{real} kernel turns into
\begin{align}
     {\cal I}_{O,\,{\rm r}}
     &= -\imath \sum_X \int \mathrm{d}\Phi_{2+|X|}(r_1,r_2,X) \, \hat\delta^{(D)}(p_1+p_2+r_1+r_2+\ell_X) \nn \\ 
     &\hspace{2cm}\times\Delta \mathbb{O} \big[{\cal M}(p_1,p_2,r_2,r_1,X)\big] 
     \, {\cal M}^*(-X,-r_1,-r_2,p_3,p_4) \nn \\
     &=-\imath\sum_X \int \mathrm{d}\widetilde\Phi_{2+|X|} \,\,  
     \Delta \mathbb{O} \hspace{-.4cm}
     \raisebox{-32pt}{
     \scalebox{0.8}{
       \kmocreal{$r_2$}{$r_1$}}} \,,
    \label{eq:kmocreal}
\end{align}
and the measuring function $\Delta \mathbb{O}$ only acts on the amplitude on the left of the unitarity cut which was introduced by the intermediate sum over states. Henceforth, we will use the abbreviated graphical notation in the last line of Eq.~\eqref{eq:kmocreal}, in which the phase space integral with measure $\mathrm{d}\widetilde\Phi_{2+X}$ is understood to be computed over the legs crossing the dashed blue line, and includes the momentum-conserving delta function. We remind the reader that we work in an all-outgoing convention for the particle momenta and note that all momenta crossing the cut flow from the left to the right in Eq.~(\ref{eq:kmocreal}) and the following. 

While this formalism can be applied fully quantum mechanically, in this work we are interested in \emph{classical} observables. This corresponds to the regime where the Compton wavelength of the external particles is the smallest length scale in the problem. Tracking the KMOC argument carefully, this feature implies that most computations boil down to simple plane-wave calculations in the classical limit. Once the dust settles, in the classical limit, the wavepackets sharply peak about their classical values of the momenta which leads to the appearance of on-shell delta functions and one arrives at a compact expression for the classical change of the observable $O$ in terms of the impact parameter $b^\mu$, conjugate to the small momentum transfer $q^\mu =p^\mu_1+p^\mu_4\sim \calO(\hbar)$,\,\footnote{We drop an $\calO(q^2)$ quantum piece in the argument of the delta functions $\hat{\delta}(x)$ and furthermore do not explicitly write the positive energy theta functions $\Theta(-p^0_1+q^0)\,\Theta(-p^0_2-q^0)$ which can be set to $1$ in the classical limit. The quantum terms originated from parametrizing the final state momenta as $p_4=-p_1+q$ and $p_3=-p_2-q$. We note that we compute all kernels in terms of fully on-shell objects in terms of specialized variables introduced in section \ref{subsec:ibp_de}. From here on out, we work in natural units $\hbar=1$, but the $\hbar$ counting can be restored from the $q$-expansion.} 
\begin{equation}
\label{eq:KMOC_DObs}
    \Delta O = \imath \int \hat{\mathrm{d}}^D q \,
    \hat\delta(-2 p_1 \cdot q)\, \hat\delta(2 p_2 \cdot q) e^{\imath b\cdot q}\,
    \left(
    \mathcal{I}_{O, {\rm v}} + \mathcal{I}_{O,{\rm r}}
    \right)\,.
\end{equation}
The KMOC analysis suggests that we ought to focus on kinematic regions where the massive particle momenta $p_i$ are large and scale like $\calO(1)$ in the classical counting and the four-momentum transfer $q$, as well as graviton loop variables that we will denote by $\ell_i$ below, scale like $\mathcal{O}(\hbar)$. In the effective field theory context, employing terminology from the ``method of regions'' \cite{Beneke:1997zp}, the classical $\hbar$ expansion is therefore equivalent to the so-called soft expansion.

Furthermore, we will also expand scattering amplitudes in $G$
\begin{equation}
    {\cal M} = {\cal M}^{(0)} + {\cal M}^{(1)} + {\cal M}^{(2)} + \cdots =  \!\!
    \raisebox{-22pt}{
      \scalebox{0.8}{
	\kmocvirtualtreenolab	
      }} +\!\!  
      \raisebox{-22pt}{
      \scalebox{0.8}{
	\kmocvirtualnlonolab	
      }} +\!\!
       \raisebox{-22pt}{
      \scalebox{0.8}{
	\kmocvirtualnnlonolab
      }} + \cdots\,,
\end{equation}
where the $L$-loop amplitude is ${\cal O}(G^{L+1})$. The impulse (kernels) have analogous expansions
\begin{align}
    \Delta p_1^\mu &= \Delta p^{\mu,(0)}_{1} + \Delta p^{\mu,(1)}_{1} + \Delta p^{\mu,(2)}_{1} + \cdots\,,\\
    \calI_{p_1}^\mu &= \calI^{\mu,(0)}_{p_1} + \calI^{\mu,(1)}_{p_1} + \calI^{\mu,(2)}_{p_1} + \cdots\,,
\end{align}
which we are going to describe in detail in section \ref{sec:KMOC_discussion_expansion}. However, it is already clear from the current statements that loop amplitudes and their unitarity cuts are essential ingredients. 

%=======================================================================
%\vspace{-0pt}
\subsection{Gravitational Impulse}
\label{subsec:impulse}
%\vspace{-0pt}
%=======================================================================
%
In this work, we discuss two observables relevant to classical gravitational scattering. The first is the gravitational impulse, $\Delta p_i^{\mu}$, which is defined as the total change in momentum of one of the particles during the collision. In the KMOC setup this is encoded by the appropriate quantum momentum operator~$\mathbb{P}_i$, which is measured asymptotically far from the collision region as follows
\begin{align}
 \Delta p^\mu_1 = 
 \langle \text{in} | S^\dagger \mathbb{P}^{\mu}_1 S | \text{in} \rangle - \langle \text{in} | \mathbb{P}^\mu_1 |\text{in} \rangle\,.
\end{align}
As summarized above, in the classical limit, this is simply a Fourier transform of the impulse kernel $\calI_{p_1}^\mu$ from momentum transfer $q$ to impact parameter space $b$
\begin{align}
    \label{eq:classical_impulse}
    \Delta p^\mu_1 & = \imath \int \hat{\mathrm{d}}^Dq 
    \,\hat{\delta}(-2p_1\cdot q)\,\hat{\delta}(2p_2\cdot q)\, e^{\imath b\cdot q}
    \, \calI^\mu_{p_1} \,,
\end{align}
which is separated into virtual and real contributions, given in terms of the amplitude as
\begin{equation}
    \calI_{p_1,\, \text{v}} 
     = 
    q^\mu \hspace{-.4cm}\raisebox{-32pt}{\scalebox{0.8}{\kmocvirtual}}\,,
    \calI_{p_1,\, \text{r}} 
     = - \imath \sum_X \int \mathrm{d}\widetilde{\Phi}_{2+|X|}\  \ell_1^\mu
    \hspace{-.4cm}
     \raisebox{-32pt}{\scalebox{0.8}{\kmocreal{$\ell_2-p_2$}{$\ell_1-p_1$}}}\,,
     \label{eq:classical_impulse_kernel}
\end{equation}
where the numerator insertions $q^\mu$ and $\ell_1$ arise from the measurement function $\Delta \mathbb{P}^\mu_1$ acting on the respective amplitudes, which extracts the momentum change of the particle 1 line. Note that relative to Eq.~\eqref{eq:kmocreal}, we have changed variables in the real contribution by shifting the massive intermediate momenta $r_i = -p_i + \ell_i$, so that all $\ell_i$ are small, $\calO(\hbar)$, in the classical expansion. The impulse on particle $2$ can be obtained by simple relabelling. 

%=======================================================================
\vspace{-0pt}
\subsection{Radiated momentum}
\label{subsec:radiated_momentum}\vspace{-0pt}
%=======================================================================
%
Another observable of interest is the total radiated momentum $\Delta R^\mu$  (which has been first computed in Ref.~\cite{Herrmann:2021lqe}) carried away in the form of gravitational waves during the scattering of two black holes. This observable is defined by measuring the momentum operator $\mathbb{R}^\mu$ of the emitted \emph{messenger particles}. As explained in Ref.~\cite{Kosower:2018adc}, this observable only receives \emph{real} contributions and its kernel is given by unitarity cuts 
\begin{align}
    \mathcal{I}_{R, {\rm r}}^\mu 
    &=-\imath \sum_X 
    \int\!\mathrm{\mathrm{d}}\widetilde{\Phi}_{2+X}\  \ell^{\mu}_X 
    \raisebox{-32pt}{\scalebox{0.8}{\kmocreal{$\ell_2-p_2$}{$\ell_1-p_1$}}}
    \label{eq:KMOC_kernel_rad}
\end{align}
Like Eqs.~(\ref{eq:classical_impulse}) and (\ref{eq:classical_impulse_kernel}),  Eq.~\eqref{eq:KMOC_kernel_rad} is valid  beyond perturbation theory. However, in the following, we will expand the radiated momentum perturbatively in $G$. The first contribution to $\Delta R^\mu$ (obtained from Eq.~\eqref{eq:KMOC_kernel_rad} by performing the Fourier transform to impact parameter space \eqref{eq:KMOC_DObs}) arises at $\calO(G^3)$. This can be understood from the fact that Bremsstrahlung of finite energy gravitons can only occur once one black holes is slightly deflected due to its gravitational interaction with the other massive object.

%================================================
%
\section{Setup and review}
\label{sec:setup_review}
%
%================================================
%
In this section, we review the relevant technology to calculate the scattering amplitudes that serve as building blocks in the KMOC formalism. Most tools were introduced in Refs. \cite{Bern:2019crd,Bern:2019nnu,Parra-Martinez:2020dzs}, in the context of classical conservative gravitational scattering, i.e.~in the potential region (c.f.~Eq.~(\ref{eq:powerCountingVelocity}). Readers familiar with Refs.~\cite{Bern:2019crd,Bern:2019nnu,Parra-Martinez:2020dzs} can skip this review. The novelties of the soft region, which includes dissipative effects, are the subject of Sec.~\ref{sec:soft_region_methods}. 

%================================================
\subsection*{Kinematics}
%================================================

Before detailing the integrand construction via generalized unitarity, we briefly review the relevant kinematics for the two-to-two scattering of massive black holes in the classical limit. This allows us to link the classical $\hbar$ expansion to the \emph{soft} (small $|q|$) expansion familiar from the method of regions \cite{Beneke:1997zp} and further motivate certain truncations in the integrand construction of the classical, conservative sector.

The four-particle scattering of massive scalars in an all-outgoing momentum convention is characterized by the following kinematic invariants,
\begin{align}
\hspace{-.5cm}
          p^2_1 = p^2_4 = m^2_1\,, 
    \quad p^2_2 = p^2_3 = m^2_2\,, 
    \quad s = (p_1{+}p_2)^2\,, 
    \quad t =q^2=(p_1{+}p_4)^2\,, 
    \quad u = (p_1{+}p_3)^2\,.
\hspace{-.5cm}
\end{align}
We work with a mostly minus metric $\eta^{\mu\nu}=\text{diag}(1,-1,-1,-1)$ and the Mandelstam invariants $s,t$, and $u$ are subject to the usual constraint
\begin{align}
    s+t+u = 2m^2_1 + 2m^2_2\,.
\end{align}
It will be useful to introduce the total mass and symmetric mass ratio
\begin{equation}
    M = m_1+m_2\,, \quad \nu = \frac{m_1m_2}{(m_1+m_2)^2}\,,
\end{equation}
as well the combinations
\begin{align}
\sigma =  \frac{s{-}m^2_1{-}m^2_2}{2m_1 m_2} 
       = \frac{p_1 \cdot p_2}{m_1 m_2}\,,
\quad
h(\sigma,\nu) = \frac{\sqrt{s}}{M}=\sqrt{1 + 2 \nu (\sigma-1)}       
\end{align}
where $\sigma$ is the relativistic factor of particle 1 in the rest-frame of particle 2 (or vice versa) and $h$ is the total energy-mass ratio. Physical particle scattering in the s-channel corresponds to the region $s>(m_1+m_2)^2$ (or $\sigma>1$), $t=q^2<0$, and $u<0$. In contrast to the massless case, for massive $2\to 2$ scattering one can define a Euclidean region where all invariants are negative and the amplitude is real. This region also plays an important role in the evaluation of Feynman integrals in section \ref{sec:soft_integrals}.

%
%================================================
\subsection{The method of regions and the classical limit}
\label{subsec:soft_expansion}
%================================================
%
We are ultimately interested in classical dynamics and we would like to only retain the minimal amount of information necessary to describe classical black holes. In this limit, the orbital angular momentum of the scattering black hole binary system is much larger than $\hbar$ and simply corresponds to the large angular momentum limit $J\gg 1$ (in natural units), which establishes a hierarchy of scales
\begin{align}
\label{eq:classical_hierarchy}
 s,|u|,m_1,m_2 \sim J^2 |t| \gg |t| = |q|^2 \,.
\end{align}
As a result, we are interested in calculating scattering amplitudes as an expansion in small $|q|$. From a heuristic analysis of scales, we perform an expansion in orders of $r_{\mathrm{s}}/|b|$, where the Schwarzschild radius is $r_{\mathrm{s}} \sim G m$ for some common mass scale $m\sim m_1+m_2$, and $|b|$ is the relative transverse distance (conjugate to the momentum transfer $q$) of the system. This implies that the relevant dimensionless expansion parameter is  $r_{\mathrm{s}}/|b|\sim Gm/|b| \sim Gm|q|$. For each additional order of Newton's constant $G$, we need to expand the amplitude up to one additional power of $|q|$. The relevant term in the $q$-expansion at tree-, one-loop, and two-loop level are $1/|q|^2, 1/|q|$, and $\log |q|$ respectively. The fact that we are interested in the non-analytic terms in the $|q|$ expansion is related to the long-distance effects in impact parameter space (Analytic terms in the $|q|$-expansion transform to $\delta$-function contributions in $b$-space).  At a given loop order (order in $G$), terms that are more subleading in $|q|$ are quantum corrections. In summary, at $\calO(G^n)$, we only need to expand the scattering amplitude of massive particles up to $\calO(|q|^{n-3})$ in the small-$q$ expansion \cite{Neill:2013wsa}, in order to extract the classical dynamics. In practice, this implies that some loop integrals can be discarded in the amplitude construction, if they are beyond the classical order.

Before moving to the actual integrand construction, it is helpful to recall the relevant kinematic scaling of external and loop momenta. We separate temporal and spatial momentum components $k = (\omega,\mathbf{k})$ to define the relevant momentum regions \cite{Bern:2019crd} 
\begin{align}
    \text{hard:}        \sim m\,     \qquad  
    \text{soft:}        \sim |q|   \,.
  \label{eq:powerCounting}
\end{align}
The classical limit is equivalent to an expansion in the dimensionless power counting variable $\lambda = |q|/m$. For convenience, instead of counting power of $\lambda$ we can count powers of $|q|$ relative to the scaling dimension of the amplitude. In terms of this scaling, all matter lines of the heavy black holes have hard momenta of $p_i \sim m \sim \calO(|q|^0)$. As mentioned above, we are interested in classical, long-distance physics mediated by graviton exchange related to the soft region, which further splits into
\begin{align}
\begin{split} 
\text{soft}\qquad
\begin{cases}
    \text{"quantum" soft:}  & \quad (\omega,\textbf{k}) \sim |q| (1,1) \\
    \text{potential:}  & \quad (\omega,\textbf{k}) \sim  |q|\, m(v,1)            \\
    \text{radiation:}  & \quad (\omega,\textbf{k}) \sim  |q|\, m(v, v)
    \end{cases}
\end{split}\,.    \label{eq:powerCountingVelocity}
\end{align}
In comparison to previous work that primarily focused on conservative dynamics \cite{Bern:2019nnu,Bern:2019crd,Parra-Martinez:2020dzs}, associated to potential modes, in the KMOC setup, we will directly work in the full soft region. This is owed to the fact that the classical $\hbar$ expansion in the KMOC formalism is intimately tied to the $q$-expansion and we do not have to make further assumptions about small velocities $v$ or instantaneous interactions. However, even in the KMOC setup, we can impose this additional velocity restriction to reproduce conservative results which serve as nontrivial cross-checks of our assembly.

%================================================
%
\subsection{Loop integrands and generalized unitarity}
\label{subsec:integrands}
%
%================================================
%
We have seen in section \ref{sec:KMOC} that the extraction of classical gravitational observables within the KMOC formalism involves virtual higher-loop scattering amplitudes as well as their unitarity cuts. Crucially, all ingredients are \emph{on shell}. This allows to employ a number of simplifying features developed in the context of the scattering amplitudes program over the last several decades. Making use of these novel tools leads to a highly simplified and streamlined construction of the relevant loop-integrands, i.e. the rational functions before the loop or phase-space integrations are performed. For example, the key technology at work is generalized unitarity \cite{Bern:1994zx,Bern:1994cg,Britto:2004nc} and color-kinematics duality \cite{Bern:2008qj,Bern:2010ue,Bern:2012uf,Bern:2017ucb, Bern:2018jmv,Bern:2019prr} which effectively reduces all gravitational calculations to the computation of sums of products of Yang-Mills tree-level amplitudes. Most details of the loop-integrand construction have already been described elsewhere \cite{Bern:2019nnu,Bern:2019crd}, so that we can limit our review to only the most relevant points and remain telegraphic otherwise. Here we give a general review of the conservative sector result, before pointing out certain additions relevant in the full soft region in section \ref{sec:soft_region_methods}. 

It is well known that loop integrands in quantum field theories can be reconstructed from their singularity structure which in turn is entirely dictated by factorization. This idea is formalized in the generalized unitarity framework which allows to construct amplitudes from their unitarity cuts. In the following, we limit our discussion to the case of two-loop contributions, which is of main interest in this work. The generalization to other loop orders is (at least conceptually) straight forward. We are therefore interested in two-loop scattering processes of two massive scalar particles that are minimally coupled to gravity.

To extract classical physics, it is not necessary to obtain the full quantum amplitude, but instead we can directly focus on the relevant pieces that contribute to long-distance interactions between the two black holes mediated by the exchange of gravitons. In the case of conservative dynamics, the gravitons mediate instantaneous interactions causing further simplifications due to the vanishing of certain potential region integrals \cite{Bern:2019crd}. In particular, one requires at least one matter line per loop, together with the absence of certain mushroom type graphs (see e.g. Fig.~\ref{fig:linMushroomWithLabels}) that will become important in the full soft region shortly. Since we are interested in long-distance effects, we can neglect all contact interactions between the two black holes. In the field theory language, this corresponds to setting to zero the four-scalar interactions. From a practical perspective, this limits the number of relevant terms that are interesting for the amplitude construction. For further discussions, see Ref.~\cite{Bern:2019crd}. 

\begin{figure}[t!]
    \centering
    \includegraphics[scale=.7]{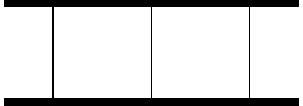}
    \includegraphics[scale=.7]{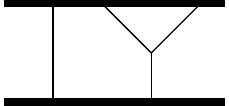}
    \includegraphics[scale=.7]{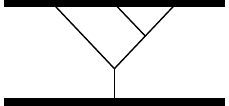}
    \includegraphics[scale=.7]{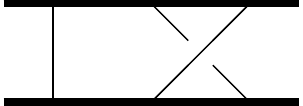}
    \includegraphics[scale=.7]{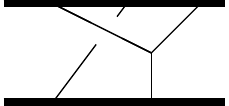}
    \includegraphics[scale=.7]{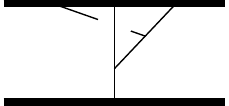}
    \raisebox{-5pt}{\includegraphics[scale=.7]{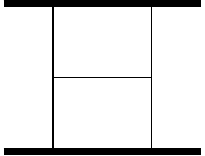}}
    \raisebox{-5pt}{\includegraphics[scale=.7]{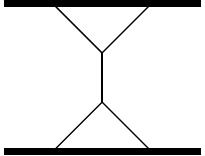}}
    \caption{Cubic diagrams relevant for the potential-region amplitude in classical GR.
    }
    \label{fig:cubic_graphs_cons}
\end{figure}

In our calculation, we use generalized unitarity in the following way. At two-loops, we write down an ansatz of cubic, local Feynman-like diagrams for all graph topologies that pass our (conservative,) long-distance, non-scaleless, classical power counting constraint. The resulting list of graphs is summarized in Figure \ref{fig:cubic_graphs_cons} \cite{Bern:2019crd,Bern:2019nnu}. For each of the diagrams, we write a yet undetermined numerator. Power-counting dictates that the two-loop numerators have to have mass-dimension 12 as is easily seen by counting derivatives in simple Feynman diagrams.
Our numerator ansatz for a given cubic graph $\Gamma$ is then written in terms of Lorentz dot products between the three independent external momenta $p_1,p_2,p_3$ and the two independent loop momenta $\ell_1,\ell_2$
\begin{align}
    N_{\Gamma} = a_{\Gamma,1} (p_1\cdot p_2)^6 + a_{\Gamma,2}(p_1\cdot p_2)^5 (\ell_1\cdot p_1) + \cdots + a_{\Gamma,n}(\ell_1\cdot \ell_2)^6\,.
\end{align}
Note that this representation of the numerator ansatz secretly includes possible contact terms of the cubic graph, where a contact term denotes any graph $\Gamma$ with one of the propagators pinched due to a numerator factor. Here, we chose not to write the numerator monomials in such a way that the contact-term stratification is manifest. (For advantages of a contact term basis, see e.g. Refs.~\cite{Bourjaily:2017wjl,Bourjaily:2020qca}.)

\begin{figure}[t!]
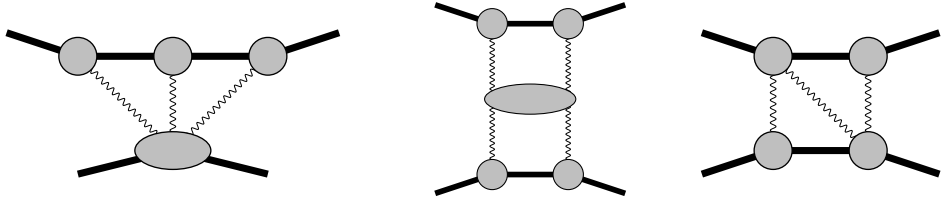

    \centering
    \raisebox{7pt}{\scalebox{1.25}{\wcut}}
    \qquad
    \hcut
    \qquad
    \raisebox{7pt}{\scalebox{1.25}{\ncut}}
    \caption{Spanning set of unitarity cuts relevant for the potential region $\calO(G^3)$.}
    \label{fig:spanningcuts}
\end{figure}

In the generalized unitarity setup, it is then the goal to fix the yet undetermined coefficients $a_{i,j}$ by matching the integrand ansatz in terms of cubic graphs against field theory cuts. This is done by taking unitarity cuts of the amplitude integrand, given in terms of products of tree-level amplitudes (on-shell functions), and equating these cuts to the cuts of the cubic diagrams. This procedure yields linear relations for the free coefficients $a_{\Gamma,j}$ that can be solved in a straightforward manner. Once we match a \emph{spanning set of cuts}, we are guaranteed to have a correct amplitude integrand. For the conservative classical impulse and radiated momentum at $\calO(G^3)$, such a spanning set of cuts is given in Figure~\ref{fig:spanningcuts}.
As mentioned before, these unitarity cuts are nothing but products of on-shell tree-level amplitudes summed over the on-shell states. These tree-level amplitudes can in principle be obtained in whichever way one can imagine, even via Feynman diagrams if necessary. For us, we make further use of recent developments in scattering amplitudes, where we can express tree-level gravity amplitudes as the square of Yang-Mills amplitudes via the celebrated BCJ double-copy procedure \cite{Bern:2008qj,Bern:2010ue,Bern:2012uf,Bern:2017ucb, Bern:2018jmv,Bern:2019prr}. How this is done in practice has already been summarized in the work of Ref.~\cite{Bern:2019crd} so we will not reiterate these steps here.

Ultimately, we write the unitarity-based $L$-loop amplitude as a sum over cubic $L$-loop graphs with classical counting
\begin{align}
\label{eq:amplitude_cubic_diags_abstract}
    \calM^{(L)}(p_1,p_2,p_3,p_4)\quad = \sum_{\Gamma\in\left\{ \substack{L-\text{loop}, \\ \text{classical}}\right\}}
    \int \prod\limits^{L}_j \widehat{\mathrm{d}}^D\ell_j \frac{N_{\Gamma}(\ell_j,p_i)}{\prod\limits_{P\in\Gamma} P(\ell_j,p_i)} \,,
\end{align}
where the numerators $N_{\Gamma}(\ell_j,p_i)$ have been determined by matching unitarity cuts and the denominator is the product of all Feynman propagators $P_{\Gamma}(\ell_j,p_i)$ of graph $\Gamma$. Concretely, for the conservative result at two-loops, this involves the diagrams in Fig.~\ref{fig:cubic_graphs_cons} whose numerators have been constraint by matching the unitarity cuts in Fig.~\ref{fig:spanningcuts}. We note that all propagators retain the full kinematic dependence and have not yet been expanded in the soft region, i.e. matter propagators are schematically of the form $1/((\ell+p_i)^2 -m^2_i)$.

%================================================
\subsection{Soft expansion, integration-by-parts and differential equations}
\label{subsec:ibp_de}
%================================================
\begin{figure}[t!]
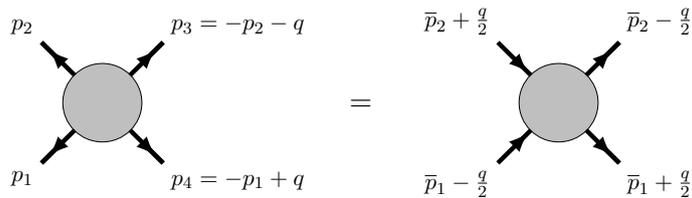

\centering
\label{fig:soft_kinematics}
    $\vcenter{\hbox{\scalebox{0.8}{\softkinA}}} 
    \quad
    =
    \quad
    \vcenter{\hbox{\scalebox{0.8}{\softkinB}}}$
    \caption{Depiction of the parameterization of external momenta useful for the soft expansion.}
\end{figure}
In this section, we briefly review the relevant tools that allow us to go from the integrands derived above, closer to the final integrated result. We first introduce special kinematic variables to facilitate the classical $\hbar$ or equivalently \emph{soft} (small |q|, or small $\lambda$) expansion in the context of the method of regions \cite{Beneke:1997zp}. All relevant definitions have already appeared in Ref.~\cite{Parra-Martinez:2020dzs}, so we are going to be brief. In order to facilitate integration, we perform an integration-by-parts reduction to a minimal set of master integrals that will be solved by differential equations.
In the following, we introduce specialized kinematics, depicted in Eq.~(\ref{eq:soft_kinematics}), 
\begin{align}
    p_1 = - \left(\overline{p}_1 - \frac{q}{2}\right)\,, \ 
    p_2 = - \left(\overline{p}_2 + \frac{q}{2}\right)\,, \ 
    p_3 =   \left(\overline{p}_2 - \frac{q}{2}\right)\,, \ 
    p_4 =   \left(\overline{p}_1 + \frac{q}{2}\right)\,,
    \label{eq:soft_kinematics}
\end{align}
tailored towards the discussion of the soft expansion of the relevant integrals. These variables have the advantage that the new vectors $\overline{p}_i$ are orthogonal to the momentum transfer $q$, $\overline{p}_i \cdot q = 0\,, $ which directly follows from the on-shell conditions $p^2_1 = p^2_4 = m^2_1$ and $p^2_2 = p^2_3 = m^2_2$. Furthermore, $s= (p_1 + p_2)^2 = (\overline{p}_1+\overline{p}_2)^2$ so that the physical region $s{>}(m_1+m_2)^2,\, q^2{<}0$ is unaltered. We also define the soft four-velocities of the two black holes $u^\mu_i = \overline{p}^\mu_i/|\overline{p}_i|$, such that $u_i^2=1$, and 
\begin{align}
\label{eq:y_and_x_def}
 y \equiv u_1\cdot u_2 = \frac{1+x^2}{2x}  
   = \sigma + \calO(q^2) \,.
\end{align}
where $x$ rationalizes various naturally appearing square-roots later on. 

Note that these soft velocities coincide with the classical four velocities of the black holes up to irrelevant corrections of $\calO(q)$ that do not affect the classical observables. As mentioned above, we are interested in the soft expansion, with the following hierarchy of scales $|\ell| \sim |q| \ll |\overline{p}_i|, m , \sqrt{s}\,,$ or equivalently $\lambda \ll 1$. Here, $\ell$ schematically represents arbitrary combinations of graviton momenta of the form $(\ell_1, \ell_2, \ell_1 \pm \ell_2, \ldots )$ and typical graviton propagators take the form $\frac{1}{\ell^2}\,,\ \frac{1}{(\ell-q)^2}\,,$ so that they have a homogeneous $|q|$-scaling and therefore do not require any non-trivial expansion. Note that we can choose a momentum routing so that graviton lines do not involve the individual momenta $\overline{p}_i$ of the external massive particles. On the other hand, matter propagators do have a non-trivial $|q|$ expansion which we express in terms of the dimensionless velocity variables $u_i$
\begin{align}
    \frac{1}{(\ell-p_i)^2-m^2_i} = \frac{1}{\ell^2 - 2\, \ell \cdot p_i}
    =
    \frac{1}{2 u_i \cdot \ell}\,\frac{1}{m_i} 
    - \frac{\ell^2 \mp \ell \cdot q}{(2 u_i \cdot \ell)^2}\,\frac{1}{m^2_i} 
    + \cdots\,, \label{eq:softExpansionMatterProp}
\end{align}
such that each order in the expansion is homogeneous in $|q|$ and the mass dependence factorizes. The matter propagators effectively ``eikonalize'' and the soft expansion to higher orders in $|q|$ can lead to raised propagator powers. Graphically, we denote these eikonalized (or linearized) matter propagators by a double-line notation, see e.g.~the diagram in Fig.~\ref{fig:linMushroomWithLabels}, to distinguish them form unexpanded propagators e.g.~in Fig.~\ref{fig:cubic_graphs_cons}.
In order to reduce tensor integrals and integrals with doubled propagators that appear unavoidably in the soft expansion of matter propagators \eqref{eq:softExpansionMatterProp}, we make use of the standard practice in collider physics and use IBP identities \cite{Tkachov:1981wb,Chetyrkin:1981qh}. These are due to the fact that  total derivatives identically vanish in dimensional regularization (see e.g.~\cite{Collins:1984xc}). By writing sufficiently many total derivatives, one obtains a set of linear relations in the space of Feynman integrals with a given set of propagators. A key insight is that such a space is in general finite dimensional \cite{Smirnov:2010hn} and solving IBPs reduces the task of computing a general integral to the evaluation of a basis of \emph{master integrals}. The most common approach to solving the system of IBP identities is Laporta's algorithm, \cite{Laporta:1996mq,Laporta:2001dd}, implemented in a variety of different packages. In the present work we use \texttt{FIRE6} \cite{Smirnov:2019qkx}. 

The soft expansion not only implements the classical $\hbar\ll J$ limit by truncating at appropriate orders in $|q|$, but also leads to an enormous simplification of the resulting integrals. Indeed, consistent with the spirit of effective field theory, the appropriate separation of scales allows us to focus on one scale at a time, here $|q|$, which essentially reduces classical gravitational scattering to a single-scale problem.

The advantage of the new soft variables $u^\mu_i$ and $q^\mu$ lies in fact that the mass dependence of a general soft (linearized) integral completely factorizes (due to the properties of the expansion of matter propagators in Eq.~\eqref{eq:softExpansionMatterProp}) and the only remaining dimensionful scale is $q^2$ which can be extracted by simple dimensional analysis. Therefore, in the soft region, the only nontrivial kinematic variable is $y=u_1\cdot u_2$ and we are going to find the values of all soft integrals by deriving and solving differential equations in $y$ here and in section \ref{sec:soft_integrals}, respectively. In order to take derivatives with respect to $y$ at the integrand level, we can express the $y$ derivative in terms of the vectors $u_1$ and $u_2$\footnote{Note that exactly the same differential operator appears in calculations of the angle-dependent cusp anomalous dimensions in gauge theory, see e.g.~\cite{Grozin:2015kna}.}
\begin{equation}
	\frac{\partial}{\partial y}=\frac{y u_1^\mu - u_2^\mu}{y^2-1}\frac{\partial}{\partial u_1^\mu}\label{eq:partialy}\,.
\end{equation}
Acting with \eqref{eq:partialy} on the set of master integrals $\vec{g}$ will produce another set Feynman integrals with the same set of propagators, which can subsequently be reduced to the master basis $\vec{g}$ to yield a differential equation
\begin{equation}
	\frac{\partial}{\partial y}\vec{g}=A(y,\epsilon)\vec{g}\label{eq:DENonCanon}\,,
\end{equation}
where $A(y,\epsilon)$ is a matrix with rational dependence on $y$ and $\epsilon = (4-D)/2$. We can use the freedom in choosing the basis of integrals and the parametrization of the kinematics to simplify the differential equation \eqref{eq:DENonCanon}. In all cases discussed in this article we are able to choose an appropriate set of \emph{canonical} master integrals $\vec{f}$ \cite{Henn:2013pwa},\footnote{Starting at three loops, generally this is no longer possible due to the presence of elliptic integrals \cite{Bern:2021dqo}.} in terms of which 
\begin{equation}
\frac{\partial}{\partial y}\vec{f}=\epsilon\left[\sum_iA_i\frac{\partial}{\partial y}\log w_i(y)\right]\vec{f}\,,
\label{eq:canonicalDE}
\end{equation}
the $\epsilon$ dependence factorizes and where $A_i$ denote matrices with constant rational entries, most of which were computed in Ref.~\cite{Parra-Martinez:2020dzs} by some of the present authors. The only missing results were the matrices for integrals Eq.~\eqref{eq:fH11}-\eqref{eq:HpureBasis} in the H family which scale as odd powers of $|q|$ before being multiplied by appropriate normalization factors, given in Appendix~\ref{app:masters_notation}. %\ref{sec:DEHodd}. 
In form \eqref{eq:canonicalDE}, the possible singularities of $\vec{f}$ are completely manifest. They are given by the zeros of the elements $\{w_i\}$---the \emph{alphabet} of the differential equation. The fact that the dimensional regularization parameter $\epsilon$ is factorized makes it straight forward to solve the system iteratively order-by-order in $\epsilon$ \cite{Henn:2013pwa}.

%================================================
%
\section{Full soft integrands and reverse unitarity}
\label{sec:soft_region_methods}
%
%================================================
%
As alluded to before, most of the tools describing classical \emph{conservative} dynamics in terms of scattering amplitudes have been successfully applied before, see e.g. Refs.~\cite{Bern:2019crd,Bern:2019nnu,Parra-Martinez:2020dzs}. Taking \emph{radiation effects} into account leads to a few novelties that we are going to discuss in this section. First of all, there are additional contributions to the integrand, which we summarize in subsection \ref{subsec:soft_integrands}. One additional feature of the KMOC framework is the presence of on-shell phase-space integrals. We employ reverse unitarity \cite{Anastasiou:2002yz,Anastasiou:2002qz,Anastasiou:2003yy,Anastasiou:2015yha}, well-known from collider physics computations, in subsection \ref{subsec:rev_unitarity}, to efficiently handle such integrals.

%================================================
\subsection{New contributions from the soft region}
\label{subsec:soft_integrands}
%================================================
%
%
\begin{figure}[t!]
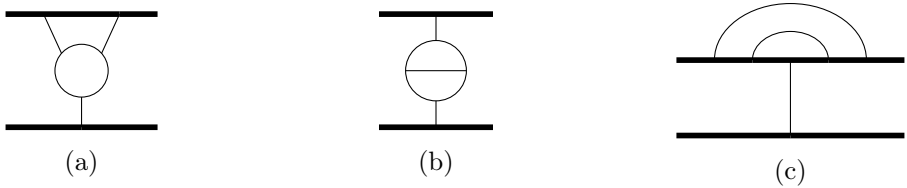

    \centering
    \begin{subfigure}{0.3\linewidth}
    \centering
    \graphquantumtriangle
    \caption{}
    \label{fig:eg_quantum_diags1}
    \end{subfigure}
    \begin{subfigure}{0.3\linewidth}
    \centering
    \graphquantumbuble
    \caption{}
    \label{fig:eg_quantum_diags2}
    \end{subfigure}
    \begin{subfigure}{0.3\linewidth}
    \centering
    \graphscaleless
    \caption{}
    \label{fig:eg_scaleless_diag}
    \end{subfigure}
    \caption{Sample diagrams that do not contribute in the classical limit. (a) and (b) are purely quantum from simple $|q|$ counting arguments, whereas (c) is naively classical but scaleless in the soft region.}
\end{figure}
In our general review of the integrand construction in subsection \ref{subsec:integrands}, we mostly discussed the conservative sector, previously presented in Refs.~\cite{Bern:2019nnu,Bern:2019crd}. Here, we are interested in going beyond conservative dynamics and taking radiation effects into account. This requires us to slightly augment the known integrand and include additional terms. The generalized unitarity strategy to determine these additional contributions, though, is the same as for the conservative result.

\begin{figure}[t!]
    \centering
    \includegraphics[scale=.7]{./figures/graph_1.pdf}
    \includegraphics[scale=.7]{./figures/graph_2.pdf}
    \includegraphics[scale=.7]{./figures/graph_3.pdf}
    \includegraphics[scale=.7]{./figures/graph_4.pdf}
    \includegraphics[scale=.7]{./figures/graph_5.pdf}
    \includegraphics[scale=.7]{./figures/graph_6.pdf}
    \raisebox{-5pt}{\includegraphics[scale=.7]{./figures/graph_7.pdf}}
    \raisebox{-5pt}{\includegraphics[scale=.7]{./figures/graph_8.pdf}}
    \newline\newline
    \includegraphics[scale=.7]{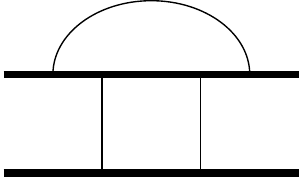}
    \includegraphics[scale=.7]{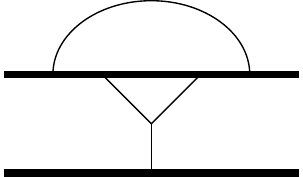}
    \includegraphics[scale=.7]{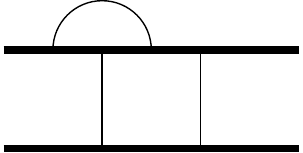}
    \includegraphics[scale=.7]{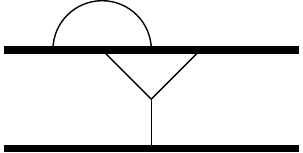}
    \includegraphics[scale=.7]{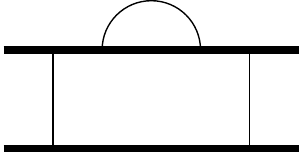}
    \includegraphics[scale=.7]{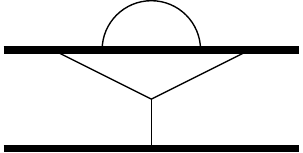}
    \!\scalebox{0.7}{\graphsixteen}
    \caption{Cubic diagrams relevant for the classical $\calO(G^3)$ impulse and radiated momentum in general relativity (including radiative contributions).
    }
    \label{fig:cubic_graphs}
\end{figure}

Compared to the conservative dynamics considered in Refs.~\cite{Bern:2019nnu,Bern:2019crd}, we have additional diagrams depicted on the second line of Figure~\ref{fig:cubic_graphs}. This is owed to the fact that there are additional terms that survive in the full soft region but are zero in the potential region. In particular, we are interested in classical physics (including radiation contributions) which, as explained in section \ref{sec:KMOC}, is encoded in the soft expansion where the momenta of the black holes scale like $\calO(m)$ and the momentum transfer and the momentum of internal graviton lines scale like $\calO(|q|)$. This leads to the following summary of $|q|$-counting rules that facilitate the classical counting (i.e. soft counting) of linearized integrals\footnote{Note that we work in dimensional regularization, where the loop measure is $d^D\ell \sim |q|^{D}$, where we work in $D=4-2\epsilon$ which yields the non-integer powers of $|q|$ in all our integrals.}
\begin{align}
    \begin{split}
        \text{graviton propagator:} &   \quad \sim |q|^{-2}\,,  \qquad 
        \text{matter propagator:}       \quad \sim |q|^{-1}\,,    \\  
        & \text{integration measure:}   \quad \mathrm{d}^4\ell \sim |q|^4\,.
    \end{split}
    \label{eq:powerCountingRule}
\end{align}         
As was pointed out in Ref.~\cite{Parra-Martinez:2020dzs}, graviton propagators scale homogeneously like $|q|^{-2}$, whereas matter propagators have a $|q|$-expansion starting at order $|q|^{-1}$. The leading order $k$ in $|q|$, $\calO(|q|^k)$ for a given two-loop graph is
\begin{align}
    k = 4L + 2 n_{v_3} -2 n_{p_g} - n_{p_m}\,,
    \label{eq:powerCountingFormula}
\end{align}
where the factor $4L$ comes from the loop measure $\prod_i^L\mathrm{d}^4\ell_i \sim |q|^{4L}$, $n_{v_3}$ is the number of three-graviton vertices of the graph, $n_{p_g}$ the number of graviton propagators, and finally $n_{p_m}$ the number of matter propagators. We call two-loop diagrams superclassical (or classically singular), when their leading order term in the $|q|$ expansion starts with $k<0$, classical when $k=0$, and quantum, if $k>0$. With these simple counting-rules, we see that the graph on  Figure~\ref{fig:eg_quantum_diags1} has $k=8+2\times 3 - 2\times 6-1 = 1>0$ which is, as advertised, quantum and we can therefore neglect such contributions. 

Another simplification, related to the soft-expansion of Feynman integrals comes from the knowledge that certain integrals become scaleless and therefore integrate to zero. There are simple rules to identify such topologies even before integration which allows us to neglect such terms in the integrand from the outset (see e.g. Figure~\ref{fig:eg_scaleless_diag}). 

\begin{figure}[t!]
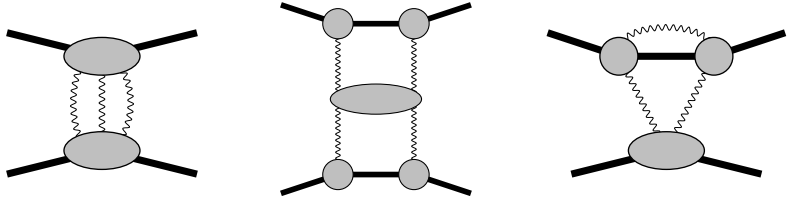

    \centering
    \raisebox{7pt}{\scalebox{1.25}{\threegravitoncut}}
    \qquad
    \hcut
    \qquad
    \raisebox{7pt}{\scalebox{1.25}{\threecomptoncut}}
    \caption{Spanning set of unitarity cuts relevant for the soft region at $\calO(G^3)$.}
    \label{fig:spanningcuts_soft}
\end{figure}

Taking into account the above rules, we find the list of cubic graphs relevant for radiative classical dynamics at $\calO(G^3)$; depicted in Figure~\ref{fig:cubic_graphs}. Compared to the conservative diagrams shown in Figure~\ref{fig:cubic_graphs_cons}, the second line is new.
These additional diagrams require us to enlarge our spanning set of cuts, compared to the conservative ones depicted in Figure~\ref{fig:spanningcuts} in order to get constraints on the new numerator ansaetze. The spanning set of cuts that allows us to fix the classical integrand in the soft region is given in Figure~\ref{fig:spanningcuts_soft}.

We note that some of the unitarity cuts also involve quantum terms and in order to match the full integrand before further classical truncation would require additional cubic graphs not listed in Figure~\ref{fig:cubic_graphs}. Since these additional terms are purely quantum, we can in principle drop them from our discussion and write the amplitude analogous to Eq.~(\ref{eq:amplitude_cubic_diags_abstract}) where the sum over cubic graphs now contains the additional contributions of the diagrams on the second line of Figure~\ref{fig:cubic_graphs} with the numerators fixed by matching the ansatz against the spanning set of cuts in Figure~\ref{fig:spanningcuts_soft}. Consistently ignoring such quantum terms in the cut matching procedure, however, is rather subtle. 

In the full soft region, completely fixing the unitarity based ansatz requires the matching of rather ``deep'' unitarity cuts with very few lines put on shell (such as the three-graviton cut on the l.h.s. of Fig.~\ref{fig:spanningcuts_soft}). Fully matching these cuts becomes increasingly cumbersome due to the addition of a multitude of quantum terms that are irrelevant for classical physics. One way to circumvent this situation is to step back from the unitarity setup and instead employ simplified gravitational Feynman rules \cite{Cheung:2016say, Cheung:2017kzx}, closely following the implementations in Ref.~\cite{Rafie-Zinedine:2018izq, Kalin:2020mvi}, to target the set of classical Feynman diagrams directly. The simplification of the gravitational Feynman rules is possible due to a judicious choice of gauge-fixing functions. The Feynman diagrams are generated by {\tt QGRAF} \cite{Nogueira:1991ex}, ignoring ghost particles which have no contributions to classical physics.\footnote{If desired ghost contributions can be easily fixed by matching the relevant unitarity cuts.} The Lorentz index contractions are carried out with an in-house code to produce numerators in terms of dot products for each diagram. We have checked that the integrand constructed via the simplified Feynman rules matches all relevant classical parts of the unitarity cuts of Fig.~\ref{fig:spanningcuts_soft}, so that we are confident in our implementation. 

Note that our integrand contains the box-bubble graph on the very right hand side of the bottom-row of Fig.~\ref{fig:cubic_graphs} which naively looks like a quantum contribution due to the internal graviton loop. However, by our $|q|$-counting arguments, this graph is of classical order by virtue of a $1/|q|$ iteration hitting a $\calO(|q|)$ quantum contribution. As we will show explicitly in subsection \ref{subsec:nnlo_impulse}, these contributions cancel in the classical observables within the KMOC formalism in a way that is similar to eikonal subtractions, see e.g.~\cite{DiVecchia:2020ymx,DiVecchia:2021ndb}.

With the relevant classical \emph{virtual} two-loop integrand at hand, we also have all relevant terms required for the \emph{real} contributions in the KMOC setup in Eqs.~(\ref{eq:kmocreal}) and (\ref{eq:KMOC_kernel_rad}). In fact there is no need to construct these cut contributions separately. Instead, we can take our virtual integrand and perform the required unitarity cut. The relevant measurement function in the form of the appropriate loop-momentum insertion for either the impulse or radiated momentum is then simply linked to the labeling of the cut. As such, we have now constructed all integrands that make an appearance in the KMOC formalism and we can turn our attention to the novelties of integrating in the full soft region as well as tools that handle these cut, or phase-space integrals. This is what we turn to next.

%

%================================================
\subsection{Soft expansion and partial fractioning \label{subsec:partialFraction}} 
%================================================

\begin{figure}[t]
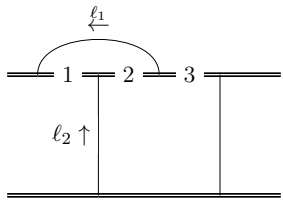

\centering
\scalebox{.8}{\linMushroomWithLabels}
\caption{Linear mushroom integral.\label{fig:linMushroomWithLabels}}
\end{figure}

There is one aspect of special soft integrals that warrants discussion. A particular feature of mushroom-type integrals (see e.g. Fig.~\ref{fig:linMushroomWithLabels}) pertains to the fact that once the matter propagators are linearized via the soft expansion of Eq.~\eqref{eq:softExpansionMatterProp}, some of the propagators in a given diagram might become linearly dependent. However, this can be addressed by partial fractioning. In the example given by  Figure~\ref{fig:linMushroomWithLabels}, the three matter propagators on the top are

These expressions are linearly dependent and partial fractioning allows us to split diagrams with all three propagators into terms with at most two of the propagators at a time
\begin{equation}
\label{eq:partial_frac_basic}
    \frac{1}{\rho_1}\frac{1}{\rho_2}\frac{1}{\rho_3}=\frac{1}{\rho_1\rho_3^2}-\frac{1}{\rho_2\rho_3^2}\,.
\end{equation}
Pictorially, this identity is expressed as a relation between mushroom-type integrals
\begin{equation}
    \vcenter{\hbox{\linMushroom}}=\vcenter{\hbox{\linMushroomPFII}}-\vcenter{\hbox{\linMushroomPFI}}\label{eq:PartialFraction}\,,
\end{equation}
where the dot represents a doubled propagator. Due to the soft expansion to higher orders, we also need to treat raised propagator powers with partial fractioning analogously to Eq.~(\ref{eq:partial_frac_basic}). The propagators on the right-hand side of \eqref{eq:PartialFraction} do not satisfy any linear relations and they can be embedded into different top-level families where the four-point vertex is blown up. For example 
\begin{align}
\vcenter{\hbox{\MushroomToEmbedd}}=\  
	(\ell_1+\ell_2)^{2} \vcenter{\hbox{\IYlin}}\,.
\end{align}
Similarly, we can proceed for all other mushroom topologies of Figure~\ref{fig:cubic_graphs}. In the soft region, these additional diagram topologies were required to match the classical part of the amplitude. However, upon soft expansion, their propagator structures overlap with existing topologies so that no new integral families are required.

The first integral on the right-hand-side of Eq.~\eqref{eq:PartialFraction} actually vanishes in dimensional regularization, because it factorize into a box integral times a matter self energy diagram which is scaleless in the soft region. More generally, non-factorizing integrals, where the loop momenta can be routed such that the integral is \emph{independent} of the momentum transfer $q$ are zero in dimensional regularization. One such example was presented in Figure~\ref{fig:eg_scaleless_diag}.
\begin{equation}
    \rho_1= 2u_1\cdot\ell_1+\imath\varepsilon\,,\quad\rho_2= 2u_1\cdot(\ell_1+\ell_2)+\imath\varepsilon\,,\quad\rho_3= 2u_1\cdot\ell_2+\imath\varepsilon\,.
\end{equation}
%

%================================================
%
\subsection{Reverse unitarity}
\label{subsec:rev_unitarity}
%
%================================================
%
We have seen in section \ref{sec:KMOC}, that classical gravitational observables, like the impulse kernel (\ref{eq:classical_impulse_kernel}), or the radiated momentum kernel (\ref{eq:KMOC_kernel_rad}) involve not only virtual amplitudes, but also certain unitarity cuts. These cut contributions include an integral over the on-shell phase space of the exchanged states. In order to efficiently evaluate such phase-space integrals, we follow our earlier letter \cite{Herrmann:2021lqe}, where we took inspiration from the enormous progress in cross-section calculations and the computation of collider physics observables where similar \emph{real} contributions appear. For some time, it has proven advantageous to handled phase-space integrals on the same footing as \emph{virtual} integrals. This idea has formally been implemented via \emph{reverse unitarity} \cite{Anastasiou:2002yz,Anastasiou:2002qz,Anastasiou:2003yy,Anastasiou:2015yha}, where one replaces on-shell delta functions and their $n$-th derivatives by the difference of (appropriate powers of) propagators with varying $\imath\varepsilon$ prescription
\begin{equation}
   \frac{2\pi \imath} {(-1)^{n}\, n!} \delta^{(n)} (z) = \frac{1}{(z-\imath\varepsilon)^{n+1}} - \frac{1}{(z+\imath\varepsilon)^{n+1}} \,. \label{eq:deltaToProp}
\end{equation}
Trading all delta functions by differences of propagators allows us to employ standard tools for \emph{loop integrals} such as dimensional regularization, IBP reduction \cite{Tkachov:1981wb,Chetyrkin:1981qh}, and (canonical) differential equations \cite{Kotikov:1990kg,Bern:1992em,Gehrmann:1999as,Henn:2013pwa,Henn:2014qga} to evaluate a minimal set of \emph{master integrals}. From a practical perspective, we can treat any on-shell delta function as a regular propagator. This is owed to the fact that integration-by-parts identities, crucial in the derivation of the differential equations, are insensitive to the Feynman $\imath\varepsilon$. The same is true for the partial fractioning discussed in subsection~\ref{subsec:partialFraction}. This significantly simplifies our computations and circumvents the difficulties in having to evaluate integrals containing derivatives of delta functions that would otherwise appear.

As will be explained in section \ref{sec:soft_integrals}, the differences between the cut integrals compared to the virtual ones arise from the following facts:
\begin{itemize}
    \item Certain cuts can break diagram symmetries of the virtual diagram.
    \item Various terms in the differential equations can be omitted because the appropriate master integrals do not have the desired unitarity cut.
    \item The boundary conditions for the differential equations for the cut master integrals change, relative to the virtual integrals. 
\end{itemize}
Of course, all the described properties of cut integrals are well known from collider physics applications and we adapt them to the gravitational setting here. To reiterate, the huge advantage of the reverse unitarity setup arises from the fact that we can directly treat the phase-space integrations for the inclusive classical observables (such as the gravitational impulse or the radiated momentum) in one go without having to perform sequential integrations over the gravitational waveform. Even for more exclusive observables, like the radiated energy spectrum, we can add one new variable at a time, which still leads to simplified integrals. A detailed discussion of such quantities is left to the future.

%================================================
%
\section{Evaluation of soft master integrals}
\label{sec:soft_integrals}
%
%================================================

In this section, we explain the evaluation of the soft master integrals relevant for the computation of radiative observables up to $\mathcal{O}(G^3)$. We first review the relevant kinematic domain and show how to compute the soft integrals at one-loop level as a warm-up exercise. Subsequently, we explain how to evaluate virtual integrals as well as two and three particle cuts at two-loop level. This completes the set of relevant master integrals for classical observables at $\mathcal{O}(G^3)$. Ultimately, our set of master integrals can be recycled to obtain analogous classical observables for different theories (such as quantum electrodynamics), or for spinning black holes once the relevant integrands are available.

%================================================
\subsection{Soft one-loop integrals: Euclidean region and analytic continuation}
\label{subsec:1loop_soft_and_kinematics}
%================================================
%
Before elaborating on the evaluation of soft master integrals via differential equations, it is illustrative to recall the kinematic dependence of soft integrals. As mentioned before, upon soft expansion, all soft master integrals have their external mass and momentum transfer ($-q^2$) dependence determined by simple dimensional analysis. The only nontrivial kinematic dependence of the integral is through the dimensionless variable $y=u_1{\cdot}u_2$ (or equivalently in terms of $x$, s.t.~$y{=}\frac{1{+}x^2}{2x}$). The $\epsilon{=}(4{-}D)/2$ dependence can sometimes be computed exactly, otherwise we work in an expansion around $\epsilon{=}0$. 

\begin{figure}[t!]
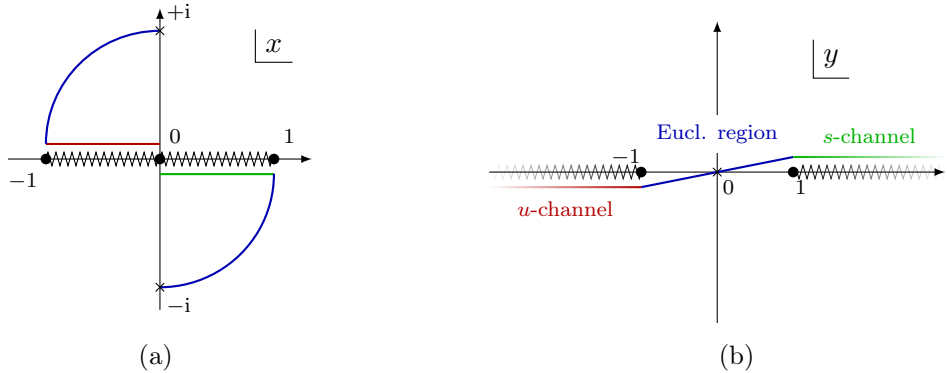

%
% The x-space figure
%
\begin{subfigure}[b]{0.5\textwidth}
	\centering
	\kinPlotX
    \subcaption{}
\end{subfigure}
%
% The y-space figure
%
\begin{subfigure}[b]{0.5\textwidth}
    \centering
%   \definecolor{lightgray}{grey}{0.2}
    \kinPlotY
\subcaption{}
\end{subfigure}
\caption{Kinematics in $x$ and $y$ space. The $x$ plane is a double cover of the $y$ plane, i.e. the two points $x$ and $1/x$ map to the same point $y$, we therefore focus on points inside the unit disc $|x|\leq 1$. Physical $s$-channel scattering (green region) corresponds to $y>1$ and $\Im(y)=0^+$, i.e. $0<x<1$ and $\Im(x)=0^-$. Physical $u$-channel scattering (red region) corresponds to $y<-1$ and $\Im(y)=0^-$, i.e. $-1<x<0$ and $\Im(x)=0^+$. There is also a Euclidean region connecting the two for real $-1<y<+1$ which corresponds to points on the unit circle $|x|=1$. Due to the double-cover property, $x=+\imath$ and $x=-\imath$ map to $y=0$. 
\label{fig:kinematics}
}
\end{figure}

The soft expansion, reviewed in section \ref{subsec:ibp_de}, is defined in a manifestly relativistic way and therefore soft integrals are genuine $D=4-2\epsilon$ dimensional Feynman integrals, although involving linearized propagators.  The manifest covariance has the benefit that the integrals are \emph{analytic functions} of $y$ and we can use analytic continuation to relate integrals in different kinematic regions. In contrast to the massless case, for massive $2\to 2$ scattering one can define a Euclidean region where all Lorentz invariants are below production threshold and the amplitude is real. As will become apparent from the explicit examples below, it is advantageous to compute the soft master integrals in the Euclidean region and then analytically continue to the desired scattering kinematics. These regions, together with the analytic continuation are summarized in Figure~\ref{fig:kinematics}. 

\noindent
As an illustrative example, consider the one-loop box family of the form (see also Ref.~\cite{Parra-Martinez:2020dzs})%
\footnote{In the following, we adopt the normalization conventions of Ref.~\cite{Smirnov:2012gma} and remove an overall factor of $\frac{\imath}{(4\pi)^2}\left(\overline{\mu}^2\right)^\epsilon {\equiv} \frac{\imath}{(4\pi)^2} \left(4\pi e^{-\EulerGamma}\mu^2\right)^{\epsilon} $ per loop, where $\mu$ is the dimensional regularization scale. To convert our results to standard Feynman integral conventions, we multiply by $\frac{i}{(4\pi)^2}\left(\overline{\mu}^2\right)^\epsilon$ per loop.}
\begin{align} 
    \vcenter{\hbox{\scalebox{0.8}{\linBoxPropLab}}} =
    G_{i_1,i_2,i_3,i_4}=\int\frac{e^{\EulerGamma\epsilon}\mathrm{d}^{D}\ell}
    {\imath\pi^{D/2}}
    \frac{1}{\rho_1^{i_1}\rho_2^{i_2}\rho_3^{i_3}\rho_4^{i_4}}
\end{align} 
where the linearized propagators are explicitly 
\begin{align}
	\rho_1=2u_1\cdot\ell + \imath \varepsilon \,,\quad
	\rho_2=-2u_2\cdot\ell + \imath \varepsilon \,,\quad
	\rho_3=\ell^2 + \imath \varepsilon \,,\quad
	\rho_4=(\ell-q)^2 + \imath \varepsilon\,.
\end{align}
There are 3 master integrals
\begin{equation} 
    f_1={} \epsilon(-q^2)\, G_{0,0,2,1}\,,      \quad             
	f_2={} \epsilon^2\sqrt{-q^2}\, G_{1,0,1,1}\,, \quad              
	f_3={} \epsilon^2\sqrt{y^2-1}(-q^2)\, G_{1,1,1,1}\,.  
\end{equation} 
The differential equation is 
\begin{equation} 
 \frac{\partial}{\partial x}\vec{f}= \frac{\epsilon}{x}
    \begin{pmatrix} 
        0 & 0 & 0 \\ 
        0 & 0 & 0 \\ 
        1 & 0 & 0 
    \end{pmatrix}\vec{f} \,.\label{eq:OneLoopDE}
\end{equation}
The system only has a single letter and can thus be integrated to all orders, using $x=-1$ as a boundary condition
\begin{equation}
\vec{f}(x)=\begin{pmatrix}
		0\\
		0\\
		\epsilon\log(-x)f_1(-1)\\
		\end{pmatrix}+\vec{f}(-1)\,.
	\label{eq:soloneloop}
\end{equation}
For the boundary conditions, we can directly evaluate the bubble and triangle integrals, using the master formula for the linearized triangle with arbitrary powers of the propagators (see e.g. Ref.~\cite{Smirnov:2012gma})
\begin{align}
&\int \frac{\mathrm{d}^D\ell}{\imath \pi ^{D/2}}
\frac{1}{[\ell^2]^{a_1}[(\ell-q)^2]^{a_2}[2 v\cdot\ell+\imath\varepsilon]^{a_3}}
=-(-1)^{a_1+a_2}(-q^2)^{D/2{-}a_1{-}a_2{-}1/2a_3}(v^2)^{-a_3}\nonumber\\
{}&\times\frac{ 
        \Gamma\left(a_3/2\right)
        \Gamma\left(D/2{-}a_1{-}a_3/2\right) 
        \Gamma\left(D/2{-}a_2{-}a_3/2\right) 
        \Gamma\left(a_1{+}a_2{+}a_3/2{-}D/2\right)}
        {2 \Gamma \left(a_1\right) \Gamma \left(a_2\right) 
        \Gamma \left(a_3\right) \Gamma \left(D{-}a_1{-}a_2{-}a_3\right)}	\,.\label{eq:genTriangle}
\end{align}
Applied to the integrals of interest, we find 
\begin{align}
\vcenter{\hbox{\bubbleWithDot}}={}&\left(-q^2\right)^{-1-\epsilon}e^{\EulerGamma  \epsilon}\frac{\sqrt{\pi } \Gamma \left(\frac{1}{2}-\epsilon \right)^2 \Gamma \left(\epsilon +\frac{1}{2}\right)}{2 \Gamma (-2 \epsilon )} \\
\label{eq:lin_triangle_result}
\vcenter{\hbox{\lintriTTNoLab}}
        \hspace{-.3cm}
        =& -
        \left(-q^2\right)^{-\frac{1}{2}-\epsilon}e^{\EulerGamma  \epsilon}  \frac{\sqrt{\pi } 
        \Gamma \left(\frac{1}{2}-\epsilon\right)^2 
        \Gamma \left(\epsilon+\frac{1}{2}\right)}{2 \Gamma (1-2 \epsilon)}\,.
\end{align}
For the boundary condition of the box integral, we resort to the method of regions. For this analysis it is convenient to choose a frame which coincides with the rest frame of $u_1$ up to $q$-corrections
\begin{equation}
u_1=(1,0,0,0)\,,\quad u_2=(\sqrt{1+v^2},0,0,v\,)\,. \label{eq:u1u2framechoice}
\end{equation}
In this frame we have $y=\sqrt{v^2+1}$. By crossing the limit $x\to -1$ corresponds to the static limit of the crossed box integral. The leading contribution in this region comes from the potential region, where the integral vanishes, the subleading contribution from the ``quantum soft'' region scales as $\mathcal{O}(v)$, so we find $f_3(-1)=0$. With this we can evaluate the crossed box integral 
\begin{equation}
\vcenter{\hbox{\linxBox}}=(-q^2)^{-1-\epsilon}e^{\EulerGamma \epsilon}
        \frac{\Gamma (-\epsilon )^2 \Gamma (1+\epsilon)}{2\Gamma (-2 \epsilon)}
        \frac{\log (x)}{\sqrt{y^2-1}}\,,\quad x>0\,.
\end{equation}
By analytic continuation, we obtain the box integral
\begin{equation}
\label{eq:soft_box_1L}
\vcenter{\hbox{\linBox}}=-(-q^2)^{-1-\epsilon}e^{\EulerGamma\epsilon}
    \frac{\Gamma (-\epsilon )^2 \Gamma (1+\epsilon)}{2\Gamma (-2 \epsilon)}
    \frac{\log (x)-\imath \pi}{\sqrt{y^2-1}}\,,\quad x>0\,.
\end{equation}
Now as we discussed before, we can use the same differential equation \eqref{eq:OneLoopDE} for the two-particle cuts. In this case the triangle and bubble functions are trivially zero because they do not have a relevant cut. This directly implies that the cut box integral is constant and given by its value at $y=1$. The integral can be directly evaluated and reduced to a $(D-2)$-dimensional Euclidean bubble integral
\begin{align}
\begin{split}
     &\hspace{-4cm}\vcenter{\hbox{\linCutBox}}
     =\frac{1}{\sqrt{y^2-1}}\int \frac{e^{\EulerGamma\epsilon}\mathrm{d}^D\ell}{\imath\pi^{D/2}} \, \frac{\hat{\delta}(2u_1\cdot\ell)\,\hat{\delta}(2u_2\cdot\ell)}{\ell^2(\ell-q)^2}
    \\
    = \frac{-\imath\pi}{\sqrt{y^2{-}1}}\int \frac{e^{\EulerGamma\epsilon}\mathrm{d}^{D-2}\bm{\ell}_\perp}{\pi^{(D-2)/2}} \frac{1}{\bm{\ell}_\perp^2(\bm{\ell}_\perp{-}\bm{q}_\perp)^2}
    & =(-q^2)^{-1-\epsilon}e^{\EulerGamma\epsilon} \frac{\imath\pi}{\sqrt{y^2{-}1}}\frac{  \Gamma (-\epsilon )^2 \Gamma (1+\epsilon )}{\Gamma (-2 \epsilon )}\,,
\end{split}    
\label{eq:Cut1LoopBox}
\end{align}
and we can check that the cutting rules are satisfied\footnote{The additional factor of $\imath$ is due to our conventions of the integral measure.}
\begin{equation}
\vcenter{\hbox{\linCutBox}}
= 2 \imath\, \Im \left[\!\!\vcenter{\hbox{\linBox}}\right]\,.
\end{equation}
The triangle integral has no $x$ dependence and is therefore completely specified by the boundary conditions at $x=-1$.

%================================================
\subsection{Virtual two-loop integrals 
\label{subsec:VirtInt}}
%================================================
%
The most involved part in our KMOC computation of the classical gravitational observables at $\calO(G^3)$ is the evaluation of the virtual two-loop soft integrals. The differential equation matrices have been constructed in Ref.~\cite{Parra-Martinez:2020dzs}, with the exception of the odd-in-$|q|$ integrals for the H family which we add in this work.\footnote{Since IBP reduction of an integral gives a sum of master integral with analytic-in-$q^2$ coefficients, integrals that scale like odd powers of $|q|$ form a decoupled system under IBP relations and differential equations.} A complete list of all master integrals, together with our conventions, is given in Appendix~\ref{app:masters_notation}.

Similarly to the one-loop discussion, it is advantageous to first evaluate all integrals in the Euclidean region and then analytically continue to the desired scattering kinematics. In the Euclidean region, the integrals are real-valued which serves as a valuable cross check on the calculation and also facilitates numerical verification against e.g. \texttt{PySecDec} \cite{Borowka:2017idc,Borowka:2018goh}.\footnote{Some pure integrals have a prefactor $\sqrt{y^2-1}$, and become purely imaginary without a real part.} 

In the most general (nonplanar) case, which the IX integral in Figure~\ref{fig:XBTopos} is an example of, the Euclidean region is $-1 < y <1$. The scattering regions are (1) s-channel: $y>1, \, \operatorname{Im}(y) = 0^+$, (2) u-channel: $y<-1, \, \operatorname{Im}(y) = 0^-$. For planar integrals, the Euclidean region is larger, given by $y<1$, and includes the u-channel scattering region, so a nontrivial analytic continuation is needed only for the s-channel scattering region.

The boundary conditions can be fixed by various methods, we discuss a set of sufficient conditions given by known single scale integrals, regularity and the method of regions.

\begin{figure}[t!]
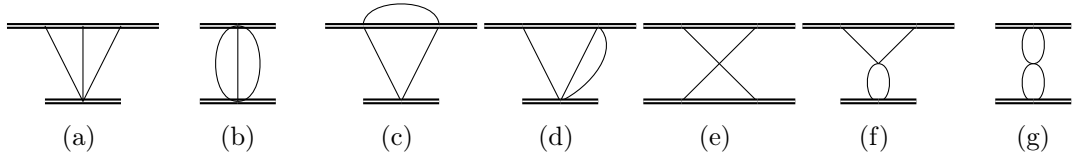

 \centering
  \begin{subfigure}[b]{0.13\linewidth}
  \centering
   \linTriTri
   \caption{\label{fig:BNDA}}
   \end{subfigure}
     \begin{subfigure}[b]{0.13\linewidth}
     \centering
   \sunriseDiag
   \caption{\label{fig:BNDB}}
   \end{subfigure}
     \begin{subfigure}[b]{0.13\linewidth}
     \centering
  \BNDIntA
   \caption{\label{fig:BNDC}}
   \end{subfigure}
     \begin{subfigure}[b]{0.13\linewidth}
     \centering
   \BNDIntB
   \caption{\label{fig:BNDD}}
   \end{subfigure}
     \begin{subfigure}[b]{0.13\linewidth}
     \centering
   \BNDIntC
   \caption{\label{fig:BNDE}}
   \end{subfigure}
     \begin{subfigure}[b]{0.13\linewidth}
     \centering
   \BNDIntD
   \caption{\label{fig:BNDF}}
   \end{subfigure}
     \begin{subfigure}[b]{0.13\linewidth}
     \centering
   \BNDIntE
   \caption{\label{fig:BNDG}}
   \end{subfigure}
 \caption{Single-scale integrals appearing as boundary values for two-loop soft integrals.}
 \vskip -.5cm
 \end{figure}

\paragraph{Single scale integrals } First, there are a handful of single-scale integrals independent of $y$, for example the sunrise integral in Figure~\ref{fig:BNDB}. These integrals are either factorizing into one-loop integrals (Figure~\ref{fig:BNDE}--\ref{fig:BNDG}), or can be be performed loop-by-loop (Figure~\ref{fig:BNDB}--\ref{fig:BNDD}), eventually reducing them to one-loop integrals which can be evaluated using the master formula \eqref{eq:genTriangle}.
Some integrals, like the double triangle in Figure~\ref{fig:BNDA}, can be computed using the trick of symmetrizing over the graviton momenta. This turns the matter propagators into delta functions and the integral becomes three-dimensional (cf. Appendix A of Ref.~\cite{Parra-Martinez:2020dzs}).

\paragraph{Regularity}
Another input is the regularity of integrals in the Euclidean region. In our kinematic parametrization, this translates to the statement that the s-channel planar integrals have to be regular at $y=-1$ or $x=-1$. At two loops, this only provides non-trivial constraints for planar integrals in the $u$-channel.  Finally, integrals odd under parity $\sqrt{y^2-1}\to -\sqrt{y^2-1}$ have to vanish at  $y=-1$. Using all these conditions, we obtain all-order boundary conditions for all integrals in the $\H$ family. 

\paragraph{Analysis of regions}
For the remaining integrals we obtain boundary conditions by the method of regions, splitting the soft region into subregions defined in Eq.~\eqref{eq:powerCountingVelocity}. We again adapt the frame defined in Eq.~ \eqref{eq:u1u2framechoice}.
As an illustrative example of how we can use the method of regions to obtain boundary values at two-loops, we consider the scalar $\RT$ integral. By naive velocity power-counting the leading contribution in the small velocity limit comes from the region where all gravitons are in the potential region. In this region the integral scales as $1/v^2$. All other regions are suppressed in velocity --- the next-to-leading contribution arises from the region where one of the gravitons is in the radiation region and scales as $v^{-1-2\epsilon}$ by naive power-counting. Therefore the value of the canonically normalized scalar $\RT$ integral at $v=0$ equals the potential-region boundary value, which has been computed in Ref.~\cite{Parra-Martinez:2020dzs},
\begin{equation}
    \epsilon^4(y^2-1)\vcenter{\hbox{\linearIII}}
   =
    \frac{\pi^2}{2}\epsilon^2-\frac{\pi^3}{12}\epsilon^3+\calO(\epsilon^4)+\calO(v^{1-2\epsilon})\,.
\end{equation}
Combining these different methods allows us to determine a complete set of boundary conditions to all orders in $\epsilon$ for the virtual soft integrals relevant at $\mathcal{O}(G^3)$. The complete velocity-dependent analytic values of the soft master integrals can then be obtained by integrating the differential equations from Ref.~\cite{Parra-Martinez:2020dzs} in conjunction with the soft boundary conditions. The explicit values of all master integrals are given in the ancillary files accompanying this work.

%================================================
\subsection{Two-particle cut integrals from sub-loop integration}
%================================================
%
A general two-particle cut integral can be evaluated by sub-loop integration. This is true for any loop order, however we focus on the two-loop case here. This can be seen as follows: all two-particle cut integrals that we encounter in the KMOC-formalism, including those with a numerator insertion, can be cast into the form 
\begin{align}
    \vcenter{\hbox{\twoPtCutGeneric}}={}&\int \frac{e^{\EulerGamma\epsilon}\mathrm{d}^D\ell}{\imath\pi^{D/2}} \, \hat{\delta}(2u_1\cdot\ell)\,\hat{\delta}(2u_2\cdot\ell)\  
    I_{\mathrm{L}}(\ell^2,y)\ I_{\mathrm{R}}((\ell-q)^2,y)\,,
\end{align}
where $I_{\mathrm{L}}$ and $I_{\mathrm{R}}$ are the sub loop integrals to the left and right of the two-particle cut, respectively. Since the momentum transfer is the only dimensionful quantity, it can be fixed by dimensional analysis
\begin{equation}
     I_{\mathrm{L}}(\ell^2,y)
        =\left[-\ell^2\right]^{\alpha_\mathrm{L}} \tilde{I}_{\mathrm{L}}(y)\,,
        \quad 
    I_{\mathrm{R}}((\ell-q)^2,y)
        =\left[-(\ell-q)^2\right]^{\alpha_\mathrm{R}} \tilde{I}_{\mathrm{R}}(y)\,.
\end{equation}
This leads to the following formula relating the sub-loop integrations to the cut 
\begin{align}
    \vcenter{\hbox{\twoPtCutGeneric}}
    =&\tilde{I}_{\mathrm{L}}\ \tilde{I}_{\mathrm{R}}
    \int \frac{e^{\EulerGamma\epsilon}\mathrm{d}^D\ell}{\imath\pi^{D/2}}\,\hat{\delta}(2u_1\cdot\ell)\,\hat{\delta}(2u_2\cdot\ell)
    \left[-\ell^2\right]^{\alpha_\mathrm{L}}\!
    \left[-(\ell-q)^2\right]^{\alpha_\mathrm{R}}\,.
\end{align}
The $\delta$-functions localize the integral to an (euclidean) integral over transverse space\footnote{See also the relevant discussion around Eq.~(\ref{eq:sudakov_param}) in appendix \ref{app:FT_collection}.}
\begin{align}
\label{eq:doubleCutMasterFormula}
\vcenter{\hbox{\twoPtCutGeneric}}
=&-\imath(2\pi)^2\frac{\tilde{I}_{\mathrm{L}}\,\tilde{I}_{\mathrm{R}}}{4\sqrt{y^2-1}}
\int\frac{e^{\EulerGamma\epsilon}\mathrm{d}^{D-2}\bm{\ell}_{\perp}}{\pi^{(D-2)/2}}
        \left[\bm{\ell}_{\perp}^2\right]^{\alpha_\mathrm{L}}
        \left[(\bm{\ell}_{\perp}-\bm{q}_{\perp})^2\right]^{\alpha_\mathrm{R}}
\\
=&-\imath\pi\frac{\tilde{I}_{\mathrm{L}}\, \tilde{I}_{\mathrm{R}}}{\sqrt{y^2-1}}
         (-q^2)^{1{-}\epsilon{+}\alpha_\mathrm{L}{+}\alpha_\mathrm{R}} 
%\nonumber\\ &\time 
    \frac{
        \Gamma \left(1{-}\epsilon {+} \alpha_\mathrm{L}\right) 
        \Gamma \left(1{-}\epsilon {+}\alpha_\mathrm{R}\right) 
        \Gamma \left(\epsilon {-} \alpha_\mathrm{L} {-} \alpha_\mathrm{R}{-}1\right)
        }
        {e^{-\EulerGamma\epsilon}
        \Gamma \left({-}\alpha_\mathrm{L}\right) 
        \Gamma \left({-}\alpha_\mathrm{R}\right) 
        \Gamma \left(2{-}2 \epsilon{+}\alpha_\mathrm{L}{+}\alpha_\mathrm{R}\right)
        }\,. \nn
\end{align}
As a concrete example we consider the two-particle-cut of the scalar III, where we have
\begin{equation}
\hspace{-.5cm}
I_{\mathrm{L}}{=}\vcenter{\hbox{\linTree}}{=}\frac{1}{-\ell^2}
\,,\,\, 
I_{\mathrm{R}}{=}\left[\!\!\vcenter{\hbox{\linBox}}\right]^*
{=}\frac{-e^{\EulerGamma\epsilon}}{\left[-(\ell{-}q)^2\right]^{1+\epsilon}}
\frac{\Gamma (-\epsilon )^2 \Gamma (1{+}\epsilon)}{2\Gamma (-2 \epsilon)}
\frac{\log x{+}\imath \pi}{\sqrt{y^2-1}}\,.
\label{eq:doubleCutBoxConstituents}
\hspace{-.4cm}
\end{equation}
Combining Eqs.~\eqref{eq:doubleCutMasterFormula} and \eqref{eq:doubleCutBoxConstituents}, we find 
\begin{align}
\vcenter{\hbox{\twoPtCutIII}}={}&(-q^2)^{-2\epsilon}
    \left[
    \frac{ e^{2\EulerGamma\epsilon}\pi^2\csc(2\pi\epsilon)\Gamma (-\epsilon )^3}
    {2\Gamma (-2 \epsilon ) \Gamma (-3 \epsilon)}
    \right]
    \frac{\log(x)+\imath\pi}{y^2-1}\,.
\end{align}
Likewise, we can evaluate all other two-particle cut integrals by our loop-by-loop integration technique, using the known one-loop building blocks and the general master formula in Eq.~\eqref{eq:doubleCutMasterFormula}. Notably, the imaginary parts of the one-loop building blocks change, depending on whether they are inserted to the left or to the right of the cut legs.

%================================================
\subsection{Triple-cut integrals from differential equations and Cutkosky rules}
%================================================
%
We first start by considering triple cuts of integrals inside the H family. It will turn out that all other triple-cut integrals can be obtained from these via differential equations. Using the cutting rules, reviewed in appendix \ref{app:cutting_rules}, we can relate the triple cut to the imaginary part for a $\H$-type integral we find
\begin{equation}
I_{\mathrm{H},\mathrm{3pt-cut}}=2\Im I_{\mathrm{H}}\,.
\end{equation}
Since we already computed the virtual integrals in subsection~\ref{subsec:VirtInt}, this allows to obtain all triple-cuts in the $\H$  family. To give a concrete example, we have
\begin{align}
\label{eq:CutNDot}
   \vcenter{\hbox{\trippleCutNDot}}={}&2\Im\left[\vcenter{\hbox{\linNDot}} \right]
   \\
   ={}&(-q^2)^{-2\epsilon-1}\frac{\pi}{\epsilon^2\sqrt{y^2-1}}
    \left\{1-2\epsilon\left[\log(1-x^2)-\log(x)\right]+\calO(\epsilon^2)\right\}\nn\,.
\end{align}
We notice that there are only four integrals which cannot be embedded into the $\H$ topology or its crossing, namely the $\RT$ and the $\IX$ and the planar and non-planar box triangles. For these integrals we can make use of a particular feature of the differential equation, which is that derivatives of these integrals are expressible in terms of integrals inside the H topology. Therefore these integrals can be computed by direct integration. As a concrete example we consider a triple cut of the $\IX$ integral. The derivative of the canonically normalized $\IX$ integral is proportional to a N-type integral, which is known from the $\H$ family
\begin{equation}
	\frac{\partial}{\partial x}\left[(y^2-1)
	\vcenter{\hbox{\trippleCutIX}}\right]
	=-\frac{1}{x}\left[\sqrt{y^2-1}\vcenter{\hbox{\trippleCutNDot}}\right]\,.
\end{equation}
For the boundary condition we make use of the method of regions. The power-counting for a cut integral is identical to the corresponding virtual integral. For the triple-cut contribution, one of the gravitons is on-shell and therefore this integral receives no contribution from the potential region. The leading behavior as $v\to 0$ is therefore dictated by the potential-radiation region which scales as $\mathcal{O}(v^{-1-2\epsilon})$ by the power-counting of Eq.~\eqref{eq:powerCounting}. As the integral appearing in the canonical differential equation is normalized by a factor of $y^2-1=v^2$ (see appendix \ref{app:masters_notation} for details) it vanishes in the static limit and using Eq.~\eqref{eq:CutNDot} we find
\begin{align}
	\vcenter{\hbox{\trippleCutIX}}={}&-\frac{1}{y^2-1}\int_1^x \frac{\mathrm{d}x^\prime}{x^\prime}\left[\sqrt{{y^\prime}^2-1}\vcenter{\hbox{\trippleCutNDot}}\right]\\
	={}&-(-q^2)^{-2\epsilon-1}\frac{\pi}{\epsilon^2(y^2-1)}
	\Big\{
	    \log(x)
	    {+}\epsilon\left[
	        \log(x)^2
	        {+}\operatorname{Li}_2(x^2)
	        {-}\frac{\pi^2}{6}
	    \right]
	    {+}\mathcal{O}(\epsilon^2)
	\Big\}\nonumber\,.
\end{align}
Similar ideas also apply to the three-particle cut integrals of III and will not be displayed explicitly. This concludes our discussion of all relevant virtual and cut master integrals required for the determination of $\calO(G^3)$ classical observables in the KMOC formalism.

%================================================
\subsection*{Integral checks}
%================================================
%
We have performed several consistency checks on our virtual and phase-space integrals. As mentioned previously, for the virtual two-loop integrals, we have performed extensive numerical checks to high precision against \texttt{PySecDec} \cite{Borowka:2017idc,Borowka:2018goh} in the Euclidean region where the integrals are real-valued. As a cross-check of our analytic continuation, we have furthermore performed numerical comparisons for scattering kinematics in $s$ and $u$-channel regions, where our analytic results agree with numerical values within numerical errors.

We also compared our results to available analytic expressions for the full integrals in the equal mass $m_1{=}m_2$ case, finding agreement for the non-analytic-in-$q$ parts for the $\H$-type integrals \cite{Bianchi:2016yiq} and ladder-integrals \cite{Smirnov:2001cm,Henn:2013woa, Heinrich:2004iq}, to the orders of $\epsilon=(4-D)/2$ available in the literature.

Furthermore, we have checked the results of our master integrals against cutting rules. For example, we have checked that Eqs.~\eqref{eq:IIIcuttingRule1} and \eqref{eq:XBcuttingRule1} hold as relations for the soft-expanded master integrals, i.e.\ with all quadratic matter propagators replaced by their linearized expressions at the leading order expansion in $|q|$ given by the first term on the r.h.s. of Eq.~\eqref{eq:softExpansionMatterProp}.

Lastly, our virtual master integrals have been checked against an independent calculation~\cite{DiVecchia:2021bdo} which we learned from private communications. Each integral has been checked to the maximum order of $\epsilon$ that has been computed in both papers.

We note that all velocity-dependent functions satisfy a \emph{first-entry condition} \cite{Gaiotto:2011dt}, where only $x$ is allowed as first symbol \cite{Goncharov:2010jf,Duhr:2011zq,Duhr:2012fh} entry. This is obvious for the one-loop integrals (which only contain $\log x$), but becomes nontrivial at two-loop order and suggests potentially further simplifications by eliminating more explicit boundary value evaluations due to this analyticity property. 

%=======================================================================
\section{Simplifications in the KMOC setup} 
\label{sec:KMOC_discussion_expansion}
%=======================================================================
We have reviewed the KMOC formalism in section \ref{sec:KMOC}, together with general formulae for the gravitational impulse kernel (\ref{eq:classical_impulse_kernel}), and the radiated momentum kernel (\ref{eq:KMOC_kernel_rad}). Here, we would like to discuss a convenient organization of these quantities as well as aspects of their perturbative expansions, before presenting their explicit results in maximal supergravity and general relativity up to $\calO(G^3)$ in section \ref{sec:result}. More concretely, we use unitarity and the cutting rules to obtain simplified KMOC formulae where certain properties of the impulse kernel, such as its reality properties or the absence of superclassical term are more manifest.

To begin the discussion, it is convenient to decompose the total impulse into its transverse, $\Delta p_\perp$, and longitudinal, $\Delta p_u$, components
\begin{equation}
\label{eq:impulse_decomp}
    \Delta p^\mu = \Delta p_\perp^\mu + \Delta p_u^\mu\,,
\end{equation}
such that $ u_i {\cdot} \Delta p_\perp {=} 0$ and $ q {\cdot} \Delta p_u {=} 0$. (For the relevant kinematic definitions, c.f.~the beginning of section \ref{subsec:ibp_de}.) Correspondingly, the impulse kernel can be written as
\begin{equation}
\label{eq:impulse_kernel_projection}
  \mathcal{I}^\mu_{p_1} =   {\cal I}_\perp \,q^\mu + 
                            \sum_{i=1,2}{\cal I}_{u_i} \, \check{u}^{\mu}_i,
\end{equation}
where we have defined \emph{dual} four-velocities
\begin{equation}
\label{eq:dual_ui}
   \check{u}^{\mu}_1 = \frac{y u_2^\mu - u_1^\mu}{y^2-1}, \qquad 
   \check{u}^{\mu}_2 = \frac{y u_1^\mu - u_2^\mu}{y^2-1},
\end{equation}
which satisfy $u_i \cdot \check{u}_j = \delta_{ij}$ and are still orthogonal to the momentum transfer $q$. Decomposing the loop momentum dependent impulse numerator in a similar fashion\footnote{In principle, there is an orthogonal direction $\varepsilon(\cdot, q,\check{u}_1,\check{u}_2)$ that, however, does not play a role.}
\begin{align}
\label{eq:ell1_decomp}
    \ell^\mu_1  = \frac{\ell_1\cdot q}{q^2} q^\mu 
                + (\ell_1\cdot u_1)\, \check{u}^{\mu}_1
                + (\ell_1\cdot u_2)\, \check{u}^{\mu}_2\,,
\end{align}
reveals that only the transverse part of the impulse has a \emph{virtual} contribution 
\begin{align}
\begin{split}
    {\cal I}_\perp =&
    \hspace{-.3cm}
    \raisebox{-32pt}{\scalebox{0.8}{\kmocvirtual}}
     - \hspace{.3cm} \imath\ \sum_X \int \mathrm{d}\widetilde{\Phi}_{2+|X|} \,\, \frac{\ell_1\cdot q}{q^2} 
    \hspace{-.3cm}
    \raisebox{-35pt}{\scalebox{0.8}{\kmocreal{$\ell_2-p_2$}{$\ell_1-p_1$}}} \,,
\end{split}
\end{align}
whereas the longitudinal part is purely \emph{real}, i.e.~it only receives contributions from the unitarity cut terms
\begin{align}
\begin{split}
   {\cal I}_{u_i} =&-\imath \sum_X \int \mathrm{d}\widetilde{\Phi}_{2+|X|}
   \,\,\ell_1\cdot u_i 
   \hspace{-.3cm}
   \raisebox{-35pt}{\scalebox{0.8}{\kmocreal{$\ell_2-p_2$}{$\ell_1-p_1$}}}\,.
\end{split}
\end{align}
Note that due to the difference in the $|a|$ scaling of $q^\mu$ and $ \check{u}^{\mu}_i$, we have to expand the longitudinal impulse kernels to one higher order in $|q|$ compared to the transverse ones. Finally, all classical observables are real\footnote{The reality properties of (possibly) complex quantities should not to be confused with our nomenclature of \emph{real}, i.e.~cut, contributions to various classical observables.} (not complex), and the various factors of $\imath$ in the KMOC setup serve this purpose. In particular, the transverse KMOC kernels need to be purely real to yield a real result after the final Fourier transform (Eq.~\eqref{eq:KMOC_DObs}), whereas the longitudinal kernels are purely imaginary. Indeed, it will serve as a nontrivial check of our computation, that all imaginary contributions to the classical observables cancel.

%=======================================================================
\noindent
\subsection{Leading and next-to-leading order impulse} 
%=======================================================================
At leading order, $\calO(G)$, the impulse kernel is given by the tree level scattering amplitude 
\begin{align}
\label{eq:impulse_kernel_LO}
\begin{split}
    \calI^{\mu,(0)}_{p_1} &= q^\mu
    \hspace{-.3cm}
    \raisebox{-32pt}{\scalebox{0.8}{\kmocvirtualtree}}\,.
\end{split}
\end{align}
There is only the \emph{virtual} contribution at this order, since the scattering amplitude starts at ${\cal O}(G)$ and the \emph{real} contribution is quadratic in the amplitude.

At next-to-leading order, $\calO(G^2)$, the impulse kernel receives both \emph{virtual} and \emph{real} contributions. The transverse component is
\begin{align}
\label{eq:nlo_transv_impulse}
    \calI^{(1)}_\perp =&
    \hspace{-.3cm}
    \raisebox{-32pt}{\scalebox{0.8}{\kmocvirtualnlo}}
    -\hspace{.3cm} \imath \int \mathrm{d}\widetilde{\Phi}_{2} \,\, \frac{\ell_1\cdot q}{q^2}  \hspace{-.3cm}
    \raisebox{-32pt}{\scalebox{0.8}{\kmocrealnlo}}\,,
\end{align}
and the longitudinal component is 
\begin{align}
    \calI^{(1)}_{u_i} =& 
    -\imath \int \mathrm{d}\widetilde{\Phi}_{2} \,\, \ell_1\cdot u_i  
    \hspace{-.3cm}\raisebox{-35pt}{\scalebox{0.8}{\kmocrealnlo}}\,.
   \label{eq:kerneluoneloop}
\end{align}
In Ref.~\cite{Kosower:2018adc} it was shown that the superclassical part of the one-loop virtual amplitude cancels at the integrand level when expanded in the classical limit. Here, we offer an alternative argument that will streamline the calculation of the kernel. The basic observation is that the cut has a horizontal flip symmetry which does not change the sign of the integral. Thus one might average over the two different labellings of loop momenta, related by $\ell_1\leftrightarrow q{-}\ell_1$
\begin{equation}
     \frac12 \left[\frac{\ell_1\cdot q }{q^2} + \frac{ (q-\ell_1)\cdot q }{q^2} \right] = \frac12\,.
     \label{eq:numsym}
\end{equation}
This means that the transverse impulse numerator insertion is in fact constant and the cut contribution can be related to the imaginary part of the amplitude via the unitarity relation in Eq.~$\eqref{eq:unitarityoneloop}$. Thus, by virtue of Eqs.~\eqref{eq:numsym} and \eqref{eq:unitarityoneloop}, all imaginary parts cancel and we find that the transverse classical impulse kernel is given by the real part of the one-loop amplitude,
\begin{equation}
    {\cal I}_\perp = \text{Re}\left[ \!\!\!\raisebox{-32pt}{\scalebox{0.8}{\kmocvirtualnlo}}\right]
    \label{eq:kernelperponeloop}
\end{equation}
We will explicitly show in section~\ref{subsec:NLO_impulse}, that at this order the superclassical pieces are contained in the imaginary part of the amplitude. Roughly, this can be understood from the fact that the imaginary part is related by unitarity (\ref{eq:unitarityoneloop}) to a cut that corresponds to the iteration of lower orders and therefore cancel in the impulse kernel.

The computation of the longitudinal kernels $\calI_{u_i}$ can also be simplified. We begin by noting that the four velocities $u_i$ in the numerator insertions in Eq.~\eqref{eq:kerneluoneloop} satisfy $u_i{\cdot} q {=} 0$. Thus, we can express them in terms of the momenta of the scattering amplitude (see the definition of the soft variables in section \ref{subsec:soft_expansion}) as $u_1{=}({-}p_1 {+} q/2)/m_1 + \calO(q^2)$ and $u_2{=}({-}p_2{-}q/2)/m_2 + \calO(q^2)$, and we can write
\begin{align}
\label{eq:impulse_num_long_decomp}
\begin{split}
    \ell_1\cdot u_1 &= \frac{-q^2}{4m_1} +  \frac{\left[(\ell_1-p_1)^2-m_1^2\right] -\ell_1^2}{2m_1} - \frac{(\ell_1-q)^2-\ell_1^2}{4m_1} =  \frac{-q^2}{4m_1} + \cdots\,,\\
    \ell_1\cdot u_2 &= \frac{+q^2}{4m_2} -  \frac{\left[(\ell_1+p_2)^2-m_2^2\right] -\ell_1^2}{2m_2} + \frac{(\ell_1-q)^2-\ell_1^2}{4m_2} = \frac{+q^2}{4m_2} + \cdots\,.
\end{split}
\end{align}
We have used the on-shell conditions $(\ell_1{-}p_1)^2{-}m_1^2 = 0$, $(\ell_2{-}p_2)^2{-}m_2^2{=}(\ell_1{+}p_2)^2{-}m_2^2 {=} 0$ and the $\cdots$ indicates contributions that pinch graviton propagators as well as quantum suppressed terms, which can be ignored.  At one-loop, the former yield short distance matter contact terms which are irrelevant for widely separated black holes. The resulting numerator is again loop-momentum independent. Using unitarity \eqref{eq:unitarityoneloop}, we find that the longitudinal impulse kernels are directly proportional to the imaginary part of the one-loop amplitude. Collecting all the ingredients we obtain a more direct relation between the impulse kernel and the scattering amplitude at this order
\begin{equation}
    \calI^{\mu,(1)}_{p_1}= q^\mu  \, \text{Re}\left[ \!\!\!
    \raisebox{-32pt}{\scalebox{0.8}{\kmocvirtualnlo}}\right] 
      + \imath\,(-q^2) \left(\frac{\check{u}^{\mu}_1}{2m_1} - \frac{\check{u}^{\mu}_2}{2m_2} \right)  \,
    \text{Im} \left[\!\!\! \raisebox{-32pt}{\scalebox{0.8}{\kmocvirtualnlo}} \right]\,,
\label{eq:impulse_kernel_nlo_all}    
\end{equation}
from which we learn that the impulse can be directly extracted from the virtual amplitude without the need of evaluating any phase space integrals.

%=======================================================================
\noindent
\subsection{Next-to-next-to-leading order impulse} 
\label{subsec:nnlo_impulse}
%=======================================================================
%
Next we discuss the simplifications at next-to-next-to-leading order, $\calO(G^3)$.
We will focus on the transverse part of the impulse and offer some comments about the longitudinal part.

%==========================================
\subsubsection{Transverse part}
%==========================================
At this order the transverse impulse kernel is given by
\begin{align}
\label{eq:impulse_NNLO_kernel}
\mathcal{I}^{(2)}_{\perp} 	
    & =  \raisebox{-22pt}{\scalebox{0.8}{\kmocvirtualnnlonolab}}
    \hspace{-.1cm}
	  - \imath \int \mathrm{d}\widetilde{\Phi}_{2} \,  \frac{\ell_1\cdot q}{q^2}
     \hspace{0cm}
     \left[\raisebox{-35pt}{\scalebox{0.8}{\kmocrealnnlolooptree}} \!\!+\!\! \raisebox{-35pt}{\scalebox{0.8}{\kmocrealnnlotreeloop}}\right] \nn \\
     &
     \hspace{3cm}
     -\imath \int \mathrm{d}\widetilde{\Phi}_{3}\,  \frac{\ell_1\cdot q}{q^2}
   \hspace{-.3cm}
   \raisebox{-35pt}{\scalebox{0.8}{\kmocrealnnlotreetree}}\,.
\end{align}
The three-particle cut in the second line always produces a real contribution. This cut enjoys the same horizontal flip symmetry as the one-loop two-particle cut. Thus one might once again average over the two different labellings of loop momenta as in Eq.~\eqref{eq:numsym}
\begin{equation}
    \int \mathrm{d}\widetilde{\Phi}_{3}\,  \frac{\ell_1\cdot q}{q^2}
   \hspace{-.3cm}
   \raisebox{-35pt}{\scalebox{0.8}{\kmocrealnnlotreetree}} = \frac12 \int \mathrm{d}\widetilde{\Phi}_{3}\,  
   \hspace{-.3cm}
   \raisebox{-35pt}{\scalebox{0.8}{\kmocrealnnlotreetree}}\,.
\end{equation}
The same considerations are valid for the real part of two-particle cuts. Any contribution and its horizontally flipped version combine to give a trivial impulse numerator upon averaging as in Eq.~\eqref{eq:numsym} such that
\begin{align}
\begin{split}
   &\int \mathrm{d}\widetilde{\Phi}_{2} \,  \frac{\ell_1\cdot q}{q^2}
     \hspace{0.2cm}
     \text{Re}\left[\raisebox{-35pt}{\scalebox{0.8}{\kmocrealnnlolooptree}} \!\!+\!\! \raisebox{-35pt}{\scalebox{0.8}{\kmocrealnnlotreeloop}}\right] \\ \\
     &=  \hspace{0.6cm}\frac12 \int \mathrm{d}\widetilde{\Phi}_{2} \,\left[\raisebox{-35pt}{\scalebox{0.8}{\kmocrealnnlolooptree}} \!\!+\!\! \raisebox{-35pt}{\scalebox{0.8}{\kmocrealnnlotreeloop}}\right] \,.
\end{split}
\end{align}
where we have dropped the restriction to the real part in the second line since the imaginary parts cancel in the sum in the absence of a nontrivial numerator. Therefore, the three-, and (real part of the) two-particle cut \emph{real} contributions combine to cancel the imaginary part of the \emph{virtual} amplitude in the impulse kernel, by virtue of unitarity \eqref{eq:unitaritytwoloop}.

The remaining pieces are the real part of the virtual amplitude together with the imaginary part of the two-particle cuts. The latter arises from the imaginary parts of the one-loop amplitudes on either side of the cut, which have opposite sign. Once again, we can simplify these by using one-loop unitarity \eqref{eq:unitarityoneloop} on the one-loop amplitude on the left of the cut
\begin{align}
\begin{split}
     \text{Im}\left[\raisebox{-35pt}{\scalebox{0.8}{\kmocrealnnlolooptree}}\right] 
     =\frac12 \int \mathrm{d}\widetilde{\Phi}_{2} \raisebox{-35pt}{\scalebox{0.8}{\iteratedtwocut}}  \,,
     \end{split}
\end{align}
where the additional phase-space integration on the r.h.s is over the newly cut legs, denoted by $\ell_3{-}p_1$ and $\ell_4{-}p_2$. The contribution with the one-loop amplitude on the right of the cut has the opposite sign. Combining both into a single term by choosing uniform labels we find our final expression for the transverse impulse
\begin{align}
\label{eq:impulse_NNLO_kernel_simp}
\mathcal{I}^{(2)}_{\perp} 	
    & =  \text{Re}\left[\raisebox{-22pt}{\scalebox{0.8}{\kmocvirtualnnlonolab}} \right]
     -\imath \int \mathrm{d}\widetilde{\Phi}_{2}^2\,  \frac{(\ell_1-\ell_3)\cdot q}{2q^2}
   \hspace{-.3cm}
  \raisebox{-35pt}{\scalebox{0.8}{\iteratedtwocut}} \,.
\end{align}
Eq.~\eqref{eq:impulse_NNLO_kernel_simp} provides a simplified prescription for the calculation of the impulse kernel, which reveals among other things, that the purpose of the (real part of) two-particle and three-particle cuts at this order is to cancel part of the imaginary part of the virtual amplitude. The only non-trivial contribution of the cuts (i.e. the \emph{real} part of the impulse kernel), take the form of the cubed of the tree amplitude, which is reminiscent of the calculation of the eikonal phase.

Eq.~\eqref{eq:impulse_NNLO_kernel_simp} also manifests the fact that the transverse impulse kernel is  real, as required by the fact that its Fourier transform is the transverse impulse which is a physical observable and hence also real.
It also facilitates exposing the cancellation of certain quantum contributions to the transverse impulse. For example, diagrams involving self-energy corrections to virtual graviton propagators are quantum corrections (i.e.\ not relevant for classical physics). Individual diagrams of this class can still be of order $q^0$, i.e.\ naively of classical order, according to the power counting rules in Eqs.~\eqref{eq:powerCountingRule} and  \eqref{eq:powerCountingFormula} arises from a cancellation between the $\calO(1/q^2)$ dependence of the single graviton exchange and the $\calO(q^2)$ quantum correction to the graviton propagator which yields a $\calO(|q|^0)$ contribution that should cancel in the classical impulse kernel. Using the identity
\begin{align}
\label{eq:bubbleBoxCancellation}
    \vcenter{\hbox{\scalebox{1}{\buboxs}}}
    +
    \vcenter{\hbox{\scalebox{1}{\buboxu}}}
    = 
    \vcenter{\hbox{\scalebox{1}{\buboxcut}}} 
    + \calO(|q|)\,,
\end{align}
we learn that such contributions are imaginary and indeed absent in Eq.~\eqref{eq:impulse_NNLO_kernel_simp}. We note that this integral only receives contributions from the "quantum soft" velocity region (see Eq.~\eqref{eq:powerCountingVelocity}). It would be interesting to explore whether the contributions from such region can be consistently dropped at an earlier stage in the calculation without spoiling some of the advantages of the full soft region computations.

In addition, Eq.~\eqref{eq:impulse_NNLO_kernel_simp} allows us to show the integrand level cancellation of superclassical terms. We first notice that all tree-amplitudes entering the iterated two-particle cut are on the same footing and are functions $\calM(\sigma,q_i^2)$ of the respective momentum transfer. Only the on-shell delta functions break the invariance of the cut under the permutation of the $q_i$'s. However at leading order in the classical expansion we have 
\begin{align}
\begin{split}
    &
    \delta(p_1^2{-}m_1^2)\,
    \delta((p_1{+}q_1)^2{-}m_1^2)\,
    \delta((p_1{+}q_1{+}q_2)^2{-}m_1^2)\,
    \delta((p_1{+}q_1{+}q_2{+}q_3)^2{-}m_1^2)
    \\
    ={}& 
    \delta(p_1^2{-}m_1^2)\,
    \delta(2p_1\cdot q_1)\,
    \delta(2p_1\cdot q_2)\,
    \delta(2p_1\cdot q_3)+\mathcal{O}(|q|)
\end{split}    
\end{align}
and likewise for the delta functions involving $p_2$. Therefore, realizing that $\ell_1{-}\ell_3{=}-q_2$ and summing over the cyclic relabellings, we find
\begin{align}
 &\int \mathrm{d}\widetilde{\Phi}_{2}^2\,  \frac{(\ell_1-\ell_3)\cdot q}{2q^2}
   \hspace{-.3cm}
  \raisebox{-35pt}{\scalebox{0.8}{\iteratedtwocut}}\nonumber\\
  ={}&-\frac{1}{3} \int \mathrm{d}\widetilde{\Phi}_{2}^2\,  \frac{(q_1+q_2+q_3)\cdot q}{2q^2}
   \hspace{-.3cm}
  \raisebox{-35pt}{\scalebox{0.8}{\iteratedtwocut}}+\mathcal{O}(q^0)\nonumber\\
={}&-\frac{1}{6} \int \mathrm{d}\widetilde{\Phi}_{2}^2\, 
   \hspace{-.3cm}
  \raisebox{-35pt}{\scalebox{0.8}{\iteratedtwocut}}+\mathcal{O}(q^0)  \,.
  \label{eq:minussuperclassical}
\end{align}
The leading superclassical part of the \emph{virtual} amplitude is purely contained in the planar double-box diagram. The same symmetrization relation can be applied to this diagram. Carefully keeping track of the $i\epsilon$ in the denominators yields
\begin{align}
    &\frac{\delta(q_1+q_2+q_3 + q)}{q_1^2q_2^2q_3^3} \frac{{\cal N}(p_i,q_i)}{3!}
    \left(\frac{1}{[(p_1{+}q_1)^2{-}m_1^2][(p_1{+}q_1+q_2)^2{-}m_1^2]} + \text{perms}(q_1,q_2,q_3)\right) \nn\\
    &\hspace{4.3cm}\times\left(\frac{1}{[(p_2{-}q_1)^2{-}m_2^2][(p_2{-}q_1{-}q_2)^2{-}m_2^2]} + \text{perms}(q_1,q_2,q_3)\right) \nn \\
    &\hspace{0.5cm}= \frac{1}{6}\frac{{\cal N}(p_i,q_i)\delta(q_1+q_2+q_3 + q)}{q_1^2q_2^2q_3^3} \delta(2p_1\cdot q_1)\delta(2p_1\cdot q_2)\delta(-2p_2\cdot q_1)\delta(-2p_2\cdot q_2)\,,
\end{align}
which generates on-shell delta functions from the four matter propagators and cancels against \eqref{eq:minussuperclassical}. Naively, there still remains a superclassical contribution in the virtual amplitude at $\calO(|q|^{-1})$, but this can be shown to be purely imaginary so it is manifestly absent in Eq.~\eqref{eq:impulse_NNLO_kernel_simp}.

An alternative, and perhaps more explicit, derivation of our simplified formulae proceeds by considering the contribution from each diagram in our integrand to Eq.~\eqref{eq:amplitude_cubic_diags_abstract}, together with the cutting rules in appendix \ref{app:cutting_rules}. We now give an explicit example of such proof focusing on the contribution of the III diagram to the impulse on the bottom matter line. The impulse numerator combines into a scalar in the sum over the two two-particle cuts as follows,
\begin{align}
\begin{split}
    & \
    \frac{\ell_1 \cdot q}{q^2} 
    \ \operatorname{Re} \Bigg[  \vcenter{\hbox{\ladderlabelsleft{$\ell_1$}{$\ell_2$}{$\ell_3$}}}  \Bigg] 
    \ + \ 
    \frac{(\ell_1+\ell_2) \cdot q}{q^2}   
    \ \operatorname{Re} \Bigg[\vcenter{\hbox{\ladderlabelsright{$\ell_1$}{$\ell_2$}{$\ell_3$}}} \Bigg] 
    \\
    = & \
    \frac{\ell_1 \cdot q}{q^2} 
    \ \operatorname{Re} \Bigg[ \vcenter{\hbox{\ladderlabelsleft{$\ell_1$}{$\ell_2$}{$\ell_3$}}}  \Bigg] 
    \ + \  
    \frac{(\ell_2+\ell_3) \cdot q}{q^2}  
    \ \operatorname{Re} \Bigg[  \vcenter{\hbox{\ladderlabelsleft{$\ell_1$}{$\ell_2$}{$\ell_3$}}}  \Bigg] 
    \\
    = \ 
    &  \
    \frac{(\ell_1 + \ell_2+\ell_3) \cdot q}{q^2} 
    \operatorname{Re} \Bigg[ \vcenter{\hbox{\ladderlabelsleft{$\ell_1$}{$\ell_2$}{$\ell_3$}}}  \Bigg] 
    \ = \operatorname{Re} \Bigg[ \vcenter{\hbox{\ladderlabelsleft{$\ell_1$}{$\ell_2$}{$\ell_3$}}} \Bigg] \, ,
\end{split}    
\end{align}
where we used $\ell_1+\ell_2+\ell_3=q$ in the last line. In the second term of the second line we used the fact that the horizontal flip symmetry of the diagram is unaffected when considering only the real part of the involved diagrams, despite the fact that the r.h.s of the cut in each diagram represents a complex conjugated amplitude. Similarly, the impulse numerator combines into a scalar in the sum over the two three-particle cuts. Here we do not need to take the real part because all these expressions are real by themselves, with tree-level expressions on both l.h.s and r.h.s of the cuts.
\begin{align}
\begin{split}
    & \frac{(\ell_1+\ell_2) \cdot q}{q^2} \vcenter{\hbox{\ladderlabelsNWSE{$\ell_1$}{$\ell_2$}{$\ell_3$}}} +
    \frac{\ell_1\cdot q}{q^2}
    \vcenter{\hbox{\ladderlabelsSWNE{$\ell_1$}{$\ell_2$}{$\ell_3$}}}
    \\ 
    = & \frac{(\ell_1+\ell_2) \cdot q}{q^2} \vcenter{\hbox{\ladderlabelsNWSE{$\ell_1$}{$\ell_2$}{$\ell_3$}}}
    +
    \frac{\ell_3\cdot q}{q^2}
    \vcenter{\hbox{\ladderlabelsNWSE{$\ell_1$}{$\ell_2$}{$\ell_3$}}}
    \\
    = &\frac{(\ell_1+\ell_2+\ell_3) \cdot q}{q^2} \
    \vcenter{\hbox{\ladderlabelsNWSE{$\ell_1$}{$\ell_2$}{$\ell_3$}}}
    =
    \vcenter{\hbox{\ladderlabelsSWNE{$\ell_1$}{$\ell_2$}{$\ell_3$}}}
\end{split}    
\end{align}
where we have again used the horizontal flip-symmetry of the cut diagram between the second term on the first and second line. Having canceled the nontrivial impulse numerators for both the three-particle cut and two-particle cut contributions, it can be seen that the imaginary parts cancel between all contributions to Eq.~\eqref{eq:classical_impulse_kernel} originating from the III diagram, using the three-term relation from Cutkosky rules given by Eq.~\eqref{eq:IIIcuttingRule1}.

%==========================================
\subsubsection{Longitudinal part}
%==========================================
%
Finally, we discuss briefly the simplification in the computation of the longitudinal part of the impulse. Recall that neither part of the longitudinal impulse kernel receives \emph{virtual} contributions, $I_{u_i\,, v} = 0$ and at this order the \emph{real} contributions are
\begin{align}
\label{eq:impulse_NNLO_kernel_long}
\mathcal{I}^{(2)}_{u_i} 	
    & = 
    \hspace{-.1cm}
	  - \imath \int \mathrm{d}\widetilde{\Phi}_{2} \,  \frac{\ell_1\cdot u_i}{q^2}
     \hspace{0cm}
     \left[\raisebox{-35pt}{\scalebox{0.8}{\kmocrealnnlolooptree}} \!\!+\!\! \raisebox{-35pt}{\scalebox{0.8}{\kmocrealnnlotreeloop}}\right] \nn \\
     &
     \hspace{3cm}
     -\imath \int \mathrm{d}\widetilde{\Phi}_{3}\,  \frac{\ell_1\cdot u_i}{q^2}
   \hspace{-.3cm}
   \raisebox{-35pt}{\scalebox{0.8}{\kmocrealnnlotreetree}}\,.
\end{align}
The subsequent Fourier transform does not flip reality properties so that we find the following is true for the longitudinal part of the impulse:
Three-particle cuts give real contributions to the impulse kernel. Every diagram and its horizontally flipped version contribute equally, due to the identity 
\begin{equation}
    [\ell - (q{-}\ell)] {\cdot} u_i = 2 \ell \cdot u_i\,.
\end{equation}
The same considerations are valid for the real part of two-particle-cut contributions, so every diagram and its horizontally flipped version contribute equally. For the imaginary parts of double-cut contributions, an extra sign difference causes cancellation between each diagram and its horizontally flipped version.

Therefore, to calculate the longitudinal impulse, we only need three-particle-cut contributions and the real part of two-particle-cut contributions. Since the double-cut integrand contains an overall factor of $i$, and only odd-in-$|q|$ master integrals are needed for the longitudinal impulse, we only need the box-triangle master integral Eq.~\eqref{eq:fRT10} with a cut on the box side as well as its horizontally flipped version.

%================================================
%
\section{Results}
\label{sec:result}
%
%================================================
%
In this section, we present the results of our computation of the two classical gravitational observables studied in this work: the impulse and the radiated momentum for the scattering of two black holes both in $\calN = 8$ supergravity and in general relativity through $\calO(G^3)$.

For the maximally supersymmetric case, the study of this scattering process was initiated in Ref.~\cite{Caron-Huot:2018ape}, and revisited in Ref.~\cite{Parra-Martinez:2020dzs} in the conservative sector. The loop integrands up to two-loops (or next-to-next-to-leading order) were constructed in Ref.~\cite{Parra-Martinez:2020dzs} by dimensionally reducing the known massless loop integrands \cite{Green:1982sw,Bern:1998ug}, and are reproduced here.

%================================================
%
\subsection{LO impulse}
\label{subsec:LO_impulse}
%
%================================================
%
Before discussing the impulse computation at loop level, for completeness, we give a lightening discussion of the leading order impulse, which is purely transverse (see Eq.~\eqref{eq:impulse_kernel_LO}) and basically the Fourier transform of the classical long-distance tree-level scattering amplitude
\begin{align}
\begin{split}
    \calM^{(0)}(p_1,p_2,p_3,p_4) & = c_t 
    \hspace{-.1cm}\vcenter{\hbox{\scalebox{.8}{\treet}}}
    \\
    \calM^{(0)}_{\calN{=}8}(p_1,p_2,p_3,p_4)& =
        (8\pi G) \frac{\left(s{-}|m_1{+}m_2 e^{i\phi}|^2\right)^2}{-t}
     \hspace{.5cm}
     {=}(8\pi G) \frac{4 m^2_1 m^2_2 (\sigma{-}\cos\phi)^2}{-t}
    \\
     \calM^{(0)}_{\rm GR}(p_1,p_2,p_3,p_4)&=    
        (8\pi G) \frac{(s{-}m^2_1{-}m^2_2)^2 {-} 2 m^2_1 m^2_2}{-t}
     {=}(8\pi G) \frac{2 m^2_1 m^2_2(2\sigma^2-1)}{-t}
\end{split}   
\end{align}
An analogous leading order analysis has already been performed in the original work of KMOC \cite{Kosower:2018adc}. The impulse kernel is given by
\begin{align}
    \calI^{(0)}_{p_1} = q^\mu\, \calM^{(0)}(p_1,p_2,p_3,p_4)
\end{align}
In maximal supergravity, we discuss the scattering of non-identical scalars and include the BPS angle $\phi$ \cite{Caron-Huot:2018ape,Parra-Martinez:2020dzs}. The factor $s-|m_1 +m_2 e^{i\phi}|^2 = 2 m_1m_2(\cosh \eta -\cos \phi)$ expresses the prefactor in terms of the relative rapidity $\eta = \text{arccosh}(\sigma)$ between the two massive states.

Upon Fourier transforming to impact parameter space, we find the leading order impulse in both theories
\begin{align}
\begin{split}
    \Delta p^{\mu,(0)}_{1,\calN{=}8} & = 
     \frac{G M^2\nu}{|b|} \frac{4(\sigma{-}\cos\phi)^2}{\sqrt{\sigma^2-1}} \frac{b^\mu}{|b|} 
     \\
     \Delta p^{\mu,(0)}_{1,{\rm GR}} & = 
     \frac{G M^2\nu}{|b|} \frac{2(2\sigma^2-1)}{\sqrt{\sigma^2-1}} \frac{b^\mu}{|b|} \,.
\end{split}     
\end{align}
Our result in general relativity agrees with the earlier expressions derived in Refs.~\cite{Portilla:1979xx,Portilla:1980uz}. 

%================================================
%
\subsection{NLO impulse}
\label{subsec:NLO_impulse}
%
%================================================
%
At next-to-leading order, the structure of the one-loop classical amplitude is as follows
\begin{align}
\begin{split}
    \calM^{(1)}(p_1,p_2,p_3,p_4) & =
    c_{\rm{II}} I_{\text{II}} 
    +
    c_{\rm{X}} I_{\text{X}}
    +
    c_{\text{tri},1} I_{\text{tri},1}
    +
    c_{\text{tri},2} I_{\text{tri},2}
    \\
    & 
    \hspace{-3cm}
   = c_{\rm{II}}  \hspace{-.3cm} \vcenter{\hbox{\scalebox{.8}{\bbox}}} 
   \hspace{-.1cm}
    {+}    
    c_{\rm{X}}  \hspace{-.3cm} \vcenter{\hbox{\scalebox{.8}{\bxbox}}} 
    \hspace{-.1cm}
    {+}
    c_{\text{tri},1}  \hspace{-.3cm} \vcenter{\hbox{\scalebox{.8}{\btriTT}}}
    \hspace{-.1cm}
    {+}
    c_{\text{tri},2}  \hspace{-.3cm} \vcenter{\hbox{\scalebox{.8}{\btriOF}}}
\end{split}    
\end{align}
where $c_i$ are the rational coefficients of the loop integrals.
As we will see, at the classical level, the structure of the amplitude reveals that there is no difference between the conservative and radiative impulse at this order. The reason is that in general $c_{\rm{II}} = c_{\rm{X}}$ so the box and crossed box integral appear in the combination $I_{\text{II}} + I_{\text{X}}$. Using the values of these soft integrals computed in Section.~\ref{sec:soft_integrals} and comparing to the values in the potential region given e.g. in Eqs.~(4.54) and (4.59) of Ref.~\cite{Parra-Martinez:2020dzs} we see that the difference between soft and potential region cancels in the sum. In addition the value of the triangle integral is identical in both regions.

%================================================
\subsubsection{$\mathcal N=8$ supergravity}
\label{subsec:sugraNLO}
%================================================
%
Let us begin our one-loop discussion with the appropriate loop integrand for the scattering of non-identical scalars \cite{Caron-Huot:2018ape,Parra-Martinez:2020dzs}
\begin{align}
    \label{eq:N8_integrand_1L}
    \calM^{(1)}(1_{\phi_1},2_{\phi_2},3_{\bar{\phi}_2},4_{\bar{\phi}_1}) 
    = - \imath (8\pi G)^2 \left(s-|m_1 +m_2 e^{i\phi}|^2\right)^4 (I_{\text{II}} + I_{\text{X}})
\end{align}
where we use the same notation as in the tree-level analysis. 
\begin{align}
\hspace{-.4cm}
\calM^{(1)}(p_1,p_2,p_3,p_4) {=} 
\overbrace{
{-}\imath\, (8\pi G)^2 16 m^4_1 m^4_2 (\sigma {-} \cos \phi)^4}^{\equiv c_{\rm{II}}} \!
    \left[ \!\!
        \vcenter{\hbox{\scalebox{.8}{\bbox}}} 
     \!{+}\!    
        \vcenter{\hbox{\scalebox{.8}{\bxbox}}} 
     \!
    \right]
\hspace{-.4cm}
\end{align}
Equipped with this integrand, we can soft expand both the $I_{\text{II}}$ and $I_{\text{X}}$ integrals and plug the resulting expressions into the impulse kernels Eqs.~(\ref{eq:nlo_transv_impulse}) and (\ref{eq:kerneluoneloop}) to obtain the transverse and longitudinal components. Let us discuss the transverse part first. From the \emph{real} term, we only get a contribution from the cut of the box integral, where we have expanded the impulse numerator in our preferred basis (\ref{eq:ell1_decomp}) and truncated at the classical order. 
On the other hand, summing the box and cross-box is equivalent to a symmetrization of the graviton loop momentum. This effectively cancels the real parts of the box and cross-box integrals and sets on-shell the two matter propagators in the box and yields a purely imaginary term
in the \emph{virtual} part of the impulse kernel
which exactly cancels the cut contribution so that we are left with 
\begin{align}
    \calI^{(1)}_{\perp} = \text{Re}\left[\calM^{(1)}(p_1,p_2,p_3,p_4)\right]=0 \,,
\end{align}
consistent with the general expectation of Eq.~\eqref{eq:kernelperponeloop}. The fact that the real parts of the one-loop amplitude is zero shows that there is no transverse deflection of the black hole orbits at one loop in maximal supergravity. Ultimately, this is due to the no-triangle property \cite{Bern:2005bb,BjerrumBohr:2006yw,Bern:2007xj,BjerrumBohr:2008ji} which was also linked to the non-precession of black hole orbits in Ref.~\cite{Caron-Huot:2018ape}. A similar cancellation of the imaginary parts in the impulse was shown in Ref.~\cite{Kosower:2018adc} for the electromagnetic impulse.

On the other hand, there is a contribution for the longitudinal impulse kernel. Naively, one could have guessed that the longitudinal impulse numerators $\ell_1\cdot u_i$, together with the cut conditions of the matter lines do not yield a contribution. However, in accord with the general expectation of Eq.~(\ref{eq:impulse_kernel_nlo_all}), due to subleading terms in the soft expansion there remains a contribution. Alternative to the general expectation from Eq.~(\ref{eq:impulse_kernel_nlo_all}), we can directly compute all diagrams in Eqs.~(\ref{eq:nlo_transv_impulse}) and (\ref{eq:kerneluoneloop}) using the explicit results for all soft integrals from the previous section. This is quite instructive and will be used at higher loops as well. Upon soft expansion and IBP reduction of the real longitudinal impulse contributions, we find
\begin{align}
\label{eq:long_impulse_kernel_nlo_n8}
    \calI_{u_1} & = - \imath \frac{(-q^2)\, c_{\rm{II}}}{4 m^2_1 m_2} 	\frac{\imath}{16\pi^2}\vcenter{\hbox{\linCutBox}}\,, 
    \quad
    \calI_{u_2}  = + \imath \frac{(-q^2)\, c_{\rm{II}}}{4 m_1 m^2_2} 	\frac{\imath}{16\pi^2}\vcenter{\hbox{\linCutBox}}\,,
\end{align}
where the double line, again, denotes linearized soft propagators. The factor $\imath/(16\pi^2)$ originates from the difference in normalization conventions between our soft integrals and standard Feynman diagrams. The value of the cut soft box is given in Eq.~\eqref{eq:Cut1LoopBox}.
Next, We perform the Fourier transform (\ref{eq:KMOC_DObs}) to impact parameter space, using the results collected in appendix \ref{app:FT_collection} to arrive at the final (purely longitudinal) result for the impulse
\begin{align}
    \Delta p^{\mu,(1)}_{1} = 
    \frac{G^2 M^4 \nu^2}{|b|^2}
    \frac{8\, (\sigma-\cos\phi)^4}
    {\left(\sigma^2-1\right)}\left[
        \frac{1}{m_1} \check{u}^\mu_1 - 
        \frac{1}{m_2} \check{u}^\mu_2
    \right]\,.
\end{align}
Note that we can replace the soft velocities in $\check{u}$ by the usual ones for free, since all superclassical pieces have canceled and we only need the leading in $q$ terms.

We would like to mention that at one loop order, the impulse is the same in the soft and potential region and receives no radiative contributions at the classical order. This is owed to the fact that soft bubble integrals (that vanish in the potential region) only contribute at higher orders in the $\hbar$ expansion and are therefore irrelevant. This also allows us to compare the one-loop impulse to the conservative result obtained from the scattering angle \cite{Parra-Martinez:2020dzs}, finding full agreement.  The extraction of the conservative impulse from the scattering angle is reviewed in appendix \ref{app:angle_impulse}, where it becomes clear that the (conservative) longitudinal impulse is completely dictated by lower-order information due to on-shell conditions.

%================================================
\subsubsection{General Relativity}
%================================================
%
Next, we consider general relativity. In principle, the same computational tools that led to all results in maximal supergravity are also applicable here. The only complication originates from a more complex loop integrand and more contributing soft master integrals. 

Just like in maximal supergravity, we begin our discussion of the one-loop impulse with the relevant integrand, which is known from e.g.~\cite{Green:1982sw,Caron-Huot:2018ape}. Notably, we find that at one loop in $D=4$ there is no distinction between the conservative result and the full soft region and it suffices to consider the following covariant diagrams
\begin{align}
\begin{split}
    \calM^{(1)}(p_1,p_2,p_3,p_4) & =
    c_{\rm{II}} I_{\text{II}} 
    +
    c_{\rm{X}} I_{\text{X}}
    +
    c_{\text{tri},1} I_{\text{tri},1}
    +
    c_{\text{tri},2} I_{\text{tri},2}
    \\
    & 
    \hspace{-3cm}
   = c_{\rm{II}}  \hspace{-.3cm} \vcenter{\hbox{\scalebox{.8}{\bbox}}} 
   \hspace{-.1cm}
    {+}    
    c_{\rm{X}}  \hspace{-.3cm} \vcenter{\hbox{\scalebox{.8}{\bxbox}}} 
    \hspace{-.1cm}
    {+}
    c_{\text{tri},1}  \hspace{-.3cm} \vcenter{\hbox{\scalebox{.8}{\btriTT}}}
    \hspace{-.1cm}
    {+}
    c_{\text{tri},2}  \hspace{-.3cm} \vcenter{\hbox{\scalebox{.8}{\btriOF}}}
\end{split}    
\end{align}
where $I_{\text{tri},i}$ is the triangle with matter propagator of mass $m_i$, and the coefficients are
\begin{equation}
\label{eq:gr_1loop_coefs}
  c_{\rm{II}} {=} c_{\rm{X}} {=} 
  -\imath(8\pi G)^2 4 m^4_1 m^4_2 (1 - 2 \sigma^2)^2\,, 
  \quad 
  c_{\text{tri},i}  {=}  -\imath(8\pi G)^2\, 3 m_1^2m_2^2\, m_i^2 (1-5\sigma^2)
\end{equation}
We could proceed with the general relation in Eq.~(\ref{eq:impulse_kernel_nlo_all}), however, here we explicitly check its validity by soft expanding Eqs.~(\ref{eq:nlo_transv_impulse}) and (\ref{eq:kerneluoneloop}). Subsequently, we insert the explicit results for all soft integrals from the previous section. 

Let us begin by discussing the transverse part of the impulse, that has in principle two contributions, one from the virtual amplitude and one from the real piece. In the impulse formula, only the box diagram contributes to the cut and by the same reasoning as in $\calN=8$, it just cancels the virtual boxes and we are left with only triangle contributions
\begin{equation}
  \calI^{(1)}_{\perp}  = 
    \frac{c_{\text{tri},1}}{m_1} \frac{\imath}{16\pi^2}
    \hspace{-.3cm} \vcenter{\hbox{\scalebox{.8}{\lintriTT}}} 
    +
    \frac{c_{\text{tri},2}}{m_2} \frac{\imath}{16\pi^2}
    \hspace{-.3cm} \vcenter{\hbox{\scalebox{.8}{\lintriOF}}}\,,
\end{equation}
where the linearized triangle is  given in Eq.~(\ref{eq:lin_triangle_result}) and we have again taken into account the standard normalization of Feynman integrals leading to the additional factor of $i/(16\pi^2)$. The transverse impulse kernel therefore reads
\begin{equation}
  \calI^{(1)}_{\perp} =
  (-q^2)^{-\epsilon }\frac{6\, \imath\pi^2 \,G^2 m^2_1 m^2_2 (m_1+m_2)(1-5\sigma^2) }{\sqrt{-q^2}}\,. 
\end{equation}
From the kernel we can easily calculate the transverse impulse via (\ref{eq:KMOC_DObs})
\begin{align}
  \Delta p^{\mu,(1)}_{1,\perp}  &= \frac{G^2 M^3\nu }{|b|^2} \frac{3\pi}{4}    \frac{(5\sigma^2-1)}{\sqrt{\sigma^2-1}} \frac{b^{\mu}}{|b|}
 \label{eq:impulseNLOGRus}
\end{align}
The remaining longitudinal impulse computation is essentially identical to the one in maximal supergravity, as only the box integral has the two-particle cut. Consequently, we simply have to replace the box coefficient $c_{\rm{II}}$ in Eq.~(\ref{eq:long_impulse_kernel_nlo_n8}) by its pure gravity counterpart (\ref{eq:gr_1loop_coefs})
\begin{align}
\label{eq:1loop_impulse_long_gr}
    \Delta p^{\mu,(1)}_{1,u} = \frac{G^2 M^4\nu^2}{|b|^2}
    \frac{2\, \, (1-2\sigma^2)^2 }
          {\, (\sigma^2-1)} 
    \left[
        \frac{1}{m_1} \check{u}^\mu_1 - 
        \frac{1}{m_2} \check{u}^\mu_2
    \right]\,.
\end{align}
so that the leading high-energy limit ($\sigma \gg 1$) of the longitudinal impulse coincides between GR and maximal supergravity. 

This concludes our one-loop calculation of the gravitational impulse within the KMOC formalism. We agree with previous results \cite{Westpfahl:1985tsl,Westpfahl_1987,Ledvinka:2008tk,Kalin:2020mvi} that can also be obtained from the scattering angle only (see appendix \ref{app:angle_impulse}), since conservative and soft region results are identical in $D=4$ up to classical order.

%================================================
\subsection{NNLO conservative impulse}
%================================================
%
Before deriving novel results in the full soft region, it turns out that we can test our two-loop setup by reproducing known results for the conservative dynamics from the KMOC point of view.  This can be done by performing the calculation in the potential region, defined in Eq.~\eqref{eq:powerCountingVelocity}. If we separate the impulse into conservative and radiative pieces
\begin{equation}
    \Delta p_1^\mu =  \Delta p_{1\,,\cons}^\mu + \Delta p_{1\,,{\rm rad}}^\mu\,,
\end{equation}
the potential region only captures the conservative contribution, $\Delta p_{1\,,\cons}^\mu $.

In the potential region, the gravitons are off-shell and therefore there cannot be real (on-shell) graviton emission. Hence, only the virtual integrals and the contribution from two-particle cuts to the \emph{real} impulse kernel in Eq.~\eqref{eq:impulse_NNLO_kernel} survive. Furthermore the "mushroom" integrals are identically zero in the potential region, and hence do not contribute to any conservative quantity. All the remaining integrals can be evaluated using the differential equations of sections \ref{subsec:ibp_de} and \ref{sec:soft_integrals}, although with modified boundary conditions appropriate for the potential region as originally described in Ref.~\cite{Parra-Martinez:2020dzs}. Reproducing the conservative impulse \cite{Kalin:2020mvi} and the scattering angle \cite{Bern:2019crd,Bern:2019nnu} constitutes a highly nontrivial check of the most complicated parts of our assembly.

%================================================
\subsubsection{$\mathcal N=8$ supergravity}
%================================================
%
The two-loop integrand of maximally supersymmetric gravity is obtained by dimensional reduction of the massless integrand \cite{Bern:1998ug} with the following result \cite{Parra-Martinez:2020dzs}
\begin{align}
\label{eq:sugra_2L_integrand}
\hspace{-.5cm}
\begin{split}
    \calM^{(2)} & = -(8\pi G)^3 16 m^4_1 m^4_2 (\sigma-\cos\phi)^4 \times 
    \\
    & \hspace{-1cm}
    \Bigg\{
     4 m^2_1 m^2_2 (\sigma{-}\cos\phi)^2 \!
    \left[
     \vcenter{\hbox{\scalebox{.7}{\intIII}}}
     \hspace{-.2cm}
     {+}
     \hspace{-.2cm}
     \vcenter{\hbox{\scalebox{.7}{\intIX}}}
     \hspace{-.2cm}
     {+}
     \hspace{-.2cm}
     \vcenter{\hbox{\scalebox{.7}{\intXI}}}
    \right]%\right.
    \\
    & \hspace{-1cm}
    %\left.
    + (-q^2)^2
    \left[
     \vcenter{\hbox{\scalebox{.7}{\intH}}}
     \hspace{-.2cm}
     {+}
     \hspace{-.2cm}
     \vcenter{\hbox{\scalebox{.7}{\intYInp}}}
     {+}
     \hspace{-.2cm}
     \vcenter{\hbox{\scalebox{.7}{\intYIFlnp}}}
     {+}
     \hspace{-.2cm}
     \vcenter{\hbox{\scalebox{.7}{\xMushroomTop}}}
     {+}
     \hspace{-.2cm}
     \vcenter{\hbox{\scalebox{.7}{\xMushroomBottom}}}
    \right]%\right. 
    \\[-5pt]
    & \hspace {6cm}   + (2 \leftrightarrow 3) \,, \hspace{4cm} \Bigg\}
\end{split}
\hspace{-.3cm}
\end{align}
where $(2\leftrightarrow 3)$ instructs to add terms with $p_2$ and $p_3$ interchanged. This integrand is the complete quantum integrand for the supergravity amplitude, and hence is valid both for conservative dynamics, as well as in the full soft region discussed below.  Note that the final four scalar diagrams are quantum suppressed in $\calN =8$ because they are accompanied by the $q^4$ prefactor and only the H-diagram survives.

Upon classically expanding the integrand (in $|q|$ or $\hbar$), subsequent IBP reduction to a set of soft master integrals, and inserting the appropriate values for both the virtual and cut pieces evaluated in the \emph{potential region}, we find the impulse kernels
\begin{align}
    \calI^{(2)}_{\perp,\text{cons}} & = -\frac{(-q^2)^{-2\epsilon}}{\epsilon} 
    \frac{16\, \pi\,  G^3\, m^2_1 m^2_2\, (\sigma{-}\cos\phi)^4} {\sqrt{\sigma^2-1}}
 \Bigg[
    \frac{(\sigma{-}\cos\phi)^2\, s}{\left(\sigma^2-1\right)^{3/2}}
    +4 m_1 m_2\, \text{arcsinh}\sqrt{\frac{\sigma-1}{2}}\,
 \Bigg]  
 \nn \\[-5pt]
    \calI^{(2)}_{u_1,\text{cons}}  & = \calI^{(2)}_{u_2,\text{cons}}=0\,.
\end{align}
Note that three particle cuts are zero in the potential region, as the internal graviton lines are never on-shell.  As expected, the superclassical terms in the transverse impulse kernel cancel between the virtual diagrams and two-particle cuts. Furthermore, the longitudinal impulse kernel does not receive a virtual contribution and vanishes due to a cancellation between the two-particle cuts of double boxes and crossed double boxes. Fourier transforming to impact parameter space yields
\begin{align}
\label{eq:nnlo_cons_impulse_neq8}
\begin{split}
\hspace{-.5cm}
 \Delta p^{\mu,(2)}_{1,\text{cons}} = 
 -\frac{ G^3M^4\nu}{|b|^3 } \frac{16\,\, (\sigma{-}\cos\phi)^4} 
 {\left(\sigma^2-1\right)} \frac{b^\mu}{|b|}
 \Bigg[
 \frac{(\sigma{-}\cos\phi)^2\, h^2(\sigma,\nu)}{\left(\sigma^2-1\right)^{3/2}}
  +4 \nu\, \text{arcsinh}\sqrt{\frac{\sigma-1}{2}}\,
 \Bigg]\,.
 \hspace{-.4cm}
\end{split} 
\end{align}

%================================================
\subsubsection{General Relativity}
%================================================
%
We can repeat a similar conservative two-loop analysis for pure gravity. The integrand is more complicated than (\ref{eq:sugra_2L_integrand}) and contains additional diagrams but has already been constructed previously in Refs.~\cite{Bern:2019nnu,Bern:2019crd}, and reproduced by the present authors using generalized unitarity. It only involves the cubic graphs in the first row of Figure~\ref{fig:cubic_graphs}. Taking said integrand, expanding it in the classical limit, reducing to a minimal set of master integrals and inserting the appropriate \emph{potential region} values of the master integrals \cite{Parra-Martinez:2020dzs} allows us to obtain the impulse kernels
\begin{align}
  \mathcal{I}^{(2)}_{\perp,\text{cons}} & =  (-q^2)^{-2\epsilon}
  \frac{2\pi G^3 m^2_1 m^2_2}{\epsilon} 
  \Bigg[ 
   s \left(16 \sigma^2 -\frac {1} {\left(\sigma^2-1\right)^2}\right)
    \\
   &
   -\frac{4}{3}m_1m_2 \,  
   \sigma \left(14 \sigma^2+25\right) 
   +8m_1m_2  \left(-4 \sigma ^4+12 \sigma ^2+3\right)  \frac{\text{arcsinh}\sqrt{\frac{\sigma-1}{2}}}
  {\sqrt{\sigma^2-1}} 
   \Bigg] \,,
   \nn\\
   \mathcal{I}^{(2)}_{u_1,\text{cons}} & = \imath(-q^2)^{\frac12 -2\epsilon}   
   \frac{12 \pi^2 \,G^3\,m_1^2 m_2^3 (m_1{+}m_2) \left(2\sigma^2-1\right)\left(5\sigma^2-1\right)}
   {\sqrt{\sigma^2-1}}\,,
   \\
   \mathcal{I}^{(2)}_{u_2,\text{cons}} & = -\imath(-q^2)^{\frac12 -2\epsilon}   
   \frac{12 \pi^2 \,G^3\,m_1^3 m_2^2 (m_1{+}m_2) \left(2\sigma^2-1\right)\left(5\sigma^2-1\right)}
   {\sqrt{\sigma^2-1}}\,.
\end{align}
Fourier transforming to impact parameter space, we obtain 
\begin{align}
\label{eq:impulseG3transCons}
\Delta p^{\mu,(2)}_{1, \perp, \text{cons}}  &=
\frac{G^3 M^4\nu}{|b|^3} \frac{2}{\sqrt{\sigma^2{-}1}} \frac{b^\mu}{|b|}
  \Bigg[
  h^2(\sigma,\nu) \left(16 \sigma^2-\frac{1}{\left(\sigma^2-1\right)^2}\right) 
  \\
  & \hspace{2cm} -\frac{4}{3} \nu \, \sigma  
  \left(14 \sigma^2+25\right)
  -8 \nu  \left(4 \sigma^4-12 \sigma^2-3\right) 
  \frac{\text{arcsinh}\sqrt{\frac{\sigma-1}{2}}}
  {\sqrt{\sigma^2-1}}
   \Bigg] \nn
\end{align}
The conservative impulse has a logarithmic divergence at high energies, corresponding to that in the scattering angle of Refs.~\cite{Bern:2019nnu,Bern:2019crd}. By comparing to the maximal supergravity result we find that the coefficient is universal in agreement with Ref.~\cite{Parra-Martinez:2020dzs}. We will come back to this point when considering the full impulse including radiation reaction.

The longitudinal impulse is
\begin{align}
\Delta p^{\mu,(2)}_{1, u, \text{cons}}  &=
    \frac{G^3 M^5 \nu^2}{|b|^3}
    \frac{3 \pi \left(2 \sigma^2{-}1\right) \left(5 \sigma^2{-}1\right)}{2\, (\sigma^2-1)} 
    \left[
    \frac{1}{m_1}\, \check{u}^{\mu}_1
    -
    \frac{1}{m_2}\, \check{u}^{\mu}_2
    \right]\,.
    \label{eq:impulseG3longCons}
\end{align}
We note that the longitudinal part of the conservative impulse does not contain new information. Its purpose at a given order in $G$ is to ensure that the energy transfer between the two particles is such that that they remain on-shell after transverse deflection at previous orders. In other words, the longitudinal impulse is the solution to the equation
\begin{equation}
    0 = (p_1+\Delta p_1)^2-m_1^2 = 
    p_1\cdot\Delta^{(1)} p_1 + \left(\Delta^{(0)}p_1\right)^2
    \label{eq:on-shell}
\end{equation}
which must be satisfied at each order in $G$. This was used in \cite{Kalin:2020fhe} to obtain the $\calO(G^3)$ longitudinal impulse in General relativity. In contrast our result in Eq.~\eqref{eq:impulseG3longCons} follows from direct calculation and Eq.~\eqref{eq:on-shell} serves as a check on our methodology.
%
%================================================
\subsection{NNLO radiative impulse }
\label{subsec:nnlo_rad_impulse_result}
%================================================

%================================================
\subsubsection{$\mathcal N=8$ supergravity}
%================================================
%
For the radiative impulse in maximal supergravity, we start from the full integrand in Eq.~(\ref{eq:sugra_2L_integrand}) obtained via dimensional reduction. Upon soft expansion of the integrand, subsequent IBP reduction to a set of soft master integrals, and inserting the appropriate values for both the virtual and cut pieces, we find the following impulse kernels
\begin{align}
    \hspace{-.5cm}
    \calI^{(2)}_\perp & =
    4 \pi\frac{(-q^2)^{-2\epsilon}}{\epsilon}  G^3 m^3_1 m^3_2 
    \Bigg[
            \left( 
               f_1(\sigma,\phi)
               \tblue{{-}\frac{4s (\sigma{-}\cos\phi)^6}{m_1 m_2(\sigma^2-1)^2}}
            \right)
    \nn \\
    &\hspace{1cm}
    +  \left(
        \sigma f_3(\sigma,\phi)
        \tblue{- 16(\sigma{-}\cos\phi)^4}
        \right) \frac{\text{arcsinh}\sqrt{\frac{\sigma-1}{2}}}{\sqrt{\sigma^2-1}}
    \Bigg]\,,
    \hspace{-.2cm}    
    \\
    \calI^{(2)}_{u_1} & = 0 \nn \\
    \calI^{(2)}_{u_2} & = \imath(-q^2)^{\frac12-2\epsilon} 8 \pi^2 G^3 m_1^3 m_2^3 \sqrt{\sigma^2-1}
    \Bigg[
        f_1(\sigma,\phi)
        +f_2(\sigma,\phi)\log \left(\frac{\sigma+1}{2}\right) 
        \nn\\
        & \hspace{5cm}
        +f_3(\sigma,\phi) \frac{\sigma\, \text{arcsinh}\sqrt{\frac{\sigma-1}{2}}}{\sqrt{\sigma^2-1}}
    \Bigg]\,,
\end{align}
where we highlight the terms that were already present in the conservative impulse in Eq.~\eqref{eq:nnlo_cons_impulse_neq8} in blue and we use $s{=}m^2_1{+}m^2_2{+}2 m_1 m_2\, \sigma$ in some terms for compactness. We also define the convenient coefficient functions that depend on $\sigma$ and the BPS angle $\phi$
\begin{align}
\label{eq:fsSUGRA}
\begin{split}
      f_1(\sigma,\phi) & = \frac{8 (\sigma-\cos \phi)^6}
                                {\left(\sigma^2-1\right)^{3/2}}\,,
      \quad 
      f_2(\sigma,\phi) =-\frac{8 (\sigma-\cos\phi)^4}{\sqrt{\sigma^2-1}} \,, 
      \\
      & 
      f_3(\sigma,\phi) = \frac{16 (\sigma-\cos\phi)^5 
                                    (\sigma^2+\sigma\cos\phi-2)}
                                {\sigma \left(\sigma^2-1\right)^{3/2}}\,. 
\end{split}
\end{align}
We would like to point out that all superclassical terms have cancelled in the impulse kernel directly. For the leading superclassical terms in the transverse impulse, this is due to the simple argument given in section \ref{sec:KMOC_discussion_expansion}. The fact that all other superclassical terms likewise cancel, serves as a further cross-check of our setup. For the longitudinal impulse, the cancellation is very simple and occurs when we add up the three-particle cuts of integrals III (planar double-box) and IX (nonplanar double-box) and combine them with the three-particle cut of the u-channel IX. 

Performing the Fourier transform to impact parameter space, and subtracting the conservative contribution in Eq.~\eqref{eq:nnlo_cons_impulse_neq8} the impulse on particle 1 is
\begin{align}
\label{eq:nnlo_impulse_rad_neq8}
\begin{split}
  \Delta p^{\mu,(2)}_{1,{\rm rad}} & = 
  \frac{G^3 M^4\nu^2}{|b|^3}
    \Bigg\{\frac{4}{\sqrt{\sigma^2-1}}\frac{b^\mu}{|b|} 
    \Bigg[
                f_1(\sigma,\phi)
    + f_3(\sigma,\phi)
         \frac{\sigma\,\text{arcsinh}\sqrt{\frac{\sigma-1}{2}}}{\sqrt{\sigma^2-1}}
    \Bigg] 
    \\
    & + \pi\, \check{u}^\mu_2
    \Bigg[
        f_1(\sigma,\phi)
        + f_2(\sigma,\phi)\log \left(\frac{\sigma+1}{2}\right)
        + f_3(\sigma,\phi) \frac{\sigma\,\text{arcsinh}\sqrt{\frac{\sigma-1}{2}}}{\sqrt{\sigma^2-1}}
    \Bigg]\Bigg\}\,.
\end{split}
\end{align}
Note that due to the absence of transverse deflection at ${\cal}O(G^2)$, the full longitudinal deflection in $\Delta p^{\mu,(2)}_1$ is purely radiative and along the $\check{u}^\mu_2$ direction. At this point we note that the same coefficient functions $f_1, f_3$ encode both the transverse and longitudinal components of the radiative impulse, except for the novel radiative term associated with $\log \frac{\sigma+1}{2}$ which arises from the H diagram. This will be compared to general relativity case in the next subsection.

The relation between the radiative contribution to the angle and the radiative impulse at this order is given in Eq.~\eqref{eq:chiradimpulse}, so we recognize
\begin{equation}
    \chi^{(2)}_{\rm rad} =  \frac{G^3 M^3\nu}{|b|^3}
    \frac{4h(\sigma,\nu)}{\sigma^2-1} 
    \Bigg[
       f_1(\sigma,\phi)
    + f_3(\sigma,\phi)
     \frac{\sigma\,\text{arcsinh}\sqrt{\frac{\sigma-1}{2}}}{\sqrt{\sigma^2-1}}
    \Bigg] 
\end{equation}
This can be compared to the result for the result with $\phi = \pi/2$ calculated using eikonal methods in  \cite{DiVecchia:2020ymx,DiVecchia:2021ndb}, finding full agreement.

%================================================
\subsubsection{General Relativity}
%================================================
%
The computation of the impulse in general relativity is rather involved, starting with the more complicated form of the integrand, whose construction was outlined in section \ref{subsec:soft_integrands}. We employ the same soft expansion, IBP and differential equation technology described above where all virtual and cut master integrals are evaluated in the soft region. Assembling all the pieces, we find the GR impulse kernels
\begin{align}
\label{eq:impulse_kernel_perp_gr_result}
    \mathcal{I}^{(2)}_{\perp} & =  4\pi\frac{(-q^2)^{-2\epsilon}}{\epsilon} 
   G^3 m^2_1 m^2_2
  \Bigg[ 
   \tblue{s \left(8 \sigma^2 -\frac {1} {2 \left(\sigma ^2-1\right)^2}\right)}
    \\
   &
     +m_1m_2\left(
   f^{\rm LS}_1(\sigma)  
   \tblue{-\frac{2}{3} \,  
   \sigma (14 \sigma^2+25)}  \right)
   \nn 
   \\
   &+m_1m_2 \left(
   \sigma f^{\rm LS}_3(\sigma)
   \tblue{- 4  \left(4 \sigma ^4-12 \sigma ^2-3\right) } \right) \frac{\text{arcsinh}\sqrt{\frac{\sigma-1}{2}}}
  {\sqrt{\sigma^2-1}} 
   \Bigg] \,,
   \nn
   \\
   \mathcal{I}^{(2)}_{u_1} & = 8\pi^2\imath(-q^2)^{\frac12 -2\epsilon}  G^3\,m_1^3 m_2^3   \left[ 
   \tblue{\frac{(m_1{+}m_2)}{m_1}
   \frac{3 (2\sigma^2-1)(5\sigma^2-1)}
   {2\sqrt{\sigma^2-1}} }\right]\,,
   \\
   \mathcal{I}^{(2)}_{u_2} & = 8\pi^2\imath(-q^2)^{\frac12 -2\epsilon} G^3\,m_1^3 m_2^3  
   \Bigg[
   \tblue{-
   \frac{(m_1{+}m_2)}{m_2} 
   \frac{3\left(2\sigma^2-1\right)\left(5\sigma^2-1\right)}
   {2\sqrt{\sigma^2-1}}}
   \\
   &
   +  \sqrt{\sigma^2-1} \left( 
     f_1(\sigma)
   + f_2(\sigma)  \log \left(\frac{\sigma +1}{2}\right)
   + f_3(\sigma)
    \frac{\sigma\,\text{arcsinh}\sqrt{\frac{\sigma-1}{2}}}{\sqrt{\sigma^2-1}}
    \right)
   \Bigg] \nn
   \,,
\end{align}
where the coefficient functions are given by
\begin{align}
\label{eq:fsGR}
\begin{split}
& f^{\rm LS}_1(\sigma) = -\frac{ (2 \sigma ^2-1)^2 (5 \sigma ^2-8)}{3 (\sigma^2-1)^{3/2}}\,,
\\
& f^{\rm LS}_3(\sigma) = \frac{2 (2 \sigma ^2-1)^2 (2 \sigma ^2-3)}{(\sigma
   ^2-1)^{3/2}} \,,
\\   
& f_1(\sigma) =\frac{210 \sigma^6-552 \sigma ^5+339 \sigma ^4-912 \sigma^3+3148 \sigma^2-3336 \sigma +1151}
{48 \left(\sigma^2-1\right)^{3/2}}\,, 
 \\
&f_2(\sigma)=-\frac{35 \sigma ^4+60 \sigma ^3-150 \sigma ^2+76 \sigma -5}{8 \sqrt{\sigma ^2-1}}\,,
\\
&f_3(\sigma) =\frac{\left(2 \sigma ^2-3\right) \left(35 \sigma ^4 -30 \sigma ^2+11\right)}{8 (\sigma^2-1)^{3/2}}\,.
\end{split}
\end{align}
After taking the Fourier transform in Eq.~\eqref{eq:KMOC_DObs} and subtracting the conservative result in Eqs.~\eqref{eq:impulseG3transCons} and \eqref{eq:impulseG3longCons} we obtain the following result for the radiative impulse in general relativity
\begin{align}
\begin{split}
  \Delta p^{\mu,(2)}_{1,{\rm rad}} & = 
  \frac{G^3 M^4\nu^2}{|b|^3}
    \Bigg\{\frac{4}{\sqrt{\sigma^2-1}}\frac{b^\mu}{|b|} 
    \Bigg[
                f_1^{\rm LS}(\sigma)
              % \tblue{ {-}\frac{s}{2\,m_1m_2(\sigma^2-1)^2}}
    + f_3^{\rm LS}(\sigma)
        %\tblue{- 2}
         \frac{\sigma\, \text{arcsinh}\sqrt{\frac{\sigma-1}{2}}}{\sqrt{\sigma^2-1}}
    \Bigg] 
    \\
    & + \pi\, \check{u}^\mu_2
    \Bigg[
        f_1(\sigma)
        + f_2(\sigma)\log \left(\frac{\sigma+1}{2}\right)
        + f_3(\sigma) \frac{\sigma\, \text{arcsinh}\sqrt{\frac{\sigma-1}{2}}}{\sqrt{\sigma^2-1}}
    \Bigg]\Bigg\}\,.
\end{split}
\end{align}
The structure is very similar to the result in maximal supergravity. However, unlike in the case of supergravity, where the longitudinal and transverse impulse was controlled by the same algebraic functions $f_i$, in general relativity, the structure is different. Interestingly, the radiative transverse impulse is captured by algebraic functions $f^{\rm LS}_i$ that purely encode leading soft (LS) dynamics of gravitons \cite{DiVecchia:2021ndb}. In hindsight, the fact that in $\calN = 8$ supergravity the leading soft theorem also controls most of the longitudinal impulse can be understood as a consequence of the no-triangle property of theories with maximal supersymmetry \cite{Bern:2005bb,BjerrumBohr:2006yw,Bern:2007xj,BjerrumBohr:2008ji}. In the absence of triangles the Weinberg soft factor exactly captures the contributions from all diagrams where the radiated gravitons are emitted from a matter leg. Thus the only new contribution can arise from the H diagram, which indeed produces the term with $\log \frac{\sigma+1}{2}$, as pointed out above.

Using Eq.~\eqref{eq:chiradimpulse} we obtain the radiative contribution to the scattering angle in general relativity
\begin{equation}
    \chi^{(2)}_{\rm rad} =  \frac{G^3 M^3\nu}{|b|^3}
    \frac{4h(\sigma,\nu)}{\sigma^2-1} 
    \Bigg[
                f_1^{\rm LS}(\sigma)
    + f_3^{\rm LS}(\sigma)
         \frac{\sigma\,\text{arcsinh}\sqrt{\frac{\sigma-1}{2}}}{\sqrt{\sigma^2-1}}
    \Bigg] \,.
\end{equation}
This result agrees with the computation by Damour in Ref.~\cite{Damour:2020tta} via a linear response formula derived in Ref.~\cite{Bini:2012ji}. This was later reproduced in \cite{DiVecchia:2021ndb} using a beautiful relation between the real part of the eikonal and the infrared divergence in its imaginary part. Such relation was proven for $\calN = 8$ supergravity by explicit computation and conjectured more generally.

%================================================
\subsection{LO radiated momentum}
%================================================
%
Besides the gravitational impulse, considered in the previous subsections, we are also able to compute the radiated momentum, both in general relativity and maximal supergravity. This observable starts at $\calO(G^3)$ and is related to the energy loss, which has been the subject of our short letter \cite{Herrmann:2021lqe} and we present them here just for completeness. In the KMOC setup, the radiated momentum can be obtained either directly by considering the expression in subsection \ref{subsec:radiated_momentum}, or from momentum conservation and the impulse on particles 1 and 2,
\begin{align}
    0=\Delta R^\mu + \Delta p^\mu_1 + \Delta p^\mu_2
\end{align}
For us, it was originally easier to obtain the radiated momentum directly, as it only involves the three-particle cut of two-loop diagrams at $\calO(G^3)$ and therefore requires fewer terms in the full soft integrand. We found the following result in $D=4$
\begin{equation}
\label{eq:rad_mom_gen}
    \Delta R^\mu = \frac{G^3 m_1^2m_2^2}{|b|^3} \frac{u_1^\mu + u_2^\mu}{\sigma+1}   \mathcal{E}(\sigma) + {\cal O}(G^4)\,,
\end{equation}
where we define 
\begin{align}
\label{eq:rad_energy}
    \frac{\mathcal{E}(\sigma)}{\pi} =
    f_1(\sigma) 
    {+} f_2(\sigma) \, \log \left(\frac{\sigma{+}1}{2}\right) 
    {+} f_3(\sigma)\,\frac{\sigma\,  \text{arcsinh}\sqrt{\frac{\sigma{-}1}{2}}}{\sqrt{\sigma^2{-}1}}\,,
\end{align}
with the theory dependent coefficient functions $f_i(\sigma)$. This analytic structure is directly inherited from the longitudinal part of the radiative impulse, computed in section \ref{subsec:nnlo_rad_impulse_result}, by momentum conservation.  As was pointed out in Refs.~\cite{Kovacs:1978eu,Bini:2020hmy}, the homogeneous mass dependence in Eq.~(\ref{eq:rad_mom_gen}) signals that the result is entirely specified by the probe limit $m_1\ll m_2$. Note that the radiated momentum in Eq.~(\ref{eq:rad_mom_gen}) is purely longitudinal and yields the energy radiated as gravitational waves. In the center-of-mass (c.m.) frame of the hyperbolic motion, we find
\begin{align}
\label{eq:Delta_E}
\begin{split}
    \Delta E^{\rm hyp} &{=} \frac{(p_1{+}p_2)\cdot \Delta R }{|p_1{+}p_2|} {=} \frac{G^3 M^4\nu^2}{|b|^3\, h(\nu,\sigma) } \ \mathcal{E}(\sigma) {+} \mathcal{O}(G^4) \,.
\end{split}    
\end{align}
From the scattering result of Eq.~(\ref{eq:Delta_E}), we obtain the energy loss for elliptic (bound) orbits via analytic continuation \cite{Kalin:2019rwq,Kalin:2019inp,Bini:2020hmy} of the result
\begin{align}
\begin{split}
\label{eq:rad_energy_elliptic}
    \Delta E^{\rm ell}(\sigma, J) &= \Delta E^{\rm hyp}(\sigma,J) - \Delta E^{\rm hyp}(\sigma,-J)\\
\end{split}\,,
\end{align}
from the physical region $\sigma>1$ to the Euclidean region $\sigma<1$, where $\sigma$ is related to the dimensionless binding energy $\overline{E}{=}\frac{h(\nu,\sigma){-}1}{\nu}{<}0$ \cite{Bini:2020hmy}
\begin{equation}
\label{eq:ell_energy_loss_more_explicit}
    \Delta E^{\rm ell}(\sigma, J) = \frac{G^3 M^7 \nu^5 (1-\sigma^2)^{\frac{3}{2}}}{J^3\, h(\nu,\sigma)^4} \, \widetilde{\mathcal{E}}^{\rm ell}(\sigma) + \mathcal{O}(G^4)\,.
\end{equation}
where $\widetilde{\mathcal{E}}$ takes the same general form as Eq.~\eqref{eq:rad_energy} \cite{Herrmann:2021lqe} and has the expected simplified $\nu$ dependence previously observed in Ref.~\cite{Bini:2020hmy}. From our perspective, this is simply inherited from the analytic continuation of the hyperbolic result.

As stated previously, the explicit result for radiated momentum in Eq.~\eqref{eq:rad_mom_gen} has been obtained in Ref.~\cite{Herrmann:2021lqe}, where the theory specific coefficient functions in general relativity are the same that appear in the impulse computation, c.f.~Eq.~\eqref{eq:fsGR}. The energy loss for a black hole scattering event can be expanded in small velocity $v{=}\frac{\sqrt{\sigma^2{-}1}}{\sigma}$ and compared to known Post-Newtonian (PN) data, finding agreement with the result known up to 2PN \cite{Kovacs:1978eu, Blanchet:1989cu,Bini:2020hmy}. We furthermore compared the small velocity expansion of the energy loss in elliptic orbits in Eq.~(\ref{eq:ell_energy_loss_more_explicit}) to the 3PN expressions for the instantaneous energy flux integrated over one orbit from Refs.~\cite{Peters:1963ux,Peters:1964zz,Wagoner:1976am,Blanchet:1989cu,Junker:1992kle,Gopakumar:1997ng,Gopakumar:2001dy,Arun:2007sg,Blanchet:2013haa} in the large eccentricity limit, i.e.~to leading order in large $J$, again finding perfect agreement with the PN data where our results overlap. After Ref.~\cite{Herrmann:2021lqe} appeared, Bini, Damour, and Geralico, informed us privately of their computation in the small velocity limit up to $\calO(v^{15})$, also agreeing with our result. Additionally, Ref.~\cite{Mougiakakos:2021ckm} verified the low-velocity limit up to $\calO(v^7)$ from a world-line EFT perspective, and Ref.~\cite{DiVecchia:2021bdo} reproduced our result with full velocity dependence using methods similar to Ref.~\cite{Herrmann:2021lqe}.

Finally, the radiated energy also appears in the tail term \cite{Bini:2017wfr,Blanchet:2019rjs,Bini:2020hmy} of the $\calO(G^4)$ radial action, which has been recently obtained by Ref.~\cite{Bern:2021dqo} by an independent computation. Comparing Eq.~(\ref{eq:Delta_E}) to that expression, we find full agreement.

%================================================
%
\subsection{Comments on universality and relation to eikonal phase}
%
%================================================
%
It is interesting to study the high energy limit of the gravitational observables considered in this work. Famously, the leading order observables are universal in this limit \cite{tHooft:1987vrq}. Similarly, the gravitational deflection angle has been observed to have universal properties at $\calO(G^3)$ \cite{Amati:1990xe,Bern:2020gjj,Parra-Martinez:2020dzs,DiVecchia:2020ymx,Damour:2020tta}, so it is natural to ask whether or not the same is true for the gravitational impulse.

Recombining the radiative impulse in Eq.~\eqref{eq:nnlo_impulse_rad_neq8} with the conservative impulse in Eq.~\eqref{eq:nnlo_cons_impulse_neq8}, we can study the high-energy ($\sigma \gg 1$) limit of the full result. The leading high-energy pieces cancel between radiative and conservative contributions, consistent with previous observations \cite{DiVecchia:2020ymx}.
\begin{align}
\label{eq:nnlo_impulse_neq8_high_energy}
   \Delta p^{\mu,(2)}_{1} {=} 
   \frac{G^3 M^4 \nu}{|b|^3} \left(
   \Big[16 (2\nu{-}1) \sigma +\calO(\sigma^0)\Big]\frac{b^\mu}{|b|} 
   +\Big[ 8\pi\, \nu (1{+}2\log 2)\sigma^3+\calO(\sigma^{2}) \Big]\check{u}^\mu_2
   \right)\,,
\end{align}
where it is interesting to note that the leading nonzero terms are independent of the BPS angle $\phi$.
On the other hand, taking the limit of our general relativity result we find
\begin{align}
\label{eq:nnlo_impulse_gr_high_energy}
\hspace{-.5cm}
   \Delta p^{\mu,(2)}_{1} {=} 
   \frac{G^3 M^4 \nu}{|b|^3} \left(
   \Big[-32 (2\nu{-}1) \sigma {+}\calO(\sigma^0)\Big]\frac{b^\mu}{|b|} 
   +\Big[\frac{35}{8}\pi\, \nu (1{+}2\log 2) \sigma^3{+}\calO(\sigma^{2}) \Big]\check{u}^\mu_2
   \right)\,.
\hspace{-.4cm}
\end{align}
The logarithmic high-energy divergence in the conservative impulse cancels, as expected \cite{DiVecchia:2020ymx, Damour:2020tta}, once radiation reaction effects are included. Interestingly, by comparing \eqref{eq:nnlo_impulse_gr_high_energy} to the maximal supergravity result \eqref{eq:nnlo_impulse_neq8_high_energy}, we find that the universality of the scattering angle (including radiation reaction) described in Ref.~\cite{DiVecchia:2020ymx} does not hold for the full impulse (both in the transverse and longitudinal directions), due to a cancellation between the leading conservative and radiative contributions. However, computing the angle from the impulse requires taking into account products of lower-order terms which restore the previously observed universality of the high-energy limit of the scattering angle \cite{DiVecchia:2020ymx, Damour:2020tta}
\begin{align}
   \chi^{(2)}_{\text{HE,GR}} = \chi^{(2)}_{\text{HE,}\calN{=}8} 
   =\frac{32 G^3 m_1^3 m_2^3 \sigma^3}{3 J^3}
   = \frac{\left[\chi^{(0)}_{\text{HE,GR}}\right]^3}{3!}\,.
\end{align}
Here, we have written the result in terms of the angular momentum $J=|b| |\bm{p}| = |b| \frac{M \nu\, \sqrt{\sigma^2-1}}{h(\sigma,\nu)}$ and note that subleading terms in the large $\sigma$ expansion differ between both theories.

It is also interesting to consider the relation between the transverse impulse, $\Delta p_{1,\perp}$ obtained in our calculation and the corresponding quantity derived from the eikonal approach in Refs.~\cite{DiVecchia:2021ndb,DiVecchia:2021bdo}.
Comparing their conjecture for the real part of the two-loop eikonal phase to the transverse impulse (best seen from Eq.~(\ref{eq:impulse_kernel_perp_gr_result})), all but one velocity dependent factors agree (up to some overall scaling due to the distinction between the two quantities). The only difference is related to the $s$-dependent term in the first line of \eqref{eq:impulse_kernel_perp_gr_result} and is due to the difference between the asymptotic impact parameter $b$ and the eikonal one, $b_e$ (see Eq.~\eqref{eq:b_eikonal}), which yields a correction proportional to $\left[\chi^{(0)}\right]^3$. Indeed, rewriting our result for the transverse impulse in terms of the eikonal impact parameter we find in $\calN=8$ supergravity 
\begin{align}
 \Delta p_{1\,,\perp}^{\mu\,{(2)}} (b_e) & =  
   \frac{G^3 M^4\nu}{|b_e|^3} \frac{4}{\sqrt{\sigma^2-1}}\frac{b_e^\mu}{|b_e|}
    \Bigg[
    f_1(\sigma,\phi)
    \nn \\
    &\hspace{1cm}
    +  \left(
        \sigma f_3(\sigma,\phi)
        - 16(\sigma{-}\cos\phi)^4
        \right) \frac{\text{arcsinh}\sqrt{\frac{\sigma-1}{2}}}{\sqrt{\sigma^2-1}}
    \Bigg]\,,
\end{align}
where the $\left[\chi^{(0)}\right]^3$ correction cancels the terms proportional to $s$ in the conservative part of the impulse of $\calN=8$ in Eq.~\eqref{eq:nnlo_cons_impulse_neq8}. In general relativity
\begin{align}
 \Delta p_{1\,,\perp}^{\mu\,{(2)}} (b_e) & =  
   \frac{G^3 M^4\nu}{|b_e|^3} \frac{4}{\sqrt{\sigma^2-1}}\frac{b_e^\mu}{|b_e|}
  \Bigg[ 
   h^2(\sigma,\nu) \left(
   8 \sigma^2 -\frac{1\tred{-(2\sigma^2-1)^3}}{2 \left(\sigma^2-1\right)^2}
   \right)
    \\
   &
     +\nu\left(
   f^{\rm LS}_1(\sigma)  
   -\frac{2}{3} \,  
   \sigma (14 \sigma^2+25)  \right)
   \nn 
   \\
   &+\nu \left(
   \sigma f^{\rm LS}_3(\sigma)
   - 4  \left(4 \sigma ^4-12 \sigma ^2-3\right) \right) \frac{\text{arcsinh}\sqrt{\frac{\sigma-1}{2}}}
  {\sqrt{\sigma^2-1}} 
   \Bigg]\,,
\end{align}
where we have denoted in red the correction due to the change of variables from $b$ to $b_e$.
Taking the high energy limit holding $b_e$ fixed restores universality in the transverse impulse
\begin{align}
 \Delta p_{1\,,\perp}^{\mu\,(2)} (b_e) = \frac{32G^3M^4\nu^2 \sigma^2}{|b_e|^3} \frac{b_e^\mu}{|b_e|} + \calO(\sigma)\,,
\end{align}
which now arises from the $\left[\chi^{(0)}\right]^3$ correction introduced by the alternative choice of impact parameter.
Note however that the impulse now grows as $\sigma^2$, rather than $\sigma$ in the transverse part of Eq.~\eqref{eq:nnlo_impulse_gr_high_energy}.

After this change of variables, we observe that, up to this order, the transverse impulse in $\calN=8$ supergravity agrees with that in Refs.~\cite{DiVecchia:2021ndb,DiVecchia:2021bdo} which is given in terms of the Real part of the eikonal phase, $\delta(b_e)$, as\footnote{This is denoted by $Q^\mu$ in Ref.~\cite{DiVecchia:2021bdo}.}
\begin{equation}
  \Delta p^\mu_{1\,\perp} (b_e) = -\frac{\partial\, \text{Re}\, \delta(b_e)}{\partial b_{e,\mu}}
\end{equation}
The same relation is true if we compare to the conjectured result for the eikonal phase in general relativity from Refs.~\cite{DiVecchia:2021ndb,DiVecchia:2021bdo}, thus proving their conjecture. It would be interesting to verify it again directly by directly calculating the eikonal phase using the results from this work.
This findings suggest a more general relation between the transverse impulse kernel and the eikonal phase which warrants further investigation.

Regarding the longitudinal part of the impulse, or the energy loss, we can study our full velocity dependent expressions in the ultra-relativistic limit $\sigma \rightarrow \infty$ of Eq.~(\ref{eq:rad_energy}). 
In $\calN=8$ supegravity, we found the result for a single BPS angle \cite{Caron-Huot:2018ape} which has the structure of Eqs.~\eqref{eq:rad_energy} and \eqref{eq:fsGR} with the appropriate $\phi$-dependent coefficient functions already defined in Eq.~\eqref{eq:fsSUGRA}. The $f_i$ in Eq.~\eqref{eq:fsSUGRA} agree with our previous expressions \cite{Herrmann:2021lqe} for $\phi=\pi/2$. As in pure gravity, the ultra-relativistic limit $\sigma \to \infty$ of the radiated momentum is controlled by the combinations $f_1$ and $-f_2+f_3/2$ 
\begin{align}
    \mathcal{E}(\sigma) = 8 \pi (1+2\log2)\sigma^3 + \calO(\sigma^2)\,,
    \label{eq:EheN8}
\end{align}
with the leading high-energy term being independent of $\phi$. Similarly, in general relativity
\begin{equation}
    \mathcal{E}(\sigma) =  \frac{35}{8} \pi  (1+2 \log2) \sigma^3 + \mathcal{O} \left(\sigma^2\right)\,.
    \label{eq:EheGR}
\end{equation}
Note that the apparent logarithmic divergence cancels in both cases. The high-energy limit of the general relativity energy loss can be compared to the prediction by Kovacs and Thorne \cite{Kovacs:1978eu}, based upon the numerical probe calculation by Peters \cite{Peters:1970mx}. Our expression agrees structurally with \cite{Kovacs:1978eu}, but disagree in the numerical coefficient. After Ref.~\cite{Herrmann:2021lqe} appeared, we were informed of the numerical computation of the high-energy coefficient by agreeing with our analytic result. 

Although the high-energy limit does \emph{not} coincide ins Eqs.\eqref{eq:EheN8} and \eqref{eq:EheGR} in its rational prefactors (8 vs. 35/8), we noted in Ref.~\cite{Herrmann:2021lqe} that the ratio of the logarithmic ($\log 2$) and non-logarithmic contributions is universal. Ultimately, it might not be too surprising that the radiated momentum depends on the theory content, since the number of massless messengers that can be radiated change between the two theories which suggests the bigger overall coefficient in maximal supergravity. Note that, in any case, our results are only valid for $\sigma \ll (GE_{\rm cm}/b)^{-1}$, beyond which perturbation theory breaks down.  For large enough $\sigma$, according to Eq.~\eqref{eq:EheGR}, the radiated energy exceeds the incoming energy, which, of course, is unphysical. Resolving this issue requires to account for destructive interference from multi-graviton emissions, which cuts off the spectrum of gravitational waves at high-frequency\footnote{We thank Gabriele Veneziano for discussions on this point.}, as explained in Refs.~\cite{Gruzinov:2014moa,Ciafaloni:2015xsr,Ciafaloni:2018uwe}.

%================================================
%
\tocless\section{Conclusions}
\label{sec:conclusions}
%
%================================================
%
In this work, we have employed the general formalism devised by Kosower, Maybee, and O'Connell (KMOC) to extract classical gravitational observables for the scattering of spinless black holes up to $\calO(G^3)$, or third Post-Minkowskian order. This framework naturally includes radiative effects and goes beyond the much-discussed conservative binary dynamics. The presence of the gravitational interaction between the two massive black holes has two key physical effects, 1) a deflection and momentum shift on the individual black holes which is related to the \emph{gravitational impulse}, and 2) the emission of gravitational Bremsstrahlung which is related to the \emph{radiated momentum}. In our previous work, we have already presented the radiated momentum in general relativity and maximal supersymmetric gravity ($\calN=8$ SUGRA). Here, we also present expressions for the impulse (which is related to the scattering angle) in both theories. 

In order to render the general KMOC framework a practical computational tool, we have incorporated a number of ideas from collider physics to handle virtual Feynman integrals together with phase-space integration. Starting from generalized unitarity to construct loop integrands from gauge-invariant on-shell quantities, we employ the method of regions to facilitate the classical expansion. The relevant Feynman diagrams can be reduced to a minimal set of \emph{master integrals} with the help of integration-by-parts identities. Using \emph{reverse unitarity} we treat virtual and phase-space integrals on the same footing. At the end of the reduction step, we are left with a small set of independent integrals. In order to evaluate the master integrals, we solve a set of (canonical) differential equations, where the main complication is reduced to the computation to the boundary values of the master integrals. Making available all analytic expressions for the soft master integrals, we assemble the classical impulse and radiative momentum observables in both GR and maximal supergravity. Our results include the full radiation effects at $\calO(G^3)$, but we have also reproduced the conservative gravitational impulse in GR, matching known results. From the impulse and the radiated momentum, we can derive the radiative scattering angle and the energy loss. Since our results are valid to all orders in the velocity, we are able to check against different regimes in the literature and compare against the Post-Newtonian computations by expanding our results in small velocity as well as against high-energy expectations. We find agreement with all known results where they overlap. 

We have compared our results to the eikonal approach in Refs.~\cite{DiVecchia:2021ndb,DiVecchia:2021bdo} and found that the transverse impulse, when written in the appropriate variables is directly connected to the eikonal phase. This shows that the conjectured relation between the real and imaginary parts of the eikonal phase, put forward in Ref.~\cite{DiVecchia:2021ndb}, is also valid in general relativity. It also suggest suggests a more general relation between the transverse impulse kernel and the eikonal phase which warrants further investigation.

For the classical quantities considered here, we performed the full phase-space integration over all intermediate particles appearing in the KMOC setup, without imposing any further restrictions (or phase space cuts). In principle, the reverse unitarity method can also be adjusted to incorporate additional measurements on the final state particles \cite{Anastasiou:2002yz,Anastasiou:2002qz,Anastasiou:2003yy}. One can envision a similar adaptation to the gravitational setup to measure more exclusive observables, such as the radiated energy spectrum or the angular distribution of the radiated momentum. These quantities depend on more scales and we leave their discussions to future work.

Besides attempting similar computations at higher orders in Newton's constant and the discussion of more exclusive observables, it would also be interesting to generalize the $\calO(G^3)$ computation for spinning observables as well as to include tidal effects. Since all relevant master integrals are known, the only remaining change requires the construction of more complicated loop integrands. Since this step is very mature and can be automatized and streamlined via generalized unitarity, it should be possible to tackle such observables in the near future.

%================================================
%
\section*{Acknowledgments}
\vspace{-.3cm}
%
%================================================

We are specially grateful to Paolo Di Vecchia, Carlo Heissenberg, Rodolfo Russo, and Gabriele Veneziano for stimulating discussions and for sharing a draft of their work \cite{DiVecchia:2021bdo} with us and for comments on our manuscript. 
We also thank Zvi Bern, Radu Roiban, Chia-Hsien Shen, and Mikhail Solon for helpful comments and collaboration on related projects; and Clifford Cheung and Rafael Porto for discussions.
E.H.\ thanks Lance Dixon and Bernhard Mistlberger for discussions about the reverse unitarity method. 
E.H.\ is supported by the U.S.\ Department of Energy (DOE) under Award Number DE-SC0009937. 
J.P.-M.\ is supported by the U.S.\ Department of Energy (DOE) under Award Number~DE-SC0011632. 
M.S.R.’s work is funded by the German Research Foundation (DFG) within the Research Training Group GRK 2044. 
M.Z.'s work is funded by the U.K.\ Royal Society through Grant URF{\textbackslash}R1{\textbackslash}20109.

%================================================
%================================================
%
\appendix
%
%================================================
%================================================

%================================================
\section{Fourier transform formulae}
\label{app:FT_collection}
%================================================
The final step of the classical impulse and radiated momentum computation involves the evaluation of Fourier transform integrals of the kind
\begin{align}
   \tilde f_\alpha(b^2) &= \imath \int \hat{\mathrm{d}}^Dq\, \, \hat{\delta}(-2m_1 u_1\cdot q) \hat{\delta}(2m_2 u_2\cdot q)\,
   e^{i q \cdot b} \, (-q^2)^{-\alpha}\,,\\
    \tilde f^\mu_\alpha(b^2) &= \imath \int \hat{\mathrm{d}}^Dq\, \, \hat{\delta}(-2m_1 u_1\cdot q) \hat{\delta}(2m_2 u_2\cdot q)\,
   e^{i q \cdot b} \, q^\mu (-q^2)^{-\alpha}\,.
\end{align}
These are simply related by differentiation $\tilde f^\mu_\alpha(b^2) = -\imath \partial \tilde f_\alpha(b^2)/ \partial b^\mu$, so we need only consider $f_\alpha(b^2)$. It is convenient to use a Sudakov decomposition of the $D$-dimensional Lorentz vector $q^\mu$,
\begin{align}
\label{eq:sudakov_param}
    q^\mu = x_1 u^\mu_1 + x_2 u^\mu_2 + \qperp^\mu \,,
\end{align}
where $\qperp^\mu$ points in the $(D-2)$-dimensional subspace transverse to $u_1,u_2$. With this parametrization the integral above becomes
\begin{align}
\begin{split}
    \tilde f_\alpha(b^2) &=   \frac{i}{4  m_1 m_2\sqrt{y^2-1}} \int\! \hat{\mathrm{d}}^{D-2}\qperp \hat{\mathrm{d}}x_1 \hat{\mathrm{d}}x_2 \hat{\delta}(x_1) \hat{\delta}(x_2) 
   e^{\imath q \cdot b}\, (-q^2)^{-\alpha} \\
  &{=} 
    \frac{\imath}{4 m_1 m_2 \sqrt{y^2-1}} \int\! \hat{\mathrm{d}}^{D-2}\qperp
    e^{-\imath \qperp \cdot \bperp} (\qperpsq)^{-\alpha}\,.
\end{split}
\end{align}
so that the delta functions localize two of the integration variables and force the momentum transfer into the $D-2$ dimensional transverse subspace. Note that the impact parameter is always transverse so $\bperpsq = -b^2 \equiv |b|^2$. The remaining  Fourier transform is elementary and we obtain
\begin{align}
\begin{split}
       f_\alpha(b^2)&=  \frac{\imath}{ m_1 m_2 \sqrt{y^2-1}} \frac{ \Gamma\left(D/2-1-\alpha\right)}
          {2^{2\alpha+2} (\pi)^{\frac{D-2}{2}}\Gamma\left(\alpha\right)} \frac{1}{|b|^{D-2-2\alpha}}\,,\\
    f_\alpha^\mu(b^2)&= - \frac{1}{ m_1 m_2 \sqrt{y^2-1}} \frac{ \Gamma\left(D/2-\alpha\right)}
          {2^{2\alpha+1} (\pi)^{\frac{D-2}{2}}\Gamma\left(\alpha\right)} \frac{b^\mu}{|b|^{D-1-2\alpha}}\,.
    \label{eq:basicFT}
\end{split}
\end{align}

%================================================
%
\section{Gravitational impulse and scattering angle}
\label{app:angle_impulse}
%
%================================================
%
The gravitational impulse $\Delta p_i^\mu$ describes the deflection of gravitationally interacting particles along the entirety of their hyperbolic trajectory. Another basic observable for such a scattering process is the scattering angle $\chi$ in the center-of-mass frame. In this appendix we describe the relation between these two observables.

Let us begin by considering a conservative scattering process. In this case, $\Delta p_{i,\, \cons}^\mu$ and $\chi_\cons$ are exactly equivalent and contain the same information. It is well known how to relate the scattering angle in the center-of-mass frame to the impulse. The incoming and outgoing momenta have components 
\begin{align}
\begin{split}
    p_1^\mu &= (E_1, \bm{p})\,,  \quad \;\;\; p_1^\mu+\Delta p_{1,\,\cons}^\mu = (E_1, \bm{p}')\,,\\
    p_2^\mu &= (E_2, -\bm{p})\,, \quad p_2^\mu+\Delta p_{2,\,\cons}^\mu = (E_2, -\bm{p}')\,,
\end{split}    
\label{eq:angle_impulse1}
\end{align}
with $|\bm{p}|=|\bm{p}'|$, such that $\Delta p_{1,\,\cons}^\mu = -\Delta p_{2,\,\cons}^\mu= (0,\bm{p'} - \bm{p})$. Thus 
\begin{equation}
    -(\Delta p_{i,\,\cons})^2 = (\bm{p'} - \bm{p})^2 
    = 4|\bm p|^2 \sin^2 \frac{\chi_\cons}{2}\,,
\end{equation}
or equivalently
\begin{align}
\label{eq:angle_impulse}
\sin \frac{\chi_\cons}{2} = \frac{\sqrt{-(\Delta p_{i,\cons})^2}}{2|\bm{p}|}\,.
\end{align}
The relation \eqref{eq:angle_impulse} can be inverted to write the impulse in terms of the scattering angle. This can be done, for instance, by solving the on-shell conditions for the final state
\begin{align}
    (p_i +\Delta p_{i,\, \cons})^2 = m_i^2\,,
\end{align}
together with momentum conservation $\Delta p_{1,\,\cons}^\mu = -\Delta p_{2,\,\cons}^\mu$ and the condition in Eq.~\eqref{eq:angle_impulse1}. One way to do this is by first decomposing the impulse in terms of the basis vectors
\begin{align}
    \Delta p^\mu_{1,\,\cons} = 
        a_1\, \frac{b^\mu}{|b|} 
        + a_2\, \check{u}^\mu_1 
        + a_3 \check{u}^\mu_2
\end{align}
and then solving for the three coefficients $a_i$ using the stated conditions. The result is
\begin{align}
\label{eq:conservative_impulse_angle}
    \Delta p^{\mu}_{1,\,\cons} &=
      \, \, 
      |\bm{p}|  \sin \chi_\cons  \, \frac{b^\mu}{|b|} + 
      |\bm{p}|  (1-\cos\chi_\cons) \left(\frac{|\bm{p}|}{m_1}\check{u}^\mu_1 {-} \frac{|\bm{p}|}{m_2} \check{u}^\mu_2 \right) \,,
\end{align}
which can be expanded perturbatively in $G$
\begin{align}
    \label{eq:impulse_from_angleLO}
    (\Delta p_{1,\,\cons}^{(0)})^\mu &= |\bm p|\, \chi^{(0)}_\cons\, \frac{b^\mu}{|b|} \\
    \label{eq:impulse_from_angleNLO}
    (\Delta p_{1,\,\cons}^{(1)})^\mu &= |\bm p|\, \chi^{(1)}_\cons\, \frac{b^\mu}{|b|}
    + \,|\bm p|\, \frac12 (\chi^{(0)}_\cons)^2 \left(\frac{|\bm{p}|}{m_1}\check{u}^\mu_1 {-} \frac{|\bm{p}|}{m_2} \check{u}^\mu_2 \right)  
    \\
    \label{eq:impulse_from_angleNNLO}
    (\Delta p_{1,\,\cons}^{(2)})^\mu &= |\bm p| \, \Big(\chi^{(2)}_\cons-\frac{1}{6} (\chi^{(0)}_\cons)^3\Big)\, \frac{b^\mu}{|b|}
    +\,|\bm p|\, \chi^{(0)}_\cons\chi^{(1)}_\cons
    \left(\frac{|\bm{p}|}{m_1}\check{u}^\mu_1 {-} \frac{|\bm{p}|}{m_2} \check{u}^\mu_2 \right)\,,
\end{align}
where $\chi^{(n)}_\cons\sim {\calO}(G^{n+1})$.
Note that magnitude of the c.o.m three-momentum $\bm{p}$ can be written in terms of the quantities used in the rest of the paper as follows
\begin{align}
   |\bm{p}| = \frac{m_1 m_2\, \sqrt{\sigma^2-1}}
           {\sqrt{2 m_1 m_2 \sigma +m^2_1+m^2_2}} = \frac{M\nu \sqrt{\sigma^2-1}}{h(\sigma,\nu)} \,.
    \label{eq:pinfty}
\end{align}
Using Eqs.~\eqref{eq:impulse_from_angleLO}--\eqref{eq:impulse_from_angleNNLO} together with Eq.~\eqref{eq:pinfty} we have checked that our computations of the gravitational impulse agree with the known results for the conservative scattering angle at ${\cal O}(G)$, ${\cal O}(G^2)$ \cite{Damour:2016gwp,Damour:2017zjx,Bjerrum-Bohr:2018xdl,Cristofoli:2019neg} and ${\cal O}(G^3)$ \cite{Bern:2019nnu,Bern:2019crd,Kalin:2020fhe}.

As a side comment, note that this formula nicely explains the structure of the conservative result in maximal supergravity, where the one-loop scattering angle $\chi^{(1)}_\cons$ is zero \cite{Parra-Martinez:2020dzs}, which can be attributed to the "no-triangle" property of this theory \cite{Caron-Huot:2018ape}. In particular, this implies that the one-loop impulse is purely longitudinal, and the two-loop impulse purely transverse, in agreement with our explicit calculation. 

Let us point out that the definition of the impact parameter $b$ is chosen in terms of the initial momenta such that it satisfies $b \cdot p_{1} = b \cdot p_{2} = 0$. This choice, however, breaks the symmetry between the initial and final state in the conservative process (i.e. time reversal invariance). Instead, one could choose to modify the impact parameter as follows 
\begin{equation}
\label{eq:b_eikonal}
      b^\mu_\eik = b^\mu - |b| \, \sin \frac{\chi_\cons}{2} \left(\frac{|\bm{p}|}{m_1}\check{u}^\mu_1 {-} \frac{|\bm{p}|}{m_2} \check{u}^\mu_2 \right) \,,
\end{equation}
such that it more symmetric between the initial and final state $ b_\eik \cdot p_i = - b_\eik \cdot (p_i +\Delta p_i)$.
Due to this modification, the magnitude changes $|b_\eik| = |b| \cos \frac{\chi_\cons}{2} $, and $b_\eik$ can be recognized as the so-called "eikonal impact parameter" that naturally arises from semiclassical considerations in the eikonal approach \cite{Bern:2020gjj}. Almost by definition, the impulse is purely transverse in these variables, that is, proportional to $b_\eik^\mu$,
\begin{equation}
    \Delta p^{\mu}_{1,\,\cons} =
      \, \, 
     2 |\bm{p}|  \sin \frac{\chi_\cons}{2}  \, \frac{b_\eik^\mu}{|b_\eik|}\,.
\end{equation}
This form of the conservative impulse is the most convenient to compare to results from the eikonal method.

More generally, radiative effects imply that the c.o.m of the binary is not an inertial reference frame, so the relation between the scattering angle and impulse is not as straight forward in the non-conservative setup. Momentum conservation $\Delta p^\mu_1 + \Delta p^\mu_2 = - \Delta R^\mu$ illustrates that the radiative dynamics are a multi-body process and that the center of mass recoils. One can still write down an analogous formula to Eqs.~\eqref{eq:impulse_from_angleLO}-\eqref{eq:impulse_from_angleNNLO} that relates the radiated momentum and radiative scattering angle to the impulse, although generically, the precise meaning of \emph{the} angle becomes somewhat obscure in the presence of radiation. Said differently, including radiation, the kinematics is that of a five-point process for which there are two $t$-channel invariants, which translates to two angles. At $O(G^3)$, however, the leading recoil effects do not yet affect the transverse part of the impulse and the angle is still given by
\begin{equation}
   \sin\chi = \frac{\sqrt{-(\Delta p_{1,\perp})^2}}{|\bm p|}\,.
\end{equation}
Alternatively, separating the angle in a conservative and a radiative piece, $\chi = \chi_\cons + \chi_{\rm rad}$, we find at order $\calO(G^3)$ the following relation
\begin{equation}
    \Delta p_{1,{\rm rad}\,,\perp}^{(2)} = |\bm p| \chi^{(2)}_{\rm rad} \frac{b^\mu}{|b|}\,.
    \label{eq:chiradimpulse}
\end{equation}

%================================================
%
\section{Review of unitarity and cutting rules}
\label{app:cutting_rules}
%
%================================================
%
In this appendix we review how unitarity and the cutting rules relate the imaginary part of \emph{virtual} amplitudes or diagrams to their unitarity cuts. This is used in the main text to simplify the calculation of the KMOC impulse kernel.
%

%================================================
\subsection*{Unitarity}
%================================================
%
As is well known, the unitarity of the S-matrix implies similar unitarity relations for S-matrix elements themselves. For the four-point amplitude these arise from writing $S= 1 + \imath T$, inserting the unitarity relation
\begin{equation}
    SS^\dagger =1 
    \quad \leftrightarrow \quad 
    2\operatorname{Im}T = -i (T-T^\dagger) = TT^\dagger
\end{equation}
between initial and final two-particle states,
 \begin{equation}
     2\operatorname{Im}  \langle p_4,\!p_3|T|p_1,\!p_2\rangle =  \langle p_4,\!p_3|TT^\dagger|p_1,\!p_2\rangle
 \end{equation}
and inserting a complete set of states as follows
\begin{align}
    &\langle p_4,\!p_3|TT^\dagger|p_1,\!p_2\rangle \!= \!\sum\limits_X \int \mathrm{d}\Phi_{2+|X|}(r_1,\!r_2,\!X)\, \langle p_4,\!p_3|T|r_1,\!r_2,\!X\rangle \langle r_1,\!r_2,\!X|T^\dagger|p_1,\!p_2\rangle \,.
\end{align}
This can be represented pictorially as follows
\begin{equation}
 2\Im \left[ \raisebox{-32pt}{\scalebox{0.8}{\kmocvirtual}} \right] 
     =  \sum_X \int \mathrm{d}\widetilde{\Phi}_{2+|X|}\ 
    \hspace{-.4cm}
     \raisebox{-32pt}{\scalebox{0.8}{\kmocreal{$\ell_2-p_2$}{$\ell_1-p_1$}}}\,,
\end{equation}
For forward scattering, the relation above is nothing but the famous optical theorem, but here we will use it as a general relation for arbitrary $q$ in perturbation theory. At one-loop it relates the imaginary part of the virtual amplitude to its two-particle cuts
\begin{equation}
   2\, \text{Im} \left[\!\!\! \raisebox{-32pt}{\scalebox{0.8}{\kmocvirtualnlo}} \right]
   = %{\cal M}_4(p_1,p_2,p_3,p_4) 
     \int \mathrm{d}\widetilde{\Phi}_{2} 
     \hspace{-.3cm}
     \raisebox{-32pt}{\scalebox{0.8}{\kmocrealnlo}}\,.
\label{eq:unitarityoneloop}
\end{equation}
At two-loops, the unitarity relation includes more terms with both two- and three-particle cuts and is given by
\begin{align}
\label{eq:unitaritytwoloop}
    2\, \text{Im}\left[
    \!\vcenter{\hbox{\scalebox{0.7}{\kmocvirtualnnlonolab}}}\right] 
    & = 
     \!\int\!\! \mathrm{d}\widetilde{\Phi}_2
    \hspace{-.2cm}\vcenter{\hbox{\scalebox{0.65}{\cutRuleAmpLoopTree}}}\hspace{-.2cm}
    + \!\int\!\! \mathrm{d}\widetilde{\Phi}_3
    \hspace{-.2cm}\vcenter{\hbox{\scalebox{0.65}{\cutRuleAmpTreeTree}}}\hspace{-.2cm}
    + \!\int\!\! \mathrm{d}\widetilde{\Phi}_2
    \hspace{-.2cm}\vcenter{\hbox{\scalebox{0.65}{\cutRuleAmpTreeLoop}}}\,.
\end{align}
Similar generalizations also hold to higher-loop order which are, however, irrelevant for the discussion in the present work. Note that these relations were crucial in order to simplify the KMOC kernels and proof the reality properties of the classical observables of interest which led to the results in section \ref{sec:KMOC_discussion_expansion}.

%================================================
\subsection*{Cutting rules}
%================================================
%
Alternatively, our computation of phase space integrals can be based on Cutkosky's cutting rules \cite{Cutkosky:1960sp}, which can be applied to individual diagrams, rather than the full amplitude.  For us, the application of the cutting rules is twofold. First, we make use of them to simplify the KMOC kernels in section \ref{sec:KMOC_discussion_expansion}. Second, we extensively utilize these rules to deduce phase-space integrals from virtual integrals. In fact, we actually use the cutting rules for \emph{soft-expanded integrals}, where massive propagators are linearized and have the form $\imath/(2 u_i \cdot \ell_i)$ while their cut versions have the form $2 \pi \theta(\ell_i^0) \delta(2u_i \cdot \ell_i)$, but the usual proofs of cutting rules, e.g.\ using Veltman's \emph{largest time equation} \cite{tHooft:1973wag}, carry through unchanged.

For illustration purposes, we consider a field theory with two massive complex scalar fields $\Phi_1$ and $\Phi_2$, and a light scalar field $\phi$, whose Lagrangian density is
\begin{equation}
  \mathcal L =  \sum_{i=1}^2 
    \left(
        \partial^{\mu} \Phi_i^\dagger \partial_\mu \Phi^{\phantom{\dagger}}_i 
        - m_i^2 \Phi_i^\dagger \Phi^{\phantom{\dagger}}_i 
    \right) 
    + \frac 1 2 \partial^\mu \phi \partial_\mu \phi 
    - \kappa \sum_{i=1}^2 \Phi_i^\dagger \Phi^{\phantom{\dagger}}_i \phi  
    - \frac{\kappa}{3!} \phi^3 \, .
\end{equation}
From the point of view of individual Feynman diagrams, the difference between this theory and the gravitational theory considered in the body of the paper is that the latter diagrams contain additional numerator, which leave the discussion below unchanged.
The Feynman vertices are always $-\imath\kappa$, for $\Phi_1^\dagger \Phi_1 \phi$, $\Phi_2^\dagger \Phi_2 \phi$, and $\phi^3$ couplings. The propagators with momentum $k$ are $\imath/(k^2-m_1^2)$, $\imath/(k^2-m_2^2)$, and $\imath/(k^2)$ for the fields $\Phi_1$, $\Phi_2$, and $\phi$, respectively.
We use thick dashed black lines to denote heavy scalar particles $\Phi_i$, and thin dashed lines to denote light scalar particles $\phi$. 
For discussing loop integration, it is convenient to introduce the notion of ``scalar integrals'' which have \emph{unit} numerators with all factors of $\imath$ removed.
For example, a simple two-loop diagram for $\Phi_1 + \Phi_2 \rightarrow \Phi_1 + \Phi_2$ scattering an the corresponding scalar integral are
\begin{equation}
\vcenter{\hbox{\DiagIIIPhiVirt}}=\imath\vcenter{\hbox{\DiagIIIVirt}}\,.
\label{eq:translateToScalar}
\end{equation}
The scalar diagram on the r.h.s. of \eqref{eq:translateToScalar} is essentially the same as the Feynman diagram on the l.h.s., except that dashed lines are replaced by solid lines to indicate that every propagator is understood to be \emph{without} the $\imath$ factor in the numerator, and every vertex is simply $\kappa$ rather than $-\imath\kappa$.

%================================================
\subsubsection*{Cutting rules and translation to scalar integrals}
%================================================
%
Having defined the virtual scalar and Feynman diagrams, we introduce unitarity cuts of Feynman diagrams, where a vertical blue dashed line highlights the cut propagators, e.g.
\begin{equation}
\vcenter{\hbox{\DiagHPhiCutCentre}}\,.\label{fig:H-cut}
\end{equation}
Every cut propagator is given by the simple replacement rule
\begin{equation}
  \frac{\imath}{k^2-m^2} \quad \longrightarrow \quad 
  2 \pi\, \theta(k^0) \, \delta(k^2 - m^2), 
  \label{eq:cutprop}
\end{equation}
which simultaneously imposes the positive energy and the on-shell condition. Instead of performing the full loop integration, in the presence of the on-shell conditions, the remaining integrals are over the Lorentz-invariant phase space of the on-shell states exchanged across the cut. According to the Cutkosky rules, the uncut propagators and vertices on the left hand side of the cut are given by their usual expressions, while those on the right are given by the complex conjugates of their usual expressions. In this notation, the ``rightmost'' cut is always the same as the uncut diagram up to certain factors, while the ``leftmost'' cut is equal to the conjugate of the uncut diagram,
\begin{align}
\begin{split}
\vcenter{\hbox{\DiagIIIPhiVirtCutRight}}&=\vcenter{\hbox{\DiagIIIPhiVirt}}=\imath\vcenter{\hbox{\DiagIIIVirt}}\,,\\
\vcenter{\hbox{\DiagIIIPhiVirtCutLeft}}&=\left(\vcenter{\hbox{\DiagIIIPhiVirt}}\right)^*=-\imath\left(\vcenter{\hbox{\DiagIIIVirt}}\right)^*\,,
\end{split}
\end{align}
When using solid lines, the propagators and vertices on either side of the cut are without any factors of the imaginary unit $\imath$, while cut propagators are still given by the right hand side of Eq.~\eqref{eq:cutprop}, $2 \pi\, \theta(k^0) \, \delta(k^2 - m^2)$. Using the cutting rules, the sum of \emph{all cuts} in a given channel is zero,
\begin{equation}
\vcenter{\hbox{\DiagHPhiVirtCutLeft}}+\vcenter{\hbox{\DiagHPhiCutCentre}}+\vcenter{\hbox{\DiagHPhiVirtCutRight}}=0
\end{equation}
Translating to scalar integrals without $i$ factors, this reads as
\begin{equation}
-\imath\left(\vcenter{\hbox{\DiagHVirt}}\right)^*+\vcenter{\hbox{\DiagHCutCentre}}+\imath\left(\vcenter{\hbox{\DiagHVirt}}\right)=0
\end{equation}
or, equivalently,
\begin{equation}
	2\Im\left(
	\vcenter{\hbox{\scalebox{.9}{\DiagHVirt}}}
	\right) = \vcenter{\hbox{\scalebox{.9}{\DiagHCutCentre}}},
\label{eq:HcuttingRule1}
\end{equation}
similar to the usual statement of the optical theorem, but with generally \emph{different} momenta on the left and on the right. For the ladder diagram, the cutting rule involves more terms,
\begin{align}
\vcenter{\hbox{\DiagIIIPhiVirtCutLeft}}+\vcenter{\hbox{\DiagIIIPhiCutLeft}}+\vcenter{\hbox{\DiagIIIPhiCutRight}}&\nonumber\\+
\vcenter{\hbox{\DiagIIIPhiCutNE}}+\vcenter{\hbox{\DiagIIIPhiCutSE}}+\vcenter{\hbox{\DiagIIIPhiVirtCutRight}}	&=0\,,\label{eq:IIIcuttingRule}
\end{align}
which translates into the following relation for scalar integrals, using diagram symmetries to combine the 6 terms into 3 terms, 
\begin{equation}
	2\Im\left(\vcenter{\hbox{\scalebox{.8}{\DiagIIIVirt}}}\right)
	=2\vcenter{\hbox{\scalebox{.8}{\DiagIIICutNE}}} + 2\Im\left(\vcenter{\hbox{\scalebox{.8}{\DiagIIICutLeft}}}\right)\,,
\label{eq:IIIcuttingRule1}
\end{equation}
where we kept an overall factor of 2 to emphasize the origin in Eq.~\eqref{eq:IIIcuttingRule}.

\noindent
The crossed double box satisfies a relation analogous to Eq.~\eqref{eq:IIIcuttingRule1},
\begin{equation}
	2\Im\left(\vcenter{\hbox{\scalebox{.8}{\DiagIXVirt}}}\right)=2\vcenter{\hbox{\scalebox{.8}{\DiagIXCutNE}}} + \Im\left(\vcenter{\hbox{\scalebox{.8}{\DiagIXCutLeft}}}\right)\,.
	\label{eq:XBcuttingRule1}
\end{equation}
A clear difference from the above equation for the planar double box, \eqref{eq:IIIcuttingRule1}, is that the double cut contribution (i.e.\ the last term) is multiplied by $(-1)$ rather than $(-2)$, because the crossed double box has only one double cut.

The u-channel double box has no double cut or triple cut. The u-channel crossed double box has only one triple cut, which evaluates to twice the imaginary part of the virtual integral, analogous to Eq.~\eqref{eq:HcuttingRule1}.

Note that Eq.~\eqref{eq:IIIcuttingRule} is also valid for field theories other than scalar $\phi^3$ theory, so Eq.~\eqref{eq:IIIcuttingRule} holds with numerators multiplying the loop integrand of every diagram in the relation, and therefore also for scattering amplitudes and their unitarity cuts!\footnote{Depending on the properties of the numerator, diagram symmetries might be broken so that one cannot directly obtain Eq.~\eqref{eq:IIIcuttingRule1}. After IBP reduction, however, we will mostly be dealing with master integrals without numerators for which the symmetry is restored.}

%================================================
%
\section{Master integrals}
\label{app:masters_notation}
%
%================================================
%
As explained in Ref.~\cite{Parra-Martinez:2020dzs} at two-loops there are three irreducible families of master integrals, the $\RT,\H$ and $\IX$ families. These families contain a total of 20 unique linearized-propagator master integrals. As explained in Ref.~\cite{Parra-Martinez:2020dzs} at two-loops there are three irreducible families of master integrals, the $\RT,\H$ and $\IX$ families. All results can be found in a computer readable format in the ancillary files of this \texttt{arXiv} submission . 
%
%
%
%================================================
\subsection*{$\RT$ and $\H$ families}
%================================================
%
\begin{figure}[t!]
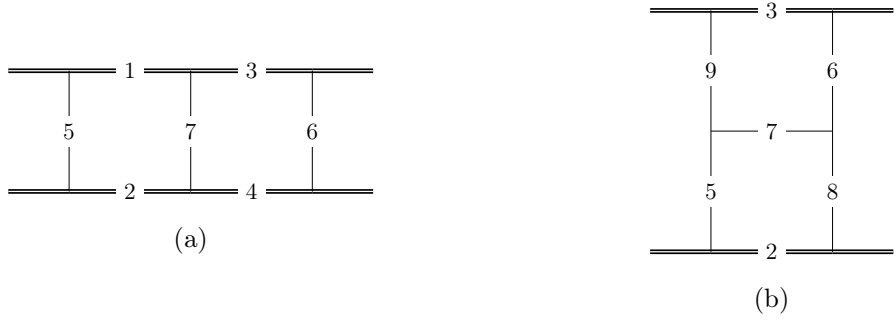

\vskip -.7cm
	\begin{subfigure}[c]{0.5\linewidth}
		\centering	%III
		$\vcenter{\scalebox{.8}{\IIIdiagAppendix}}$
		\subcaption{}
	\end{subfigure}
	\begin{subfigure}[c]{0.5\linewidth}
	    \centering	%H
	    $\vcenter{\scalebox{.8}{\HdiagAppendix}}$
	    \subcaption{}
	\end{subfigure}
\caption{Propagators for $\RT$ (a) and $\H$ family (b).\label{fig:PropsIIIandH}}
\end{figure}
\noindent
We first consider generic integrals of the form
\begin{equation}
I_{i_1,i_2,\dots, i_9}=\int\frac{\mathrm{d}^{D}\ell_1}{(2\pi)^D}\int\frac{\mathrm{d}^{D}\ell_2}{(2\pi)^D}\frac{1}{\tilde \rho_1^{i_1}\tilde \rho_2^{i_2}\cdots\tilde \rho_9^{i_9}}\,.
\end{equation}
Where the propagators are
\begin{alignat}{3}
&\tilde \rho_1=(\ell_1-p_1)^2-m_1^2\,,\qquad\quad &&\tilde \rho_2=(\ell_1+p_2)^2-m_2^2\,,\qquad\quad && \tilde \rho_3=(\ell_2-p_4)^2-m_1^2\,,\nonumber\\
&\tilde \rho_4=(\ell_2+p_3)^2-m_2^2\,,\qquad\quad&&\tilde \rho_5=\ell_1^2\,,\qquad\quad&&\tilde \rho_6=\ell_2^2 \,,\nonumber\\
&\tilde \rho_7=(\ell_1 + \ell_2 - q)^2\,,\qquad\quad&&\tilde \rho_8=(\ell_1 - q)^2\,,\qquad\quad&&\tilde \rho_9=(\ell_2 - q)^2\,.\label{eq:TwoLoopPropsFull}
\end{alignat}
The scalar III double-box integral is $I_{\RT} = I_{1,1,1,1,1,1,1,0,0}$, the scalar H double-box is ${I_{\H} = I_{0,1,1,0,1,1,1,1,1}}$.
In the soft region, we construct an expansion of the integrand around small $|\ell_i| \sim |q|$, where the leading order parts of $\tilde \rho_i$, denoted by $\rho_i$ are given by
\begin{alignat}{3}
&\rho_1=2\, \ell_1\cdot u_1\,,\qquad\qquad 
&&\rho_2=-2\, \ell_1 \cdot u_2\,,\qquad\qquad 
&& \rho_3=-2\, \ell_2\cdot u_1\,,\nonumber\\
&\rho_4=2\, \ell_2\cdot u_2\,,\qquad\qquad
&&\rho_5=\ell_1^2\,,\qquad\qquad
&&\rho_6=\ell_2^2 \,,\nonumber\\
&\rho_7=(\ell_1 + \ell_2 - q)^2\,,\qquad\qquad
&&\rho_8=(\ell_1 - q)^2\,,\qquad\qquad
&&\rho_9=(\ell_2 - q)^2\,.
\label{eq:TwoLoopPropsSoft}
\end{alignat}
The labeling of the propagators is depicted in Figure~\ref{fig:PropsIIIandH} and the soft integrals are defined with the following normalization conventions
\begin{equation}
G_{i_1, \, i_2, \dots , i_9} = \int \frac {\mathrm{d}^D \ell_1 \, e^{\EulerGamma \epsilon}}{\imath \pi^{D/2}} \int \frac {\mathrm{d}^D \ell_2 \, e^{\EulerGamma \epsilon}}{\imath \pi^{D/2}}
\frac{1}{\rho_1^{i_1} \rho_2^{i_2} \dots \rho_9^{i_9}} \,.
\label{eq:HIIItwoloopGtilde}
\end{equation}
A pure basis for the $\RT$ family is given by
\begin{small}
\begin{align}
f_{\RT,1}={}& \epsilon ^2 (-q^2) G_{0,0,0,0,1,2,2,0,0}\,,\label{eq:fRT1}\\ 
f_{\RT,2}={}&\epsilon ^4 \sqrt{\relfbar^2-1}\, G_{0,1,1,0,1,1,1,0,0}\,, \label{eq:fRT2}\\ 
f_{\RT,3}={}&\epsilon ^3 (-q^2) \sqrt{\relfbar^2-1}\, G_{0,1,1,0,2,1,1,0,0}\,,\\ 
f_{\RT,4}={}&-\epsilon ^2 (-q^2) G_{0,2,2,0,1,1,1,0,0} +\epsilon ^3 \relfbar \, (-q^2) G_{0,1,1,0,2,1,1,0,0}\,,\\ 
f_{\RT,5}={}&\epsilon ^3\sqrt{\relfbar^2-1}\,  (-q^2) G_{1,1,0,0,1,1,2,0,0}\,,\\ 
f_{\RT,6}={}&\epsilon ^3 (1-6 \epsilon)\, G_{1,0,1,0,1,1,1,0,0}\,, \label{eq:fRT6} \\ 
f_{\RT,7}={}&\epsilon ^4\left(\relfbar^2-1\right) (-q^2)  G_{1,1,1,1,1,1,1,0,0}\,,\\
f_{\mathrm{\RT,8}}={}&\epsilon ^3 \sqrtmQSq G_{1,0,0,0,1,1,2,0,0}\,,\\ 
f_{\mathrm{\RT,9}}={}&\epsilon ^3 \sqrtmQSq G_{0,2,1,0,1,1,1,0,0}\,,\\ 
f_{\mathrm{\RT,10}}={}&\epsilon ^4\sqrt{\relfbar^2-1} \sqrtmQSq G_{1,1,1,0,1,1,1,0,0}\,.\label{eq:fRT10}
\end{align}
A pure basis for the $\H$ family is given by
\begin{align}
f_{\H,1}={}&\epsilon ^2 (-q^2) G_{0,0,0,0,0,0,1,2,2}\,, \label{eq:HpureBasis1} \\ 
f_{\H,2}={}&\epsilon ^2 (1-4 \epsilon )\, G_{0,0,2,0,1,0,1,1,0}\,,\\
f_{\H,3}={}&\epsilon ^2 (-q^2)^2 G_{0,0,0,0,2,1,0,1,2}\,,\\ 
f_{\H,4}={}&\epsilon ^4 (-q^2) G_{0,1,1,0,1,1,0,1,1}\,,\\ 
f_{\H,5}={}&\epsilon ^4 \sqrt{\relfbar^2-1}\, G_{0,1,1,0,0,0,1,1,1}\,,\\ 
f_{\H,6}={}&\epsilon ^3 \sqrt{\relfbar^2-1}\, (-q^2) G_{0,1,1,0,0,0,2,1,1}\,,\\ 
f_{\H,7}={}&-\epsilon ^2 (-q^2) G_{0,2,2,0,0,0,1,1,1} +\epsilon ^3 \relfbar \, (-q^2) G_{0,1,1,0,0,0,2,1,1}\,,\\  
f_{\H,8}={}&\frac{\epsilon ^2 (4 \epsilon -1)}{\sqrt{\relfbar^2-1}}\left[(2 \epsilon -1) G_{0,1,1,0,0,1,1,0,1}+\relfbar\, G_{0,2,0,0,0,1,1,0,1}\right] \,,\\
f_{\H,9}={}&\epsilon ^4\sqrt{\relfbar^2-1}\,  (-q^2)^2 G_{0,1,1,0,1,1,1,1,1}\,,\\ 
f_{\H,10}={}&-\epsilon ^4 (-q^2) G_{-1,1,1,-1,1,1,1,1,1} + \frac{1}{2} \epsilon^2 (2 \epsilon -1)\, G_{0,0,0,0,1,1,0,1,1}\nonumber\\
&+2\epsilon ^4 \relfbar\,  (-q^2) G_{0,1,1,0,1,1,0,1,1} + \epsilon  (3 \epsilon -2) (3 \epsilon -1)\, (-q^2)^{-1} G_{0,0,0,0,1,1,1,0,0}\,,\\
f_{\H,11} = {}& \frac{1}{\sqrt{-q^2}} \epsilon ^2 (2 \epsilon -1) (3 \epsilon -1) G_{0,1,0,0,1,1,1,0,0}\,, \label{eq:fH11} \\
f_{\H,12} = {}& \epsilon ^3\sqrt{-q^2} G_{0,2,1,0,1,1,1,0,0}\,,\\
f_{\H,13} = {}&\epsilon ^3 (2 \epsilon -1)\sqrt{-q^2} G_{0,1,0,0,1,1,0,1,1}\,,\\
f_{\H,14} = {}& \frac{1}{\sqrt{-q^2} }\epsilon ^2 (2 \epsilon -1)^2 G_{0,1,0,0,0,1,1,0,1}\,,\\
f_{\H,15} = {}& \epsilon ^3\sqrt{-q^2} G_{0,1,2,0,1,0,1,1,0}\,,\\
f_{\H,16} = {}&(\relfbar-1) \epsilon ^3(-q^2)^{5/2} G_{0,2,1,0,1,1,1,1,1}+ \frac{12  \epsilon ^3 (2 \epsilon -1) (3 \epsilon -1)}{(2 \epsilon +1)\sqrt{-q^2}} G_{0,1,0,0,1,1,1,0,0}\notag\\
&+ \frac{16 \epsilon ^4\sqrt{-q^2}}{ (\relfbar+1) (2 \epsilon +1)} G_{0,1,2,0,1,0,1,1,0} + \frac{16  \relfbar \epsilon ^3 (2 \epsilon -1)^2}{(\relfbar+1) (2 \epsilon +1)\sqrt{-q^2}} G_{0,1,0,0,0,1,1,0,1}\notag\\
&-\frac{4 (\relfbar+1) \epsilon ^4\sqrt{-q^2}}{ (2 \epsilon +1)} G_{0,2,1,0,1,1,1,0,0}\,.
\label{eq:HpureBasis}
\end{align}
\end{small}

%================================================
\subsubsection*{Differential equation for odd-in-$|q|$ $\H$ master integrals }
%\label{sec:DEHodd}
%================================================
%
We give the missing matrices appearing in the differential equations Eq.~\eqref{eq:canonicalDE}, for the odd-in-$|q|$ master integrals Eqs.~\eqref{eq:fH11}-\eqref{eq:HpureBasis} that were not previously written out in Ref.~\cite{Parra-Martinez:2020dzs}. To be more precise, these are integrals that scale like half-integer powers of $|q|=\sqrt{-q^2}$, \emph{before being multiplied by normalization factors} like $\sqrt{-q^2}$.
\begin{align}
\begin{split}
A_{\H,0}^{(\rm o)}={}&\left(
\begin{array}{cccccc}
 0 & 0 & 0 & 0 & 0 & 0 \\
 0 & -2 & 0 & 0 & 0 & 0 \\
 0 & 0 & 0 & 0 & 0 & 0 \\
 0 & 0 & 0 & 0 & 0 & 0 \\
 0 & 0 & 0 & 0 & 2 & 0 \\
 0 & -8 & -4 & 0 & 0 & 0 \\
\end{array}
\right)\,,\qquad
A_{\H,+1}^{(\rm o)}=\left(
\begin{array}{cccccc}
 0 & 0 & 0 & 0 & 0 & 0 \\
 3 & -2 & 0 & 0 & 0 & 0 \\
 0 & 0 & 0 & 0 & 0 & 0 \\
 0 & 0 & 0 & 0 & 0 & 0 \\
 0 & 0 & 0 & -2 & -2 & 0 \\
 -24 & 16 & 0 & -16 & -16 & -2 \\
\end{array}
\right)\,,\\
A_{\H,-1}^{(\rm o)}={}&\left(
\begin{array}{cccccc}
 0 & 0 & 0 & 0 & 0 & 0 \\
 -3 & 6 & 0 & 0 & 0 & 0 \\
 0 & 0 & 0 & 0 & 0 & 0 \\
 0 & 0 & 0 & 0 & 0 & 0 \\
 0 & 0 & 0 & 2 & -2 & 0 \\
 24 & 0 & 8 & 16 & 16 & 2 \\
\end{array}
\right)\,,
\end{split}
\end{align}
where the subscript $0$ indicates that the coefficient matrix is associated with $\mathrm{d}\log x$ and the subscripts $\pm 1$ refer to $\mathrm{d}\log (1\mp x)$.

%================================================
\subsection*{$\IX$ family}
%================================================
%
\begin{figure}[t!]
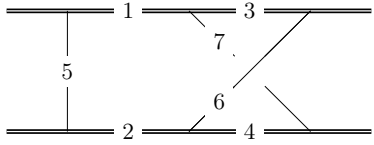

	\centering
	\vskip -.8cm
    $\vcenter{\scalebox{.8}{\IXdiagAppendix}}$
	\caption{\label{fig:XBTopos}Propagators for the $\IX$ family}
\end{figure}
\noindent
We first consider generic integrals of the form
\begin{equation}
I_{i_1,i_2,\dots, i_9}=\int\frac{\mathrm{d}^{D}\ell_1}{(2\pi)^D}\int\frac{\mathrm{d}^{D}\ell_2}{(2\pi)^D}\frac{1}{\widetilde{\rho}_1^{i_1}\widetilde{\rho}_2^{i_2}\cdots \widetilde{\rho}_9^{i_9}}\,.
\end{equation}
Where the propagators are, as depicted in Figure~\ref{fig:XBTopos}
\begin{small}
\begin{alignat}{3}
&\tilde \rho_1=(\ell_1-p_1)^2-m_1^2\,,\qquad\quad 
&&\tilde \rho_2=(\ell_1+p_2)^2-m_2^2\,,\qquad\quad 
&& \tilde \rho_3=(\ell_2-p_4)^2-m_1^2\,,\nonumber\\
&\tilde \rho_4=(\ell_1+\ell_2-q-p_3)^2-m_2^2\,,\qquad\quad
&&\tilde \rho_5=\ell_1^2\,,\qquad\quad
&&\tilde \rho_6=\ell_2^2 \,,\nonumber\\
&\tilde \rho_7=(\ell_1 + \ell_2 - q)^2\,,\qquad\quad
&&\tilde \rho_8=(\ell_1 - q)^2\,,\qquad\quad
&&\tilde \rho_9=(\ell_2 - q)^2\,,
\label{eq:XBPropsFull}
\end{alignat}
\end{small}
and the scalar non-planar double-box integral is $I_{\IX} = I_{1,1,1,1,1,1,1,0,0}$.
The small-$|q|$ expansion consists of integrals of the form
\begin{equation}
G_{i_1, \, i_2, \dots , i_9} = \int \frac {\mathrm{d}^D \ell_1 \, e^{\EulerGamma \epsilon}}{\imath \pi^{D/2}} \int \frac {\mathrm{d}^D \ell_2 \, e^{\EulerGamma \epsilon}}{\imath \pi^{D/2}}
\frac{1}{\rho_1^{i_1} \rho_2^{i_2} \dots \rho_9^{i_9}} ,
\label{eq:XBtwoloopGtilde}
\end{equation}
where the leading order parts of the propagators are
\begin{small}
\begin{alignat}{3}
&\rho_1=2\, \ell_1\cdot u_1\,,\qquad\qquad 
&&\rho_2=-2\, \ell_1 \cdot u_2\,,\qquad\qquad 
&& \rho_3=-2\, \ell_2\cdot u_1\,,\nonumber\\
&\rho_4=-2\,(\ell_1+\ell_2)\cdot u_2\,,\qquad\qquad
&&\rho_5=\ell_1^2\,,\qquad\qquad
&&\rho_6=\ell_2^2 \,,\nonumber\\
&\rho_7=(\ell_1 + \ell_2 - q)^2\,,\qquad\qquad
&&\rho_8=(\ell_1 -q)^2\,,\qquad\qquad
&&\rho_9=(\ell_2 - q)^2\,.
\label{eq:TwoLoopPropsSoftXI}
\end{alignat}
\end{small}
A pure basis of master integrals is given by
\begin{small}
\begin{align}
f_{\IX,1}={}&\epsilon ^2(-q^2) G_{0,0,0,0,2,2,1,0,0}\,,\\ 
f_{\IX,2}={}&\epsilon ^4\sqrt{\relfbar^2-1}\,  G_{0,0,1,1,1,1,1,0,0}\,,\\ 
f_{\IX,3}={}&\epsilon ^3(-q^2) \sqrt{\relfbar^2-1}\,  G_{0,0,1,1,2,1,1,0,0}\,,\\
f_{\IX,4}={}&\epsilon ^2(-q^2) G_{0,0,2,2,1,1,1,0,0} + \epsilon ^3(-q^2) \relfbar  G_{0,0,1,1,2,1,1,0,0}\,,\\ 
f_{\IX,5}={}&\epsilon ^4\sqrt{\relfbar^2-1}\,   G_{0,1,1,0,1,1,1,0,0}\,,\\ 
f_{\IX,6}={}&\epsilon ^3(-q^2)\sqrt{\relfbar^2-1}\,   G_{0,1,1,0,1,1,2,0,0}\,,\\ 
f_{\IX,7}={}&\epsilon ^2(-q^2) G_{0,2,2,0,1,1,1,0,0} -\epsilon ^3(-q^2) \relfbar \,   G_{0,1,1,0,1,1,2,0,0}\,,\\ 
f_{\IX,8}={}&\epsilon^3(1-6\epsilon) G_{1,0,1,0,1,1,1,0,0}\,,\\ 
f_{\IX,9}={}&\epsilon ^3(-q^2)\sqrt{\relfbar^2-1}\,   G_{1,1,0,0,1,1,2,0,0}\,,\\ 
f_{\IX,10}={}&\epsilon ^4(-q^2) (\relfbar^2-1) G_{1,1,1,1,1,1,1,0,0}\, ,\\
f_{\IX,11}={}&\epsilon ^3\sqrtmQSq G_{1,0,0,0,1,1,2,0,0}\,,\\ 
f_{\IX,12}={}&\epsilon ^3\sqrtmQSq G_{0,2,1,0,1,1,1,0,0}\,,\\ 
f_{\IX,13}={}&\epsilon ^3\sqrtmQSq G_{0,0,2,1,1,1,1,0,0}\,,\\ 
f_{\IX,14}={}&\epsilon ^4\sqrtmQSq\sqrt{\relfbar^2-1}  G_{1,0,1,1,1,1,1,0,0}\,,\\ 
f_{\IX,15}={}&\epsilon ^4\sqrtmQSq\sqrt{\relfbar^2-1}  G_{1,1,1,0,1,1,1,0,0}\,.
\end{align}
\end{small}

%================================================
\subsection*{Values of the master integrals in the Euclidean region}
%================================================
%
In this appendix we give the explicit values of the soft integrals in the Euclidean region up to order $\epsilon^2$. Higher orders in $\epsilon$ can be found in the ancillary files. For convenience, we choose to write all integrals in terms of the $x$ variable, c.f.~Eq.~\eqref{eq:y_and_x_def} and subsection \ref{subsec:1loop_soft_and_kinematics}, Fig.~\ref{fig:kinematics}. Even though not manifest in this representation, the integrals are purely real or imaginary (depending on whether or not the integral is normalized by $\sqrt{y^2-1}$) in the Euclidean region $x=e^{\imath\theta}\,,\theta\in (0, -\pi/2] \cup [\pi/2, \pi)$. Furthermore, all functions satisfy the first-entry condition \cite{Gaiotto:2011dt}, where only $x$ is allowed as first symbol \cite{Goncharov:2010jf,Duhr:2011zq,Duhr:2012fh} entry. 
\begin{align}
 f_{\text{H,1}}={}&\frac{\pi ^2 \epsilon ^2}{6}-1\,,\\
f_{\text{H,2}}={}&\frac{\pi ^2 \epsilon ^2}{24}-\frac{1}{4}\,,\\
f_{\text{H,3}}={}&1-\frac{\pi ^2 \epsilon ^2}{6}\,,\\
f_{\text{H,4}}={}&0\,,\\
f_{\text{H,5}}={}&0\,,\\
f_{\text{H,6}}={}&-\frac{1}{2} \epsilon  \log (-x)+ \epsilon ^2 \left(\text{Li}_2(-x)+ \text{Li}_2(x)-\frac{\pi^2}{12}\right.\\
&\left.+ \log (-x) \left[ \log (1-x^2)-\frac{1}{2} 
\log (-x) \right]\right)\,, \nn \\
f_{\text{H,7}}={}&\frac{1}{12} \epsilon ^2 \left(6 \log ^2(-x)+\pi ^2\right)+\frac{1}{2}\,,\\
f_{\text{H,8}}={}&\frac{1}{2} \epsilon  \log (-x)- \epsilon ^2 \left(\text{Li}_2(-x)+ \text{Li}_2(x) -\frac{\pi^2}{12} \right.\\
&\left.+ \log (-x) \left[\log (1-x^2)-\frac{1}{2} 
\log (-x) \right] \right)\,, \nn \\
f_{\text{H,9}}={}&0\,,\\
f_{\text{H,10}}={}&0\,,\\
f_{\text{H,11}}={}&-\frac{1}{2} \pi ^2 \epsilon ^2\,,\\
f_{\text{H,12}}={}&-\frac{1}{4} \pi ^2 \epsilon ^2\,,\\
f_{\text{H,13}}={}&\frac{\pi ^2 \epsilon ^2}{2}\,,\\
f_{\text{H,14}}={}&\frac{\pi ^2 \epsilon ^2}{4}\,,\\
f_{\text{H,15}}={}&\frac{\pi ^2 \epsilon ^2}{4}\,,\\
f_{\text{H,16}}={}&0\,,\\
f_{\text{III,5}}={}&-\frac{3}{4} \epsilon  \log (-x)\,,\\
f_{\text{III,6}}={}&-\frac{1}{6} \pi ^2 \epsilon ^2\,,\\
f_{\text{III,7}}={}&-\frac{1}{2} \epsilon ^2 \log ^2(-x)\,,\\
f_{\text{III,10}}={}&0\,,\\
f_{\text{IX,10}}={}&\frac{1}{4} \epsilon ^2 \log (-x) (\log (-x)+\imath \pi )\,,\\
f_{\text{IX,14}}={}&0\,.
\end{align}
The remaining functions are related to these as follows
\begin{align}
    f_{\RT,1}={}&f_{\IX,1}=f_{\H,1}\,,\\
    f_{\RT,2}={}&f_{\IX,5}=(f_{\IX,2}|_{x\to-x})=f_{\H,5}\,,\\
    f_{\RT,3}={}&f_{\IX,6}=(f_{\IX,3}|_{x\to-x})=f_{\H,6}\,,\\
    f_{\RT,4}={}&-f_{\IX,7}=(-f_{\IX,4}|_{x\to-x})=f_{\H,7}\,,\\
    f_{\RT,8}={}&-f_{\H,11}\,,\\
    f_{\RT,9}={}&f_{\IX,12}=(-f_{\IX,13}|_{x\to-x})=f_{\H,12}\,,\\
    f_{\IX,8}={}&f_{\RT,6}\,,\\
    f_{\IX,9}={}&f_{\RT,5}\,,\\
    f_{\IX,15}={}&f_{\RT,10}\,.
\end{align}

%================================================
%    References (& /Document)
%================================================
\bibliographystyle{JHEP}
\bibliography{jhep_refs.bib}
\end{document}